\documentclass[12pt,epsfig]{article}

\def\mathnew{\mathsurround=0pt}
\def\simov#1#2{\lower .5pt\vbox{\baselineskip0pt \lineskip-.5pt
       \ialign{$\mathnew#1\hfil##\hfil$\crcr#2\crcr\sim\crcr}}}
\def\simg{\mathrel{\mathpalette\simov >}}
\def\siml{\mathrel{\mathpalette\simov <}}

\def\msun{M_\odot}

\def\gr{$\gamma$-ray~}
\def\swift{\textit{Swift~}}
\def\fermi{\textit{Fermi~}}

\def\gbm{\textit{GBM~}}
\def\lat{\textit{LAT~}}

\def\vareps{\varepsilon}
\def\eps{\epsilon}

\def\noind{\noindent}

\def\beq{\begin{equation}}
\def\enq{\end{equation}}
\def\bea{\begin{eqnarray}}
\def\ena{\end{eqnarray}}
\def\bec{\begin{center}}
\def\enc{\end{center}}

\def\blist{\begin{list}{$\bullet$}{\itemsep 0.0in \parsep 0.0in}}
\def\elist{\end{list}}
\def\bitem{\begin{list}{\arabic{enumi}.}{\usecounter{enumi} \itemsep 0.0in \parsep 0.0in}}
\def\eitem{\end{list}}
\def\cm{\hbox{~cm}}
\def\s{\hbox{~s}}
\def\erg{\hbox{~erg}}

\def\GeV{\hbox{~GeV}}
\def\MeV{\hbox{~MeV}}
\def\keV{\hbox{~keV}}
\def\eV{\hbox{~eV}}

\def\part{\partial}

\def\mpl{m_{Pl}}

\def\h75{h_{75}}
\def\Omh75{\Omega h^2_{75}}

\def\fun#1#2{\lower3.6pt\vbox{\baselineskip0pt\lineskip.9pt
  \ialign{$\mathsurround=0pt#1\hfil##\hfil$\crcr#2\crcr\sim\crcr}}}
\def\VEV#1{\left\langle #1\right\rangle}

\def\jcap{Jour. Cosmology and Astro-Particle Phys.}
\def\apjs{Astrophys. J. Supp.}
\def\aj{Astron. J.}
\def\nat{Nature}
\def\na{New Ast.}
\def\nar{New~Ast. Rev.}
\def\nup{Nucl. Phys.}
\def\cmp{Comm. Math. Phys.}
\def\prl{Phys. Rev. Lett.}
\def\pl{Phys. Lett.}
\def\rmp{Rev. Mod. Phys.}
\def\ijmp{Int. Jour. Mod. Phys.}
\def\mpl{Mod. Phys. Lett.}
\def\pr{Phys. Rev.}
\def\prd{Phys.Rev.D}
\def\araa{Annu. Rev. Astron. Astrophys.}
\def\aap{Astron. Astrophys.}
\def\aaps{Astron. Astrophys. Supp.}
\def\aa{Astron. Astrophys.}
\def\pasj{Pub. Astr. Soc. Japan}
\def\physrep{Phys. Rep.}

\usepackage[dvipsnames,usenames]{color}
\usepackage{graphicx}
\usepackage{amssymb}
\usepackage{amsmath}
\usepackage{epsfig}
\usepackage{amssymb}
\usepackage{graphicx}
\usepackage{graphics}

\begin{document}

\setcounter{figure}{0}

\Large
\centerline{\bf Gamma Ray Bursts}
\centerline{\bf and their links with Supernovae and Cosmology}
\normalsize
\centerline{\bf Peter M\'esz\'aros$^1$ and Neil Gehrels$^2$}
\noindent
$^1${Center for Particle and Gravitational Astrophysics,
Dept. of Astronomy \& Astrophysics and Dept. of Physics,
Pennsylvania State University, University Park, PA 16802, USA}\\
\noindent
$^2${Astrophysics Science Division, NASA Goddard Space Flight Center,
Greenbelt, MD 20771 USA}

\begin{abstract}
Gamma-ray bursts are the most luminous explosions in the Universe,
whose origin and mechanism is the focus of intense interest. They
appear connected to supernova remnants from massive stars or the merger 
of their remnants, and their brightness makes them temporarily detectable
out to the larges distances yet explored in the Universe.  After pioneering
breakthroughs from space and ground experiments, their study is entering
a new phase with observations from the recently launched \fermi satellite,
as well as the prospect of detections or limits from large neutrino and 
gravitational wave detectors. The interplay between such observations and 
theoretical models of gamma-ray bursts is reviewed, as well as their 
connections to supernovae and cosmology.
\end{abstract}


\section{Introduction}
\label{sec:intro}

Roughly once a day, somewhere within our Hubble horizon a Gamma-ray burst (GRB) 
occurs, which for the next few seconds or tens of seconds completely overwhelms
the gamma-ray flux from the rest of the Universe, including the Sun. 
In fact, the GRB prompt electromagnetic energy output during tens of seconds 
is comparable to that of the Sun over $\sim {\rm few} \times 10^{10}$ years, 
or to that of our entire Milky Way over a few years; and their X-ray and
optical afterglow over the first day after the outburst can outshine the brightest 
quasars, as well as supernovae, making them potentially important probes of the 
distant Universe. Since the discovery of their X-ray afterglows by the {\it Beppo-SAX} 
satellite in 1997 and the subsequent detection of their optical counterparts, we 
have measured these objects out to the farthest cosmological distances. Thanks to
triggers and measurements from the \swift \cite{Gehrels+04swift} and \fermi
\cite{Michelson+10fermi} \cite{Meegan+09GBM} satellites, we have now detailed 
multi-wavelength data for many hundreds of bursts, and redshifts for over 200 
of them, and this data set will continue to grow with the continuation of 
\swift and \fermi, and the possible upcoming  {\it SVOM} mission \cite{Paul+11svom}.

GRBs are thought to arise either when a massive star ($\simg 25\msun$) undergoes 
core collapse, or possibly when a double neutron star or a neutron star and a 
black hole binary merges \cite{Woosley+06araa}. The first scenario applies to 
so-called {\it long} GRBs (LGRBs), whose $\gamma$-ray light-curve is lasts for 
$t_\gamma\simg 2\s$, while the second scenario is the likeliest one so far for {\it short} 
GRBs, whose \gr light-curves last $t_\gamma \siml 2\s$ \cite{Kouveliotou+93} (for the 
latter, this short duration refers to photons at $\vareps\simg 100 \keV$; some ``short" 
bursts, at softer energies, have tails lasting as much as 100 s 
\cite{Gehrels+09araa,Vedrenne+09book}).
In either scenario, it appears inevitable that a compact core object of a few
to several solar masses forms, whose radius is of order of the Schwarzschild radius
for this mass, $r_g\sim 10^6 (M/3\msun)\cm$, over a timescale comparable to a few
dynamic (free-fall) times, which is likely to be a black hole. Accretion of
residual infalling gas leads, if the core is fast rotating (guaranteed for a binary),
to an accretion disk whose inner radius is $r_0\sim 3 r_g\sim 10^7\cm$, and the
typical variability timescale of accretion is $t_0\sim (2GM/r_0^3)^{-1/2}\sim
10^{-3}\s$.  The bulk of the gravitational energy, of order a solar rest mass
or $10^{54}\erg$ is, as in SNe, rapidly radiated as thermal neutrinos ($E_{\nu,th}
\sim 10\MeV$), and some amount is radiated as gravitational waves. A smaller fraction, 
of order $E_j\sim 10^{51}-10^{52} \erg$ is converted into a fireball of equivalent 
blackbody temperature $T_0\sim$ few MeV. This energy eventually emerges as the burst, in
the form of a jet. However, for purposes of the dynamics one can generally use (see below)
the isotropic equivalent energy $E_0=E_j(4\pi/\Omega_j) \sim 10^{53}E_{53}\erg$,
which with a nominal total burst duration $t_b\sim 10\s$ implies a nominal isotropic
equivalent luminosity $L_\gamma \sim E_0/t_b =10^{52}L_{52}\erg\s^{-1}$ (if most of
that energy is emitted as $\gamma$-rays). The number density of photons at $r_0$ is 
roughly given by $L_\gamma =4\pi r_0^2 cn_\gamma \vareps$ where $\vareps\sim kT\sim\MeV$,
and the ``compactness parameter (roughly the optical depth of a photon with energy
$\simg m_e c^2$ against $\gamma\gamma \to e^+e^-$ pair production) is
\beq
\ell' \sim \tau_{\gamma\gamma}\sim n_\gamma \sigma_T r_0
 \sim \frac{\alpha \sigma_T L_\gamma}{4\pi r_0 c\vareps} \sim 10^{15},
\label{eq:taugam1}
\enq
where $\sigma_T$ is the Thomson cross section and $\alpha$ is the fraction of the
luminosity above $m_e c^2$. This creates a fireball of gamma-rays,
electron-positron pairs and hot baryons, where most of the entropy and pressure is
in the photons and leptons. The optical depth is huge, and the radiation pressure
far exceeds gravity, so the fireball expands and becomes relativistic.
A simple lower limit on the expansion bulk Lorentz 
factor follows from the observations of photons up to $\simg$ GeV energies, in some
bursts.  Such photons, trying to escape the source, would collide against softer 
photons and pair produce, $\gamma+\gamma\to e^+ + e^-$, degrading the spectrum to 
$\siml 0.5\MeV$. However the pair production threshold is angle dependent, and 
pair production is avoided if
\beq
\vareps_1 \vareps_2 (1-\cos\theta) \leq 2 m_e^2 c^4.
\label{eq:ggthresh}
\enq
In a relativistically moving jet, causality implies that only photons within angles 
$\theta\siml 1/\Gamma$ can interact, so with $\vareps_1 \sim 30\GeV,~\vareps_2\sim \MeV$,
and $\cos\theta\sim 1-\theta^2/2$, we see that this implies
\beq
\Gamma \simg \sqrt{(\vareps_1 /m_e c^2)(\vareps_2 /m_e c^2)}/2 \simg 10^2.
\label{eq:Gammalim1}
\enq
A more general constraint on $\Gamma$ is obtained by considering the typical
photon spectral distribution  in GRBs, which is a broken power law ``Band" function 
$n(\vareps)\propto \vareps^{-\beta}~{\rm ph}\cm^{-3}\MeV^{-1}$, where $\beta\simeq 1$ 
or $\beta\simeq 2$ for $\vareps$ below or above a break frequency $\vareps_{br}\sim \MeV$
\cite{Band+93,Fishman+95cgro}. The $\gamma\gamma$ optical depth at each energy $\geq 
m_ec^2$ depends on the optical depth to target photons at $\leq  m_ec^2$ satisfying 
the threshold condition (\ref{eq:ggthresh}), in the jet comoving frame. This optical 
depth can be shown to be $\propto \Gamma^{-6}$, so for increasingly high $\Gamma$ 
the source becomes optically thin to increasingly higher energy photons. The result
is that typical GRBs, even if the highest energy photons observed are only 100 MeV,
require bulk Lorentz factors in excess of $\Gamma\sim 250$ \cite{Baring+97}.

If the entire burst energy is released in impulsively, injecting an energy
$E_0$ in a timescale $t_0$ inside a radius $r_0$, with the numbers comparable
to those above, the initial entropy per baryon is $\eta\sim E_0/M_0c^2$, where 
$M_0$ is the baryon load of the fireball. If the pressure is mainly due to radiation
and pairs, the inertia is due to baryons (``baryonic dynamics" regime) the bulk 
Lorentz factor initially accelerates as $\Gamma(r)\sim (r/r_0)$, e.g. \cite{Meszaros06}.
After the baryons have become non-relativistic in their own frame, the expansion
changes to a coasting behavior at a saturation radius $r_{sat}\sim r_0\eta$, and 
the fireball continues to expand freely with $\Gamma\simeq \eta\simeq$ constant. 
The observationally estimated values are $\Gamma\sim \eta\sim 10^2-10^3$,
so the baryon load is typically $10^{-5}-10^{-6}\msun$. The behavior is similar
if the energy and mass input is spread out over accretion times (i.e. outflow
feeding or ejection times) of $t_b\sim 10-100\s$, as inferred for ``long" GRB.
On the other hand, if the fireball pressure, or rather stress tensor, is dominated
by magnetic fields, the dynamic behavior is different; depending on the symmetries
of the fields, the acceleration behavior can range from $\Gamma\propto r^{1/3}$,
e.g. \cite{Drenkhahn02,Metzger+11grbmag,Meszaros+11gevmag} to $\Gamma\propto r^\eta$ 
where $1/3\leq \eta \siml 2/3$, at least when the outflow is one-dimensional
\cite{Komissarov+09maggrb,Narayan+10numjet}. This regime is referred to as 
magnetically dominated, or Poynting dominated dynamics.

In practice, the outflow is inferred to be jet-like, rather than isotropic, with
an  average solid angle $\VEV{\Omega_j}/4\pi\sim 1/500$ or $\VEV{\theta_j}\sim 1/30$
\cite{Frail+02collim,Gehrels+09araa}. In the case of core collapse (``long") GRBs, 
this can be due to the outer parts of the star providing a massive barrier, which is 
best pierced along the centrifugally lightened rotation axis, along which the fireball 
escapes.  The stellar envelope provides a sideways pressure which channels the jet. 
However, as long as the jet opening angle $\theta_j$ exceeds $1/\Gamma\sim 10^{-2}$
(i.e. $\simg 0.5^o$), which is generally the case, the expansion occurs as if
it were isotropic: causality prevents the gas to have any knowledge of what
happens outside an angle $1/\Gamma$. For compact binary mergers, the data on 
jet opening angles is much sparser, but the average value may not be too 
different \cite{Fong+12-111020sgrb}.

\section{Observations}
\label{sec:obser}

\begin{figure}[htb]
\begin{minipage}{0.7\textwidth}
\includegraphics[width=1.0\textwidth,height=4.0in,angle=0.0]{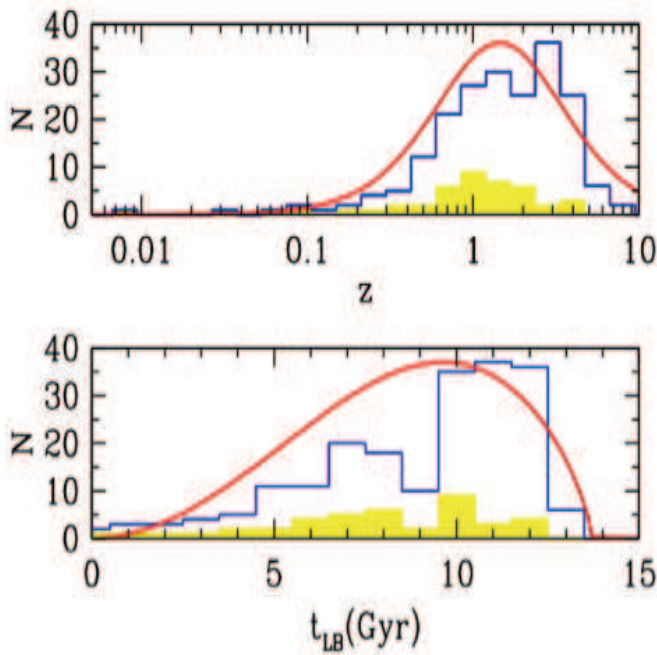}
\end{minipage}
\hfill
\hspace{2mm}
\begin{minipage}[t]{0.26\textwidth}
\vspace*{-1.5in}
\caption{Redshift distribution and cosmic look-back time of GRBs.
The Swift GRBs are in blue, the pre-Swift GRBs in yellow and the co-moving
volume of the universe is the red curve.  The GRBs roughly follow the co-moving
volume \cite{Gehrels+09araa}.  }.
\end{minipage}
\label{fig-Swift-z}
\end{figure}

The \swift mission, launched in November 2004, finds bursts and observes the prompt 
phase with the Burst Alert Telescope (BAT).  The afterglow is then observed with the 
X-Ray Telescope (XRT) and the UV Optical Telescope (UVOT).   Measurements of the redshift 
and studies of host galaxies are typically done with large ground-based telescopes which 
receive immediate alerts from the spacecraft when GRBs are detected.   Swift has, by far, 
the largest number of well-localized bursts, afterglow observations and redshift 
determinations.  As of 1 April 2012, BAT has detected 669 GRBs (annual average rate of 
$\sim$ 90 per year).  Approximately 80\% of the BAT-detected GRBs have rapid repointings 
(the remaining 20\% have spacecraft constraints that prevent rapid slewing).  Of those, 
virtually all long bursts observed promptly have detected X-ray afterglow.  Short bursts 
are more likely to have negligible X-ray afterglow, fading rapidly below the XRT 
sensitivity limit.  The fraction of rapid-pointing GRBs that have UVOT detection is 
$\sim 35\%$.  Combined with ground-based optical observations, about $\sim 60\%$ of 
\swift GRB have optical afterglow detection.   There are so far (mid 2012) about 
200 \swift GRBs with redshifts, compared with 41 in the pre-Swift era.
{The redshift distribution of \swift GRBs is shown in Fig. \ref{fig-Swift-z}.}

The \fermi mission, launched in June 2008, has two instruments, the Gamma-ray Burst Monitor 
(GBM) and the Large Area Telescope (LAT).  The GBM has scintillation detectors and covers 
the energy range from 8 keV to 40 MeV.  It measures spectra of GRBs and determines their 
position to $\sim 5^o$ accuracy.  The LAT is a pair conversion telescope covering the energy 
range from 20 MeV to 300 GeV.  It measures spectra of sources and positions them to an 
accuracy of $< 1^o$.   The GBM detects GRBs at a rate of $\sim$ 250 per year, of which on 
average 20\% are short bursts.  The LAT detects bursts at a rate of $\sim$ 8 per year.

\section{The photon spectrum: standard picture}
\label{sec:rad}

The expansion converts internal energy into bulk kinetic energy, so that the
gas cools in its own rest frame and soon becomes an inefficient radiator.
In the absence of dissipation, at the photospheric radius $r_{ph}$ where the flow 
becomes optically thin to scattering (which for baryon dominated dynamics usually 
occurs above the saturation radius $r_{sat}$) the escaping radiation would carry 
only a small fraction of the burst kinetic energy, and might be expected to have a 
quasi-black-body spectrum \cite{Paczynski86,Shemi+90}. This motivated the fireball 
shock model, where the bulk kinetic energy is reconverted by shocks into random 
particle energy, and thence into non-thermal radiation, at radii beyond the 
scattering photosphere, where the flow is optically thin; this could be at either 
external shocks \cite{Rees+92fbal,Meszaros+93impact}, where the jet interacts with 
external (e.g.interstellar) matter, or at internal shocks \cite{Rees+94is}
occurring within the jet, at radii intermediate between the photosphere and the
external shock radius. These radii can be expressed as
\bea
r_{ph} &\simeq (L\sigma_T /4\pi m_pc^3\eta^3) 
   \sim 4\times 10^{12} L_{\gamma,52} \eta_{2.5}^{-3}\cm, ~~~~~~~\cr
r_{dis}& \simeq  \Gamma^2 c t_v 
  \sim 3\times 10^{13} \eta_{2.5}^2 t_{v,-2}~\cm,~~~~~~~~~~~~~~~~~~~~~~\cr
r_{dec} & \simeq (3E_0/4\pi n_{ext}m_pc^2\eta^2)^{1/3}
  \sim 2 \times 10^{17}E_{53}^{1/2}n_0^{-1/2}\eta_2^{2/3}\cm.
\label{eq:3radii}
\ena
Here the photospheric radius assumes baryonic dynamics \cite{Rees+94is} (for magnetic 
dynamics see \cite{Meszaros+11gevmag}).  The dissipation (internal shock)
radius $r_{dis}$ follows from the relativistic relation between observer time
$t$, the radius $r$ and Lorentz factor, $r\simeq \Gamma^2 ct$. Considering two
shells of matter ejected at time intervals comparable to the variability timescale
of ejection $t_v$, with Lorentz factors differing by $\Delta\Gamma \sim \Gamma$,
and the deceleration radius $r_{dec}$ where external shocks start follows from the
energy conservation assumption, $E_0\simeq (4\pi/3)r_{dec}^3 n_{ext} m_pc^2 \Gamma^2$, 
when the swept up matter has been shock-heated to an energy comparable to the
explosion energy (with $\Gamma\sim\eta$, e.g. \cite{Meszaros06} for details).

In the (collisionless) shocks, the particles are reheated to thermal energies 
comparable to the pre-shock relative kinetic energies per particle. The internal 
shocks are semi-relativistic (since $\Delta\Gamma\sim \Gamma$, the relative Lorentz 
factor $\Gamma_{rel}=(1/2) [(\Gamma_1/\Gamma_2)+(\Gamma_2/\Gamma_1)]$ is 
semi-relativistic), and this results in a shock luminosity $L_{sh}=\eps_{sh}L_0$
where $L_0$ is initial kinetic luminosity $L_0=E_0\eta/t_b {\dot M}c^2$ and
the shock dissipation  efficiency $\eps_{sh}\siml 0.1$. Particles repeatedly
bounce across the shock, scattered by magnetic irregularities whose energy density 
can be assumed to build up to some degree of equipartition with the proton thermal 
energy,  so the comoving magnetic field is ${B'}^2/8\pi \sim \eps_B 4\Gamma_{rel}^2 m_pc^2$
(primed quantities are in comoving frame),
where $\eps_B\leq 1$. The repeated crossing Fermi accelerates the particles to a
relativistic power law \cite{Rees+92fbal,Rees+94is}; the minimum electron (comoving)
random Lorentz factor is $\gamma_{e,m} \sim \eps_e (m_p/m_e)\Gamma_{rel} \gg 1$
(for internal shocks $\Gamma_{rel}\sim 1$, while for external $\Gamma_{rel}\sim \eta$)
and one expects a power law $N(\gamma_e)\propto \gamma_e^{-p}$ above that, with
$p\sim 2-2.3$. One also expects something similar for the protons in the flow.

\noind
{\bf i) Internal shock prompt radiation}:
The electrons in the internal shock will emit synchrotron and inverse Compton (IC) 
radiation, leading to non-thermal broken power law photon spectrum, roughly similar to
the observed ``Band" spectra \cite{Meszaros+93impact,Rees+94is}. For reasonable
values of $L,\eta,t_v,\eps_B,\eps_e$ the synchrotron peak energy (observer frame)
corresponding to the minimum $\gamma_{e,m}$ are comparable to the observed
Band spectral break energies,
\beq
\vareps_{sy,m} \sim \vareps_{br} \sim 1~\eps_B^{1/2}\eps_e^{3/2}\frac{L^{1/2}_{\gamma,52}}
                                           {\eta^2_{2.5} t_{v,-2}}~\MeV.
\label{eq:esybr}
\enq
Eq. (\ref{eq:esybr}) assumes the randomized kinetic luminosity of internal shocks $L_{sh}$ 
to be related to their $\gamma$-ray luminosity through $L_\gamma=\eps_e L_s$, which 
is true in the fast cooling regime where cooling time is shorter than the
dynamic time $t'_{sync} \ll t'_{dyn}$ \cite{Sari+98spectra}, which in internal 
shocks is true \cite{Waxman97grbfbal}. In this fast cooling regime, for a typical 
Fermi electron index $p\simeq 2$ the photon spectral index above $\vareps_{br}$ is 
expected to be $n(\vareps) \propto \vareps^{-(p/2)-1}\propto \vareps^{-2}~ 
{\rm ph}\cm^{-3} \s^{-1}$, as typical for the high energy branch of the canonical 
Band spectrum. The synchrotron model predicts below $\vareps_{sy,m}$ a low energy branch
$n(\vareps)\propto \vareps^{-2/3}$, which through superposition of maxima for various
parameters could fit the observed Band average low energy branch $n(\vareps)\propto 
\vareps^{-1}$ \cite{Meszaros+93impact} (but flatter spectra are a problem see below). 

\noind
{\bf ii) External shock radiation and afterglow}: At $r_{dec}$ the relativistic ejecta has 
used up about half its initial energy in sweeping up an amount $M_{sw}\sim M_{ejecta}/\eta$
of external material, driving a forward shock into the external gas and a reverse shock
into the ejecta. This occurs at an observer time 
\beq
t_{dec,0}\sim r_{dec}/(2c\eta^2)\simeq  10~(E_{53}/n_0)^{1/3}\eta_{2.5}^{-8/3}~\s.
\label{eq:tdec0}
\enq
The forward shock is initially highly relativistic, $\Gamma_{sh}\sim \eta$, so
$\Gamma_{rel,fs}\sim\eta$ and the synchrotron spectrum is in the hard X-rays or 
gamma-rays. The reverse shock builds up slowly and for usual conditions becomes
semi-relativistic, $\Gamma_{rel,rs}\sim 1$ at the deceleration time, when it has 
crossed the ejecta. For this reason its $\gamma_{e,M}$ is smaller than that of
the forward shock electrons, and the reverse shock spectrum peaks in the optical
or UV \cite{Meszaros+93multi,Meszaros+97ag}.
Beyond $r_{dec}$ the expansion continues but it is increasingly slowed down due 
to increasing amount of swept up matter. In the adiabatic approximation the bulk 
Lorentz factor changes from being $\sim\eta\sim$ constant to a power law decline 
behavior given by $E_0\propto r^3 \Gamma^2$, or
\beq
\Gamma \sim \eta(r/r_{dec})^{-3/2}.
\label{eq:Gammaad}
\enq
In both the forward and the reverse shock one expects again Fermi acceleration
of electrons to a power law distribution leading to synchrotron and inverse Compton
radiation, but the synchrotron break energy becomes softer in time as the Doppler boost
decreases in accordance with eq.(\ref{eq:Gammaad}). This leads to an afterglow
\cite{Meszaros+97ag} progressing from X-rays through optical to radio lasting
from minutes to days to months, with fluxes decaying as power laws in time.
This prediction was indeed confirmed by observations with the Beppo-SAX satellite
of the X-ray \cite{Costa+97} and with ground telescopes of the optical 
\cite{Vanparadijs+97} afterglow of GRB 970228, soon followed by the first confirmation 
of a cosmological redshift \cite{Metzger+97} and a radio detection \cite{Frail+97}
for GRB 970508. The amount of data on, and understanding of, afterglows has since 
increased enormously, see e.g. \cite{Gehrels+09araa,Vedrenne+09book}. 

The external reverse shock gas, most luminous at $t\sim t_{dec}$, is in pressure
equilibrium with forward shock gas, and having a higher particle density and
smaller energy per electron than the forward shock, its synchrotron spectrum 
peaks in the O/UV. This was predicted to lead to an observable prompt optical 
emission \cite{Meszaros+93multi,Meszaros+97ag}, later detected with robotic ground 
telescopes such as ROTSE \cite{Akerlof+99} triggered by spacecraft, the number
of such detections being now several dozen \cite{Gehrels+09araa}.

The external forward shock is expected to also give rise to an IC component, in 
particular a synchrotron-self-Compton (SSC) from upscattering its own synchrotron 
photons \cite{Meszaros+93multi,Meszaros+94gev}, which would appear in the GeV range. 
Such GeV emission was detected already by EGRET \cite{Hurley+94-GRB940217}, and
more recently by the Fermi LAT, e.g. \cite{Abdo+09-080916}. This is discussed
in \S \ref{sec:gev}.

\section{Prompt MeV emission: issues and developments}
\label{sec:prompt}

Issues arise with the radiation efficiency of internal shocks, which is
small in the bolometric sense (5-10\%), unless the different shells have
widely differing Lorentz factors \cite{Spada+00is,Beloborodov00is,Kobayashi+01is}.
The MeV efficiency is also substantially affected by IC losses 
\cite{Papathanassiou+96is,Pilla+98is}, in the BATSE range being typically $\sim 1-5\%$, 
both when the MeV break is due to synchrotron \cite{Kumar99,Spada+00is,Guetta+01is} 
and when it is due to inverse Compton \cite{Panaitescu+00ssc}.

The synchrotron interpretation of the GRB radiation is the most attractive;
however, a number of effects can modify the simple synchrotron spectrum.
One is that the cooling could be rapid, i.e.  when the comoving synchrotron
cooling time $t'_{sy}= 9m_e^3 c^5/ 4e^4 B'^2 \gamma_e) \sim 7\times
10^8/B'^2\gamma_e ~{\rm s}$ is less than the comoving dynamic time
$t'_{dyn}\sim r/2c\Gamma$, the electrons cool down to
$\gamma_c= 6\pi m_e c /\sigma_T B'^2 t'_{dyn}$ and the spectrum above
$\nu_c\sim \Gamma (3/8\pi)(eB'/m_e c)\gamma_c^2$ is $F_\nu \propto \nu^{-1/2}$
\cite{Sari+98spectra,Ghisellini+00grb-emis}. 

The radiative efficiency  issue has motivated investigating various alternatives, e.g. 
relativistic turbulence in the emission region \cite{Narayan+09turb,Kumar+09turb}. This
assumes that relativistic eddies with Lorentz factors $\gamma_r\sim 10$ exist in
the comoving frame of the bulk $\Gamma\gtrsim 300$ flow, and survive to undergo at
least $\gamma_r$ changes over a dynamic time, leading both to high variability
and better efficiency. Various constraints may however pose difficulties
\cite{Lazar+09turb}, while numerical simulations \cite{Zhang+09turb} indicate  that
relativistic turbulence would lead to shocks and thermalization, reducing it to
non-relativistic.

The synchrotron spectral interpretation faces a problem from the observed distribution of 
Band low energy spectral indices $\beta_l$ (where $N_\vareps\propto \vareps^{\beta_l}$ 
below the spectral peak), which has a mean value $\beta_l\sim -1$, but for a fraction 
of bursts this slope reaches values $\beta_l > -2/3$ which are incompatible 
with a low energy synchrotron asymptote $\beta_l=-2/3$ \cite{Preece+00batse}. Possible 
explanations include synchrotron self-absorption in the X-ray \cite{Granot+01sync} or 
in the optical range up-scattered to X-rays \cite{Panaitescu+00ssc}, low-pitch angle 
scattering  or jitter radiation \cite{Medvedev00jitter,Medvedev06jitter}, 
or time-dependent acceleration \cite{Lloyd+02timeres}, where low-pitch angle diffusion 
might also explain high energy indices steeper than predicted by isotropic scattering. 

Pair formation can become important \cite{Rees+94is,Papathanassiou+96is,Pilla+98is}
in internal shocks or dissipation regions occurring at small radii, since a high 
comoving luminosity implies a large comoving compactness parameter $\ell' \gg 1$.
Pair-breakdown may cause a continuous rather then an abrupt heating and lead
to a self-regulating moderate optical thickness pair plasma at sub-relativistic
temperature, suggesting a comptonized spectrum \cite{Ghisellini+00grb-emis}. Copious 
pair formation in internal shocks may in fact extend the photosphere beyond the
baryonic photosphere value (\ref{eq:3radii}). Generic photosphere plus internal shock
models \cite{Meszaros+00phot,Meszaros+02phot,Ryde05} which includes the emission of a 
thermal photosphere as well as a non-thermal component from internal shocks outside of it,
subject to pair breakdown, which can produce both steep low energy spectra, preferred
breaks and a power law at high energies. A moderate to high scattering depth can
lead to a Compton equilibrium which gives spectral peaks in the right energy range
\cite{Peer+04waxb}. Pair enrichment of the outflow (due to back-scatter $\gamma\gamma$
interactions) can in general affect both the radiative efficiency and the spectrum
\cite{Madau+00pair,Thompson+00pair,Meszaros+01pair,Beloborodov02pair,Thompson06pair}.

\subsection{Photospheric models}
\label{sec:phot}

In the synchrotron interpretation the observed peak frequency is dependent on the bulk 
Lorentz factor, which may be random, and since the observed peaks appear to concentrate
near 0.2-1 MeV \cite{Gehrels+09araa}, the question can be posed whether this is indeed
due to synchrotron, or to some other effect. An alternative is to attribute a
preferred peak to a black-body at the comoving pair recombination temperature
in the fireball photosphere \cite{Eichler+00thermal}. In this case a steep low energy
spectral slope is due to the Rayleigh-Jeans part of the photosphere, and the
high energy power law spectra and GeV emission require a separate explanation
\cite{Meszaros+00phot}. A related explanation has been invoked \cite{Thompson94},
considering scattering of photospheric photons off MHD turbulence in the coasting
portion of the outflow, which up-scatters the adiabatically cooled photons up to 
the observed break energy and forms a power law above.

For a photosphere occurring at $r<r_{sat}$, which in a baryon-dominated model requires 
high values of $\eta$, the radiative luminosity in the observer frame is undiminished, 
since $E'_{rad}\propto r^{-1}$ but $\Gamma\propto r$ so $E_{rad}\sim$ constant, or 
$L_{ph}\propto r^2 \Gamma^2 {T'}^4 \propto$ constant, since $T'\propto r^{-1}$. 
However for the more moderate values of $\eta$ the photosphere occurs at $r>r_{sat}$, 
and whereas the kinetic energy of the baryons is constant $E_{kin}\sim E_0\sim$ constant 
the radiation energy drops as $E_{rad}\propto (r/r_{sat})^{-2/3}$, or $L_{ph}\sim L_0
(r_{ph}/r_{sat})^{-2/3}$ \cite{Meszaros+93impact,Meszaros+00phot}. This weakening
of the photospheric luminosity leads again to a lowered efficiency, as well as a lower
peak energy than observed. However, if the photosphere is dissipative (due to
shocks or other dissipation occurring at or below the photosphere) then a high
efficiency is regained, and the thermal peak photon energies are in the range of 
observed Band peaks \cite{Rees+05phot}.
An important aspect is that Compton equilibrium of internal
shock electrons or pairs with photospheric photons lead to a high radiative
efficiency, as well as to spectra with a break at the right preferred energy
and steep low energy slopes \cite{Rees+05phot,Peer+05peak,Peer+06phot}. It also leads 
to possible physical explanations for the Amati \cite{Amati+02episo} or Ghirlanda 
\cite{Ghirlanda+04corr} relations between spectral peak energy and burst fluence 
\cite{Rees+05phot,Thompson+07phot}.

\subsection{Magnetic models}
\label{sec:mag}

An alternative set of models for the prompt emission assume that this is due to
magnetic reconnection or dissipation processes, or else to the external shock.
Magnetic models fall into two categories, one where baryons are absent or dynamically 
negligible, at least initially \cite{Usov94,Drenkhahn02,Drenkhahn+02,Lyutikov+03grbmag},
and another where the baryon load is significant, although dynamically sub-dominant 
relative to the magnetic stresses \cite{Thompson94,Meszaros+94ext,Thompson06pair}.
These scenarios would in all cases still lead to an external shock, whose radius 
would be again given by $r_{dec}$ in equ. (\ref{eq:3radii}), with a standard forward 
blast wave, but possibly a  weaker or absent reverse shock \cite{Meszaros+94ext,
Meszaros+97ag}, due to the very high Alfv\'en (sound) speed in the ejecta.  For the 
same reason, internal shocks may be prevented from forming in magnetized outflows 
However, this depends on the magnetization parameter $\sigma$; if not too large.
reverse shocks \cite{Zhang+05magrev,Giannios+08revshock,Narayan+11magshock} or
internal shocks might still form \cite{Fan+04magis}, although with different
strengths and radiation characteristics. In fact, ``internal" dissipation  regions
may form due to magnetic reconnection, at radii comparable but differing from 
$r_{dis}$ of eq.(\ref{eq:3radii}), where electric fields due to reconnection (instead 
of a Fermi mechanism) leads to particle acceleration, and a high radiative efficiency 
is conceivable.

A hybrid dissipation model, entitled ICMART  \cite{Zhang+11icmart} 
involves a hybrid magnetically dominated outflow leading to semi-relativistic turbulent
reconnection. Here a moderately magnetized $\sigma=({B'}^2/4\pi\rho' c^2) \lesssim
100$ MHD outflow undergoes internal shocks as $\sigma \to 1$, leading to turbulence
and reconnection which accelerates electrons at radii $r\gtrsim 10^{15}$ cm. These
involve fewer protons than usual baryonic models, hence less conspicuous photospheres, 
and have significant variability, and the efficiency and spectrum are argued to 
have advantages over those in the usual synchrotron internal shock models. 

The baryon-free Poynting jet models resemble pulsar wind models, except for
being jet-shaped, as in AGN baryon-poor models.  The energy requirements of
GRB (isotropic-equivalent luminosities $L_\gamma \gtrsim 10^{52}$ erg s$^{-1}$)
require magnetic fields at the base in excess of $B\sim 10^{15}$ G, which can
be produced by shear and instabilities in an accreting torus around the black hole (BH).
The energy source can be either the accretion energy, or via the magnetic coupling 
between the disk and BH, extraction of angular momentum from the latter occurring via 
the Blandford-Znajek mechanism \cite{Blandford+77znajek}. The stresses in
in this type of model are initially magnetic, involving also pairs and photons,
and just as in purely hydro baryon-loaded models they lead to an initial Lorentz
factor growth $\Gamma\propto r$ up to a pair annihilation photosphere
\cite{Meszaros+97poynting}.  This provides a first radiation component, typically
peaking in the hard X-ray to MeV, with upscattering adding a high energy power law.
Internal shocks are not expected beyond this photosphere, but an external shock
provides another IC component, which reaches into the GeV-TeV range.  

The baryon-loaded magnetically dominated jets have a different acceleration
dynamics than the baryon-poor magnetic jets or the baryon dominated hydrodynamic
jets: whereas both the latter accelerate initially as $\Gamma \propto r$ and
eventually achieve  a coasting Lorentz factor $\Gamma_f \sim L_\gamma /{\dot M}c^2$,
the baryon-loaded magnetically dominated jets have a variety of possible
acceleration behaviors, generally less steep than the above. In the simplest
treatment of a homogeneous jet with transverse magnetic field which undergoes
reconnection,  the acceleration is $\Gamma\propto r^{1/3}$ \cite{Drenkhahn+02,
Meszaros+11gevmag}, while in inhomogeneous jets where the magnetic field and the
rest mass varies across the jet the average acceleration ranges from
$\Gamma\propto r^{1/3}$ to various other power laws intermediate between this
and $\Gamma \propto r$ \cite{McKinney+12magphot,Metzger+11grbmag}.
Few calculations have been made \cite{Giannios+07photspec} of the expected (leptonic)
spectral signatures in the simpler magnetized outflow photospheres, typically in 
a one-zone steady state approximation, showing that a Band-type spectrum can
be reproduced.

\section{GeV-TeV phenomenology and models}
\label{sec:gev}

\begin{figure}[htb]
\begin{minipage}{0.8\textwidth}
\includegraphics[width=1.0\textwidth,height=3.0in,angle=0.0]{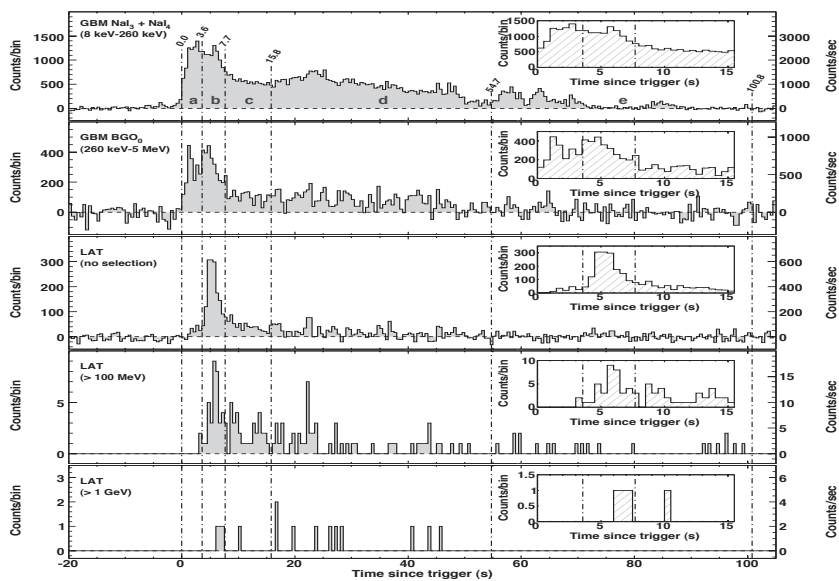}
\end{minipage}
\hspace{2mm}
\begin{minipage}[t]{0.17\textwidth}
\vspace*{-1.5in}
\caption{Light curves of GRB080916C showing the \gbm (top two curves) and 
the \lat (bottom three curves) energy ranges  \cite{Abdo+09-080916}.}.
\end{minipage}
\label{fig-080916-lc}
\end{figure}

The first \fermi  GRB observations, starting in late 2008, soon yielded a number of
surprises. One of the first bright objects showing radically new features was GRB 080916C 
\cite{Abdo+09-080916}, in which  the GeV emission started only with a second pulse,
which was delayed by $\sim 4$ s relative to the first pulse, which visible only in MeV
(Fig. \ref{fig-080916-lc}).

\begin{figure}[htb]
\begin{minipage}{0.75\textwidth}
\includegraphics[width=1.0\textwidth,height=3.0in,angle=0.0]{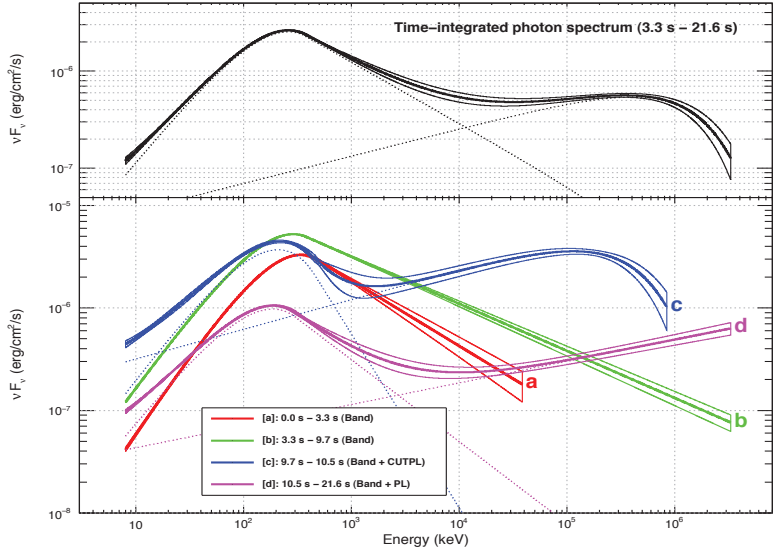}
\end{minipage}
\hspace{5mm}
\begin{minipage}[t]{0.2\textwidth}
\vspace*{-1.5in}
\caption{Spectra of GRB090926A from \fermi at four different time intervals,
a= [0.0-3.3s], b= [3.3-9.7s], c= [9.7-10.5s], d= [10.5-21.6s]
\cite{Ackermann+11-090926}.}
\end{minipage}
\label{fig-090926A-spec}
\end{figure}

The spectra of GRB 080916C consisted of simple Band-type broken power laws, the first
pulse having a soft high energy index disappearing at GeV, but the second and
subsequent pulses having harder high energy indices reaching well into the GeV range.
There was no evidence for a second spectral component (such as expected from
inverse Compton or hadronic effects).  The peak energy of the Band function evolved
from soft to hard and back to soft, but in this as well as in other \fermi LAT bursts,
the GeV emission persisted in afterglows typically lasting $\gtrsim 1000$ s.
On the other hand, in a few bursts, such as GRB090902B \cite{Abdo+09-090902},
a second spectral component did indeed appear, at $5\sigma$ significance, and also 
a lower energy power law extension whose significance is lower but suggestive, 
Another burst with a high energy second component was GRB 090926A 
\cite{Ackermann+11-090926}, this one showing a clear cut-off or turnover to the
high energy power law (Fig. \ref{fig-090926A-spec}).

A significant advance from \fermi LAT was the discovery of the first GeV {\it short} 
burst, GRB 090510 \cite{Ackermann+10-090510}, whose general behavior (including
a GeV delay) was qualitatively similar to that of long bursts. Several more
short bursts have been discovered since with the \fermi LAT.

The \fermi-LAT extended emission, if one ignores various details, 
has a relatively simple interpretation in terms of conventional forward 
shock leptonic synchrotron models (i.e. relying on accelerated electrons or 
$e^+e^-$ pairs) \cite{Ghisellini+10grbrad,Kumar+09gevfs}. Such models provide 
a natural delay between an assumed prompt MeV emission (assumed implicitly to
come from, e.g. internal shocks or other ``inner" mechanisms) and the GeV emission 
from the external shock, which starts after a few seconds time delay. 
However, taking into account more carefully the constraints provided by the 
Swift MeV and X-ray observations, and considering carefully  the accompanying
inverse Compton (IC) scattering and Klein-Nishina effects, it is clear that at least
during the prompt emission, there must be a subtle interplay between the shorter
lasting mechanism providing the MeV radiation and the mechanism or emission
region responsible for the bulk of the longer lasting GeV radiation 
\cite{Corsi+10-090510,DePasquale+09-090510ag,He+11-090510}. One general shortcoming
of these early studies was a postponement of addressing the interaction of the 
GeV emission with a specific, self-consistent model of the prompt emission, 
including the radiative inefficiency in an implicit internal shock assumption. 

\begin{figure}[htb]
{
\begin{minipage}{0.5\textwidth}
\includegraphics[width=1.0\textwidth,height=2.5in,angle=0.0]{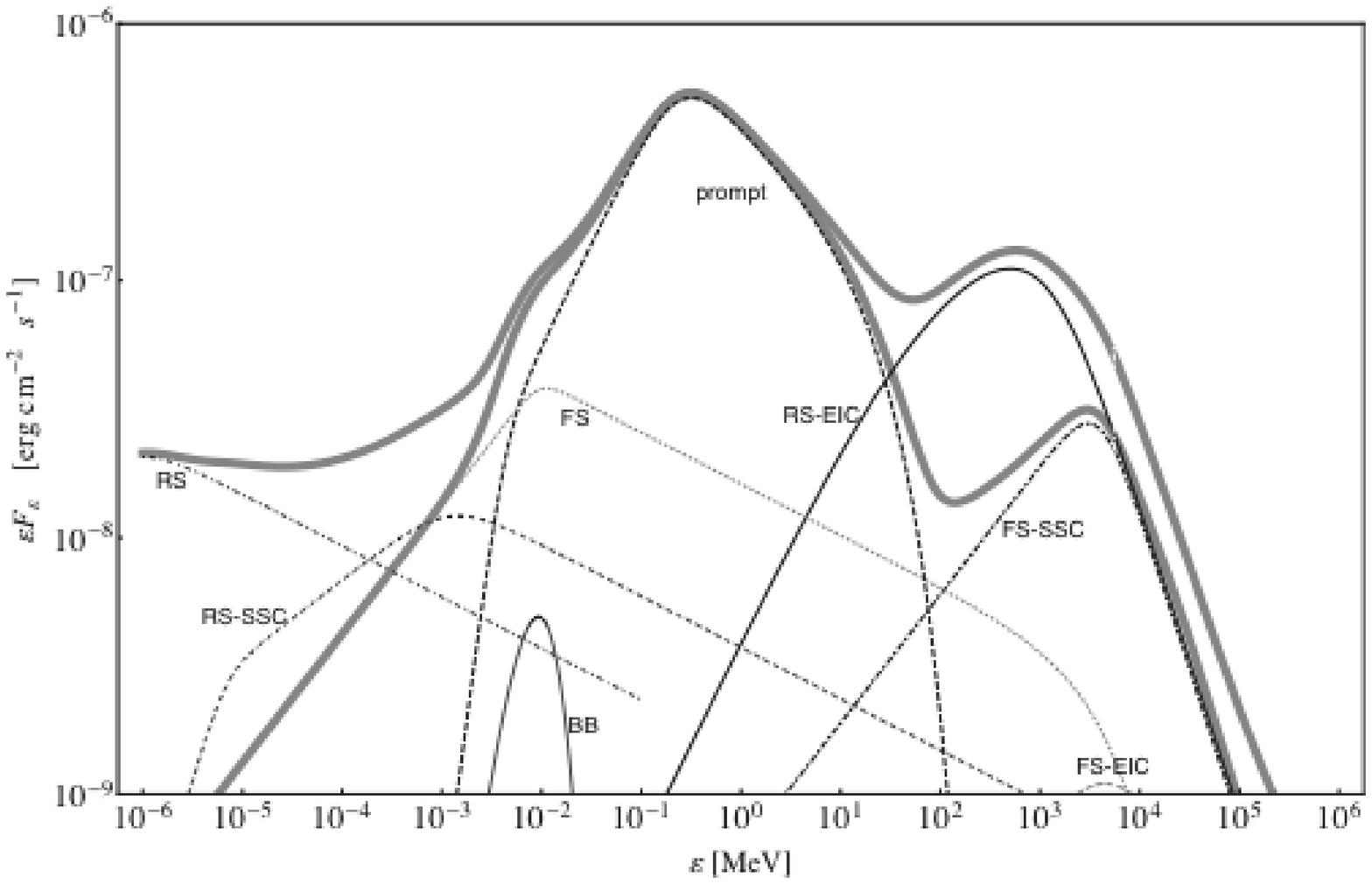}
\end{minipage}
\begin{minipage}[t]{0.5\textwidth}
\vspace*{-1.3in}
\includegraphics[width=1.0\textwidth,height=2.5in,angle=0.0]{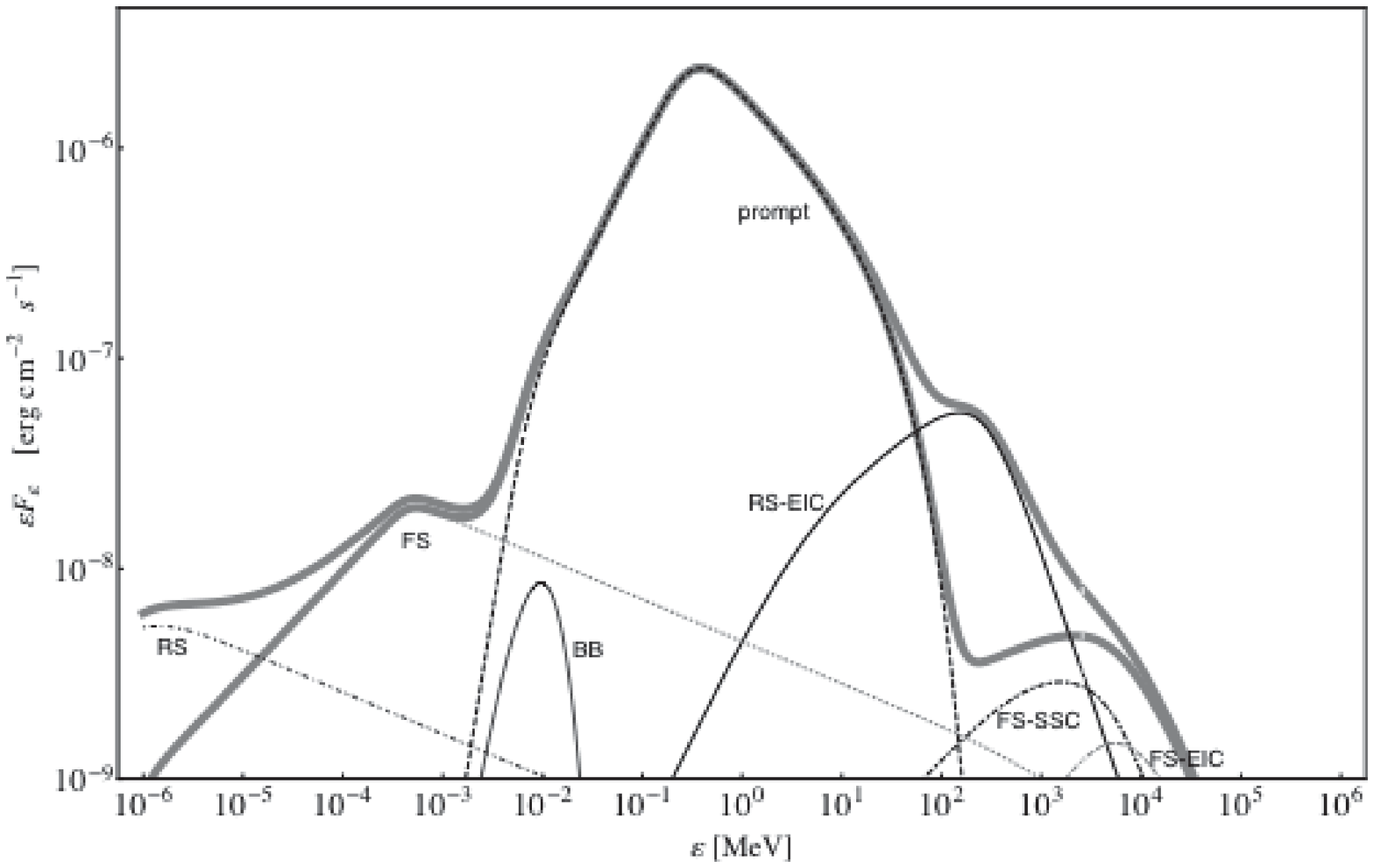}
\end{minipage}
\caption{A magnetically dominated leptonic model where the MeV Band spectrum is due 
to photospheric emission, there are no internal shocks, and the external reverse and
forward shock upscatter the MeV photospheric spectrum into the GeV range. Parameters 
are typical for \fermi LAT GRBs, but in some cases lead (left) to a two-component 
spectrum, while in others (right panel) it can be fitted as a single Band spectrum 
extending to the GeV range \cite{Veres+12mag}.}
\label{fig-ver12-two-one}
}
\end{figure}

A resolution of this problem is possible if the prompt MeV Band spectrum is due to 
an efficient dissipative photosphere (baryonic, in this case) with an internal shock
upscattering the MeV photons at a lower efficiency, giving the delayed GeV
spectrum \cite{Toma+11phot}. Alternatively, for a magnetically dominated outflow, 
where internal shocks may not occur, an efficient dissipative photospheric Band
spectrum can be up-scattered by the external shock and produce the observed
delayed GeV spectrum \cite{Veres+12mag}. 
Depending on the parameters, the combined
spectrum can look like a two-component or a single Band spectrum 
(Fig. \ref{fig-ver12-two-one}).
On the other hand, a delayed GeV spectrum can also be expected  in hadronic 
models, which assume the co-acceleration, along with the electrons, of protons 
which undergo electromagnetic cascades and synchrotron losses along with their 
secondaries \cite{Razzaque10-090510,Asano+09-090510,Murase+12reac,Asano+12grbhad}
see \S \ref{sec:had}.

The \lat data show that a fraction of GRB are emitting (in their own rest frame) 
photons in the energy range of at least up to $30-90$ GeV. A partial list of 
Fermi LAT detections \cite{Omodei+11fermigrb} of maximum observed photon 
energies and redshifts ($E_{\gamma,obs},z$) is 
$(13.2,4.35),(7.5,3.57),(5.3,0.74),(31.3,0.90),
(33.4,1.82)$, $(19.6,2.10)$, $(2.8,0.897),(4.3,1.37)$.
This list shows that (i) even $z>4$ bursts can produce $E_\gamma >10$
GeV photons at the observer, and (ii) some $z\sim 1$ bursts can produce $E_\gamma >
30$ GeV photons at the observer. 
This is highly encouraging for the planned large Cherenkov Telescope Array (CTA), 
as described in recent reviews \cite{Bouvier+11-CTA,Funk+12-CTA}. The CTA detection 
rate is estimated \cite{Bouvier+11-CTA} to be $0.7-1.6$ per year, based on
the rate of Swift triggers (while GBM triggers on Fermi are more frequent, their
positional accuracy is poorer). This rate is affected by uncertainties in the fraction 
of bursts which emit in the GeV range, relative to those emitting below 100 MeV
\cite{Guetta+11-latgbm,Beniamini+11-latgbm}. E.g., as of February 2011, in 2.5 years,
Fermi LAT detected 4 bursts at energies $>10$ GeV (or 20 at $>0.1$ GeV) out of
some 700 bursts detected by Fermi GBM at $E<100$ MeV. This very small fraction
of the total ($\lesssim 1\%$) of course is in part due to the size constraints
under which space detectors must operate. 

In the standard internal shock model of prompt emission, the intra-source
$\gamma\gamma$ absorption typically prevents photons in excess of a few GeV
to emerge \cite{Papathanassiou+96is,Pilla+98is}, unless the bulk Lorentz
factor is above $\sim 700$ \cite{Razzaque+04gev}.
For photospheric models of the prompt emission, e.g. \cite{Beloborodov10pn},
photons in excess of 10 GeV can escape the source from radii $r_{\gamma\gamma}
\sim 10^{15}$ cm, and such radii are also inferred phenomenologically from
one-zone analyses of the Fermi data on GRB. However, most of the GeV emission
occurs during the afterglow, which is good for ground-based TeV Cherenkov 
telescopes, whose reaction time can be slower. Indeed, the GeV emission can 
last up to $\sim 1000 \s$, far more than the $\sim 2-50 \s$ of the MeV emission.
In the standard external shock scenario, the compactness parameter is smaller
than in the internal shock, 
and inverse Compton scattering is expected to lead to multi-GeV and TeV photons
\cite{Meszaros+94ext,Meszaros+94gev}, the details depending on the electron
distribution slope and the radiative regime (e.g. slow or fast cooling). This
scenario is thought to be responsible for the afterglows of GRB \cite{Meszaros+97ag},
and is also thought to be responsible for the extended GeV emission observed
by LAT so far \cite{Ghisellini+10grbrad,Kumar+09gevfs,Wang+10kn,Zhang+11-latgrb}, etc.
Of course, propagation in the intergalactic medium from high redshifts leads to
additional $\gamma\gamma \to e^\pm$ interaction with the extragalactic background
light, or EBL  \cite{Coppi+97ebl,Finke+10ebl,Primack+11ebl}, the threshold for
which depends on the photon energy and the source redshift.

Thus, if TeV emission is produced, it is mainly expected to be detectable from
$z\lesssim 0.5$, while the 10-30 GeV emission should be (and is) detectable from 
higher redshifts. Thus, the GeV detectability is dictated by the source physics, 
the source rate and the immediate source environment. The source rate, based on 
MeV observations, is well constrained \cite{Gehrels+09araa}, while the near-source 
environment effects can be reasonably parametrized (e.g. \cite{Gilmore+10gamgam}). 
The source physics, however, has large uncertainties. E.g. in an external shock model 
the simple synchrotron self-Compton (SSC) model can be additionally complicated by
the scattering of photons arising at other locations well inside the external shock, 
e.g. from the photosphere \cite{Toma+11phot}, or from an inner region energized by 
continued central engine activity \cite{Wang+06gevflare}.  Similar uncertainties 
about the soft photon source and location would affect hadronic cascade models.  
The observed GBM high energy spectral slopes are in many cases steep enough 
not to expect much GeV emission from their extrapolation \cite{Omodei+11fermigrb},
while in other cases the LAT spectrum shows a cutoff or turnover, e.g. in GRB 090926B 
\cite{Ackermann+11-090926}. Nonetheless, all things considered, the estimate 
\cite{Bouvier+11-CTA} of $0.7-1.6$ CTA detections per year appears to be a conservative 
lower limit.

\section{Hadronic models}
\label{sec:had}

If GRB jets are baryonic, or magnetically dominated but with non-negligible baryon
load, the charged baryons should be co-accelerated with the electrons in any
shocks or reconnection zones, and hadronic processes would lead to both
secondary high energy photons and neutrinos. Monte Carlo codes have been developed
to model hadronic effects in relativistic flows, including $p,\gamma$ cascades,
Bethe-Heitler interactions, etc.  E.g. \cite{Asano+09grb,Asano+09-090510} used 
such a code to calculate the photon spectra from secondary leptons resulting from
hadronic interactions following proton acceleration in the same shocks that accelerate 
primary electrons in GRBs. The code uses an escape probability formulation
to compute the emerging spectra in a steady state, and provides a detailed
quantification of the signatures of hadronic interactions, which can be compared
to those arising from purely leptonic acceleration. Spectral fits of the \fermi \lat
observations of the short GRB 090510 were modeled by \cite{Asano+09-090510} as electron
synchrotron for the MeV component and photohadronic cascade radiation for the
GeV distinct power law component. 

\begin{figure}[htb]
{
\begin{minipage}{0.5\textwidth}
\includegraphics[width=1.0\textwidth,height=2.5in,angle=0.0]{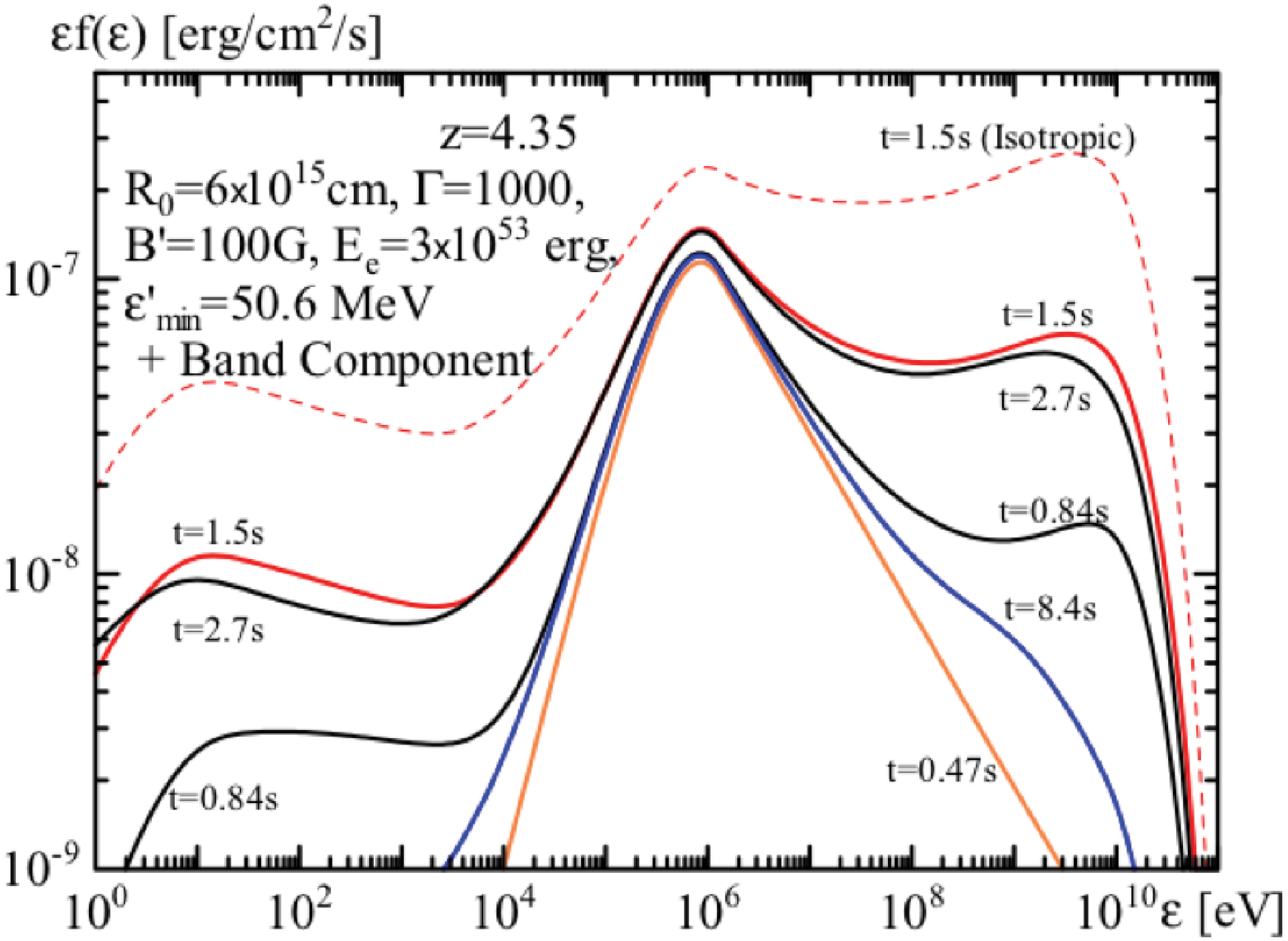}
\end{minipage}
\begin{minipage}[t]{0.5\textwidth}
\vspace*{-1.3in}
\includegraphics[width=1.0\textwidth,height=2.5in,angle=0.0]{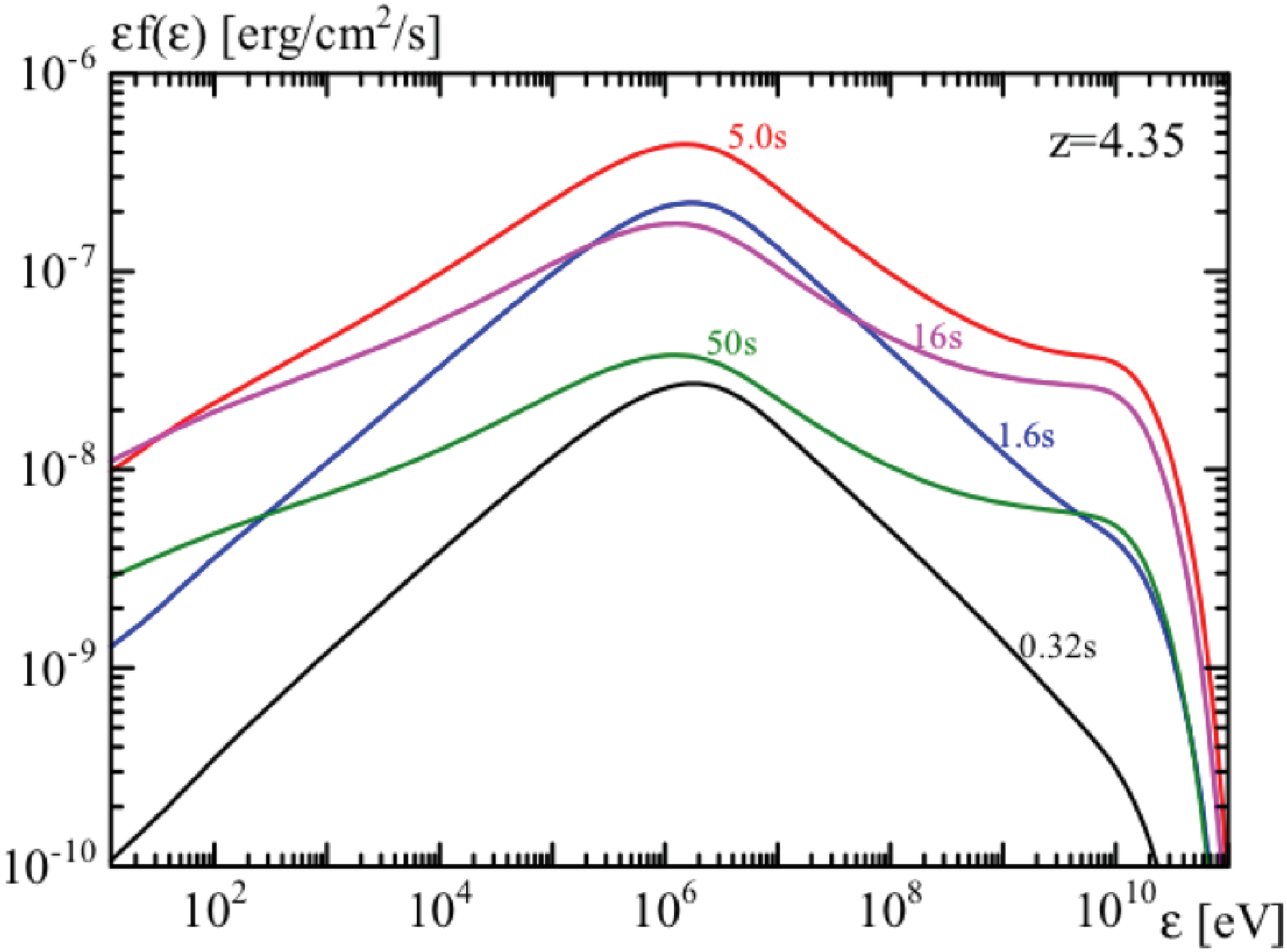}
\end{minipage}
\caption{Temporal evolution of the observable spectral photon flux for typical \fermi 
LAT parameters, from Monte Carlo simulations. Left: a purely leptonic two-zone model 
with a photospheric (MeV)  Band component and upscattering into the GeV range by a 
shock further out \cite{Asano+11grbtemp}. Right: a one-zone hadronic model, where
electron synchrotron produces the Band MeV spectrum and hadronic cascade secondaries
produce the GeV spectrum, as well as a low energy component \cite{Asano+12grbhad}.  }
\label{fig-asa11-12-lephadtime}
}
\end{figure}

Since acceleration as well as cascade development can take some time, in principle
even one-zone models might result in GeV-MeV photon delays. E.g.  \cite{Razzaque10-090510} 
assumes for GRB 090510 the prompt MeV to be electron synchrotron and the GeV to be 
proton synchrotron, whose cooling time cranks down the photon energy into the GeV range 
on a few second delay timescale, the electron plus proton synchrotron merging into a 
single Band function with the approximate spectral slope of the GeV photons.
A more recent one-zone hadronic calculation \cite{Asano+12grbhad} shows that even
when proton synchrotron is not important, hadronic cascade development leads to
a second GeV component, with time delays comparable to the observed ones
(Fig. \ref{fig-asa11-12-lephadtime}, right panel). Similar delays can, however,
be also obtained in purely leptonic two-zone photosphere plus external shock 
models \cite{Asano+11grbtemp} (Fig. \ref{fig-asa11-12-lephadtime}, left panel).

Hadronic interactions can also have implications for a low energy photon power law 
below the Band function, perhaps resulting in a GRB optical prompt flash, as
discussed by \cite{Asano+10optex}. For the usual Band MeV spectrum produced by 
conventional leptonic mechanisms, the acceleration of hadrons leads to secondaries 
whose radiation produces both a high energy ``extra" GeV component and a prompt 
bright optical emission from secondary synchrotron. This might explain, e.g. the 
observed ``naked eye" 5th magnitude flash of GRB 080319B, e.g. \cite{Racusin+08nakedeye}.

Hadronic binary collisions in baryon-loaded jets can  also be important, both for
efficient energy dissipation and for shaping the photon spectrum.
This is because the baryons will consist of both protons ($p$) and neutrons ($n$), 
especially if heavy elements are photo-dissociated. The protons are coupled to
the radiation during the acceleration phase but the neutrons are carried along only
thanks to nuclear ($p,n$) elastic collisions, whose characteristic timescale at some
point becomes longer than the expansion time. At this point the $p$ and $n$ radial
relative drift velocity $v$ approaches $c$, leading to the collisions becoming inelastic, 
$p+n \to \pi^+,\pi^0$, in turn leading to positrons, gamma-rays and neutrinos
\cite{Bahcall+00pn}. Such inelastic $(p,n)$ collisions can also arise in jets where
the bulk Lorentz factor is transversely inhomogeneous \cite{Meszaros+00gevnu},
e.g. going from large to small as the angle increases, as expected intuitively
from a jet experiencing friction against the surrounding stellar envelope.
In such cases, the neutrons from the slower, outer jet regions can diffuse
into the faster inner regions, leading to inelastic $(p,n)$ and $(n,n)$  collisions
resulting again in pions. An interesting consequence of either radial or tangential
$(n,p)$ drifts is  that the decoupling generally occurs below the scattering photosphere,
and the resulting positrons and gamma-rays deposit a significant fraction of the
relative kinetic energy into the flow, reheating it \cite{Beloborodov10pn}.
Internal dissipation below the photosphere has been advocated, e.g.
\cite{Rees+05photdis} to explain the MeV peaks as quasi-thermal photospheric
peaks \cite{Ryde+11phot,Peer11fermirev}, while having a large radiative efficiency. 
Such internal dissipation is naturally provided by $(p,n)$ decoupling, and numerical 
simulations \cite{Beloborodov10pn} indicate that a Band spectrum and a high
efficiency is indeed obtained, which remains the case even when the flow
is magnetized up to $\vareps_B =2$ \cite{Vurm+11phot}, while keeping the dynamics
dominated by the baryons. These numerical results were obtained for nominal cases
based on a specific radial $(n,p)$ velocity difference, although the phenomenon
is generic.

The photon spectral signatures of a magnetically dominated, baryon loaded
leptonic plus hadronic  GRB model involving nuclear collisions has been calculated
by \cite{Meszaros+11col}. This uses a realistic transverse structure of a fast
core-slow sheath. The analytical results indicate that the transverse neutron
collisions become most effective, resulting in GeV photons at radii from which
the observer-frame time delay relative to the photospheric MeV photons is
appropriate to explain the observed \fermi time lags.
The purely leptonic (SSC, EIC) time delays and spectral components of such
a baryon-loaded magnetic model, in the absence of drifts and transverse gradients, 
have been calculated  by \cite{Bosnjak+12delay}, leading to delays in the 
range observed by {\it Fermi}.

A hadronic model which attempts to self consistently produce the GeV radiation,
the MeV Band spectrum and the low energy (optical) power law is discussed in
\cite{Murase+11reac}. The  protons accelerated in the shocks or magnetic reconnection 
regions result in hadronic cascades which produce, as in \cite{Asano+10optex},
the GeV and optical power laws, while the cooled leptonic secondaries are 
re-accelerated via Fermi 2nd order mechanism in the MHD turbulent waves produced
by the same shocks or reconnection regions, leading self-consistently to an
MeV Band spectrum.

\section{Gravitational waves from GRB}
\label{sec:gw}

GRB may be also sources of gravitational waves (GWs). The most likely such sources  
are short GRBs \cite{Centrella11gwrev}, if these indeed arise from  merging compact 
objects \cite{Gehrels+09araa}. The rates in advanced LIGO and VIRGO may be 
at least several per year \cite{Leonor+09-ligogrb}. Long GRBs are more speculatively
as sources, since in the favored core collapse scenario the collapse may be more
chaotic \cite{Fryer+02gwcol}. They may, nonetheless, be weakly detectable as GW sources, 
especially if the core collapse breaks up into substantial blobs \cite{Kobayashi+03gwgrb},
or if they go through a magnetar phase leading to a bar \cite{Corsi+09mag}. More recent, 
detailed numerical calculations of collapsar (long) GRBs lead to GW prospects which 
range from pessimistic \cite{Ott+11gwcoll} to modest \cite{Kiuchi+11gwcoll}.

\section{High energy neutrinos from GRBs}
\label{sec:nu}

High energy ($10^9\eV \siml E_\nu \siml 10^{18}\eV$ neutrinos may be expected 
from baryon-loaded GRBs if sufficient 
protons are co-accelerated in the shocks \cite{Waxman11-crnu}. The most widely considered 
paradigm involves proton acceleration and $p\gamma$ interactions in internal shocks, 
resulting in prompt $\sim 100$ TeV HENUs \cite{Waxman+97grbnu,Murase+06grbnu}. Other 
interaction regions considered are external shocks, with $p\gamma$ interactions on 
reverse shock UV photons leading to EeV HENUs \cite{Waxman+00nuag}; and pre-emerging 
or choked jets in collapsars resulting in HENU precursors \cite{Meszaros+01choked}. 
Also, for baryonic dominated GRBs, a neutrino component may arise from
photospheric $p\gamma$ and $pp$ interactions \cite{Murase+08photonu,Wang+09photonu}.
An EeV neutrino flux is also expected from external shocks in very massive Pop. III
magnetically dominated GRBs \cite{Gao+11pop3nu}. Current IceCube observations
\cite{Ahlers+11-grbprob,Abbasi+11-ic40nugrb,Abbasi+12grbnu-nat} are putting significant 
constraints on the internal shock neutrino emission model, with data from the full 
array still to be analyzed.
One caveat is that, since the above analysis, several groups 
\cite{Hummer+11nu-ic3,Li11-nu-ic3,He+12grbnu} have recalculated the GRB internal shock 
neutrino production in greater detail, including multi-pion and Kaon production in the 
$p\gamma$ interactions, and allowing for various astrophysical uncertainties including
different values of the Lorentz factor and the accelerated proton to electron ratio
$L_p/L_e=1/f_e$. The conclusion from these revised calculations is that  the current
IceCube (IC40+IC59) measurements need to be extended for another 4 to 9 years
for obtaining a strong constraint.

\section{Progenitors and Supernova Connection}
\label{sec:prog}

Including the collimation correction, the GRB electromagnetic emission is
energetically quite compatible with an origin in, say, either compact mergers
of neutron star-neutron star (NS-NS) or black hole-neutron star (BH-NS) binaries 
\cite{Paczynski86,Eichler+89ns,Narayan+92merg,Meszaros+92tidal}, or with a core
collapse (hypernova or collapsar) model of a massive stellar progenitor
\cite{Woosley93col,Paczynski98,Macfadyen+99col}, which would be related to but 
much rarer than core-collapse supernovae \cite{Woosley+06araa,Gehrels+09araa}.
While in both scenarios the outcome could be, at least temporarily, a massive 
fast-rotating ultra-high magnetic field neutron star (a magnetar) 
\cite{Wheeler+00grbmag,Metzger+11grbmag}, the mass of the resulting central object
exceeds substantially the Chandrasekhar mass and is is expected to lead, sooner or 
later, to the formation of a central black hole. The latter will be fed through a 
(seconds to minutes) accretion episode from the surrounding disrupted
core stellar matter, which provides the energy source for the ejection of
relativistic matter responsible for the radiation. This inference appears
to be confirmed by numerical simulations, for NS-NS or NS-BH mergers 
\cite{Ruffert+99merg,Rosswog05nsbh,Rezzolla+11sgrb-mag} as well as for
collapsar models \cite{Macfadyen+99col,Macfadyen+01sncol,Zhang+04jetnum}.

The above numerical  simulations also indicate that (1) the compact object merger 
accretion disks are less massive and the accretion episode (when the  disk is not 
highly magnetized) lasts less than a few seconds, compatible with the observations of 
the canonical short GRBs (SGRBs); and (2) in the collapsar models the accretion
lasts for tens of seconds or more, compatible with the durations of canonical
long GRBs (LGRBs). 

The observations of LGRBs indicate that they are generally located in active 
star-forming environments, usually in blue, small or not too massive, gas-rich 
galaxies \cite{Paczynski98grbsfr,Woosley+06araa,Gehrels+09araa}, which is where one 
expects massive stars to be present. Progenitor stars more massive than  $\sim 25-28\msun$, 
following core collapse, are expected to result in a black hole (BH) central remnant, 
either directly or through an intermediate neutron star (NS) phase 
\cite{Fryer+99,Fryer06col}.
Such BH core collapse events, if the core is sufficiently fast rotating, can lead
to a fall-back fed accretion disk powering a relativistic jet, which is able to escape 
a star of $\sim 10^{11}\cm$ \cite{Macfadyen+01sncol,Zhang+04jetnum,Woosley11prog}. This 
radius corresponds to those of Wolf-Rayet (WR) stars, which are thought to arise from more
massive $M>25\msun$ progenitors, whittled down by wind mass loss prior to core collapse.
The high rotation rate, which favors the wind mass loss and also the formation of a
longer lasting accretion disk, is thought to be enhanced when the star arises
in a metal-poor environment \cite{Woosley+06araa}, which in fact seems to
characterize many LGRB host galaxies \cite{Stanek+06,Gehrels+09araa}.

The massive core collapse model of LGRBs is confirmed by the fact that
LGRBs are, in some cases, demonstrably associated with type Ib/c supernovae, 
whose explosion is, to within errors, contemporaneous with the GRB 
\cite{Galama+98-980425,Hjorth+03-030329sn,Dellavalle11grbsn,Hjorth+11grbsn}.
The SNe Ib/Ic are generally thought to have WR progenitors, whether they are
associated with GRB or not; only a small fraction of order few \% of SNe Ib/c 
appear to be associated with LGRBs \cite{Soderberg+07sne,Soderberg+10hn,Hjorth+11grbsn}.
However, the SNe Ib/c associated with GRBs, as well as a good fraction of those
not associated, are classified as hypernovae (HNe), meaning that they have unusually
broad spectral lines indicating a semi-relativistic envelope ejecta ($v/c\sim 1$),
and inferred isotropic energies $E_{HN,iso}\sim 10^{52.5}\erg$, as opposed to the 
average SNe with $v/c\sim 0.1$ and $E_{SN,iso}\sim 10^{51}\erg$ \cite{Paczynski98grbhn,
Waxman+99-1998bw,Nomoto+10hn,Thielemann+11sn,Lazzati+12hn}.

For short gamma-ray bursts (SGRB) the most widely favored candidates are mergers 
of neutron star binaries (DNS) or neutron star-black hole (BH) binaries,
which lose orbital angular momentum by gravitational wave radiation and undergo a merger.
This second progenitor scenario has only now begun to be tested thanks to the Swift
detection of short burst afterglows \cite{Gehrels+05sgrb,Berger+05-050724,Berger+06sgrb,
Nakar07sgrb,Berger+07sgrb}. The SGRBs are found both in evolved (elliptical) galaxies 
\cite{Gehrels+05sgrb} and in galaxies with star formation 
\cite{Gehrels+09araa,Berger11sgrb}, in proportions compatible with that 
expected for an old population such as neutron stars.
While neutron stars are expected, and found, in young star-forming galaxies,
massive young stars are not expected, and not found in old population ellipticals.
And, indeed, no supernova has ever been found exploding at the same time and location
as a SGRB \cite{Gehrels+09araa}. Of course, neutron stars are the product of supernovae, 
which can have occurred much earlier than the burst, from progenitors whose initial mass 
was is $8 \msun \siml M_\ast \siml 25\msun$. A kick is imparted to the NS at the 
supernova event, so the NS can wander off significant distances from its birth site
(many Kpc, even outside the host galaxy). If the NS was born in a binary and/or later 
became a binary, the time between the initial explosion and the eventual merger can 
range up to $10^8-10^9$ years, and is very unlikely to be less than $10^6$ years 
\cite{Coward+12dnsmerg,Kiel+10dns-col-popsynt}. Only a handful of SGRBs have yielded
reliable lightcurve breaks suitable for determining a jet opening angle. The latest
measurements and comparison to previous data \cite{Fong+12-111020sgrb} indicate an
an average $\theta_j\sim 5^o$ (comparable to the LGRB average value, although
there is one outlier at $25^o$). This is interesting, since for DNS mergers there is no 
stellar envelope (as for LGRBs) to provide collimation, at most there would be a wind; 
however, such narrow jets would be compatible with twisted or hoop-stress collimated 
MHD jets from DNS mergers \cite{Rezzolla+11sgrb-mag,Shibata+11sgrb-mag}.

\section{High redshift GRBs}
\label{sec:hiz}

Long GRBs are astonishingly bright, both in gamma-rays and at longer wavelenghts.
In the optical, typical brightnesses are $\sim 18$th magnitude a few hours after the 
trigger (and some have been detected in the 5th-10th magnitude range seconds after 
trigger), while a Milky-Way-type galaxy has $\sim 32$nd magnitude at a redshift $z=8$. 
In fact, GRBs vie with galaxies for the record on the highest confirmed redshift 
measurements, e.g. GRB080913 at $z=6.7$ \cite{Greiner+09-z6.7}, GRB090423 at $z=8.2$ 
\cite{Tanvir+09-z8.2,Salvaterra+09-z8.2} (through spectroscopy), while GRB 090429B 
has a photometric redshift of $z\simeq 9.4$ \cite{Cucchiara+11-z9}.  It is possible 
that even much more distant objects than these have already been detected in the 
gamma-ray and X-ray detectors of {\it Swift} and {\it Fermi}, although for such $z>9$
objects a specific (optical/IR or other) redshift signature is extremely difficult and
noisy, so redshift diagnostics are increasingly harder to obtain in this range.  
The above discoveries do, however, indicate that the prospect of eventually reaching 
into the realm of  Pop. III objects is becoming increasingly realistic.  

Population III  GRBs at $10\siml z \siml 20$ may result, as they do at lower redshifts,  
from massive, $M>25-30\msun$ metal-poor stars whose core collapses  to a black hole
\cite{Bromm+06hizgrb}.
However, the mass of Pop. III stellar progenitors could be as high as $\sim 1000\msun$,
leading to $100-500 \msun$ black holes \cite{Komissarov+10pop3}, although the Pop. III
masses are a subject of debate (and could be much lower \cite{Stacy+09pop3,Norman10pop3}.  
For extremely massive black holes, the jets are likely to be 
Poynting-dominated, e.g. powered by the Blandford-Znajek mechanism. The expansion 
dynamics and the radiation arising from such very massive Poynting jet GRBs was discussed 
by \cite{Meszaros+10pop3}. At typical redshifts $z\sim 20$ this implies a ``prompt"
emission extending to $\lesssim$ 1 day which should be detectable by \swift or
\fermi, being most prominent initially around 50 keV due to the jet pair photosphere,
followed after a similar time interval by an external shock synchrotron component
at a few keV and an inverse Compton component at $\gtrsim$ 70 GeV \cite{Toma+11pop3}.
Both the `prompt' emission and the longer-lasting afterglows \cite{Toma+11pop3}
of such Pop. III GRBs should be detectable with the BAT or XRT on {\it Swift} or
the GBM on {\it Fermi}. On {\it Swift}, image triggers may be the best way to detect
them, and some constraints on their rate are provided by radio surveys.
They are expected to have GeV extensions as well, but redshift determinations need 
to rely on L-band or K-band spectroscopy.

\section{The Cosmology Connection}
\label{sec:cosm}

Since GRBs are seen out to the largest redshifts yet measured, and for periods
of hours to days they can outshine any other objects at those distances, their
potential usefulness as tools for cosmology has been intriguing for some time. 

The simplest way, to use them as distance markers, is unfortunately not straightforward. 
This is because they are not good ``standard candles", which could be used e.g.
in a Hubble-type diagram to compare flux against redshift to compare against 
cosmological models to deduce a closure parameter, or an acceleration rate. Even
the collimation-corrected average gamma-ray energy $E_{\gamma,j}\sim 10^{51}$ ergs 
has too much variance to be of direct use as a standard candle. The error is still
almost twice as large as that obtained from SNIa, at least at redshifts $z\siml 1.5$,
which is the most important region for dark energy studies. The hope is, still, that
one or more of the various empirical correlations between observed spectral quantities 
could lead to a calibration, as the Phillips relation does for supernovae, which 
could turn GRBs into an effective distance ruler.

One such empirical correlation is between the photon spectral peak energy $E_{pk}$ and 
the apparent isotropic energy $E_{iso}$ \cite{Amati+02episo,Amati+06episo}, namely
$E_{pk} \propto E_{iso}^{\alpha}$ with $\alpha\sim 1/2$ (Amati relation).
Other correlations are between $E_{pk}$ and the peak luminosity $L_{pk}$ 
\cite{Yonetoku+04-EpkLum} (Yonetoku relation); or for bursts where the jet opening 
angle is known, between $E_{pk}$ and the collimation corrected gamma-ray energy of 
the jet $E_{\gamma,jet}$ \cite{Ghirlanda+04-EpkEjet,Ghirlanda07grbcosm} (Ghirlanda
relation); or between $E_{pk}$ and $L_{\gamma,jet}$ \cite{Dai+04cosm};
or between $E_{pk}$, $E_{iso}$ and the light curve break time $t_{br}$ 
\cite{Liang+06-EisoEpktbrk} (Liang-Zhang relation); or between the X-ray luminosity
at break time $L_{X,br}$ and the break time $t_{br}$ \cite{Cardone+10grbcosm}. 
Of course, these correlations are of interest in themselves as possible constraints 
on the radiation mechanism or emission region, and various interpretations have been
made, e.g. \cite{Zhang+02break,Rees+05photdis,Thompson+07phot,Ghirlanda+12grbcorr}. 
However, for cosmology only the tightness of the empirical correlation is what 
matters. Observational biases could, of course, pose problems \cite{Nakar+05corr,
Ghirlanda+12grbcosm}, and circularity issues may be a concern; the latter, however 
can in principle be minimized by restricting oneself only to directly observed 
quantities \cite{Graziani11grbcosm}.

While such correlations make it possible to extend Hubble diagrams up to redshifts 
$z\siml 8-9$ with GRBs, their usefulness for deriving cosmological parameters is 
still limited.  The main reason is that the matter (dark plus baryonic) 
and the vacuum energy densities evolve as $\rho_m \propto (1+z)^3,~ 
\rho_V \propto (1+z)^{3(1+w)}$, respectively,   where $w$ is the equation of state 
multiplier, $p=w\rho$, for vacuum $w=-1$ being the currently favored value. At present 
the vacuum energy dominates the dynamics, $\Omega_{V,0}=(\rho_{V,0}/\rho_{cr,0}\simeq 
0.7$ while $\Omega_{m,0}\simeq 0.3$. Going back in time or up in redshift, the vacuum 
energy grows slowly (or not at all, if $w=-1$), while the matter energy grows fast, so
that for redshifts $z > z_V=[(\Omega_{V,0}/\Omega_{m,0})^{-1/3w} -1] ~ \sim~ 0.5$
the vacuum energy becomes negligible compared to that of matter. Thus, the information 
about the vacuum-related acceleration and a vacuum equation of state is gleaned mainly 
from data at $z\siml 0.5$, where the SNIa with redshifts greatly outnumber the GRBs 
with redshift.
Some recent papers using GRBs in Hubble diagrams are, e.g. \cite{Liang+10grbcosm}
and \cite{Demianski+11grbcosm}, the latter including 109 GRBs with known redshifts
calibrated with 567 SNIa.  While strongly suggestive, and increasingly interesting,
the sample is still small compared to SNe Ia, especially at low $z$, and the dispersion 
remains larger then for SNe, so the usefulness of GRBs as statistical distance 
indicators remains to be seen. 
\begin{figure}[htb]
\begin{minipage}{0.63\textwidth}
\includegraphics[width=1.0\textwidth,height=3.0in,angle=0.0]{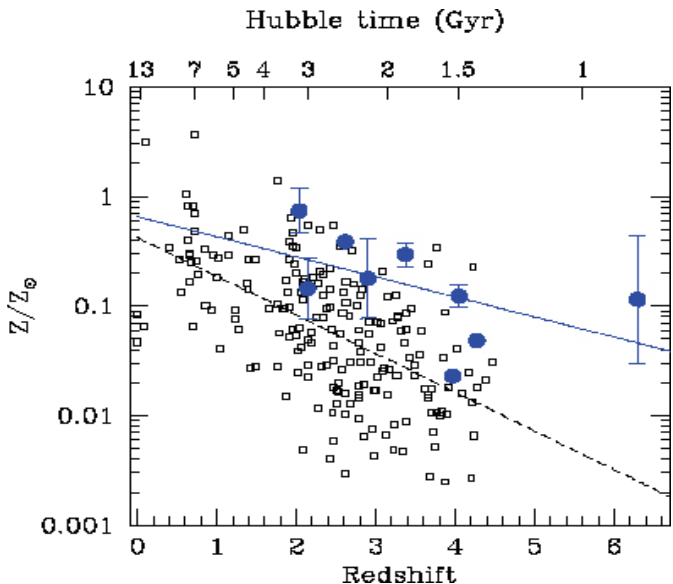}
\end{minipage}
\hfill
\begin{minipage}[t]{0.32\textwidth}
\vspace*{-1.5in}
\caption{Redshift evolution of the metalicity relative to solar values, for
GRBs shown with blue dots and QSOs show with open circles.  The GRB metalicity
is on average $\sim$5 times larger than in QSO. These are based on damped Lyman alpha
(DLA) spectral features.  The upper horizontal x-axis indicates the age of the
Universe (Hubble time) \cite{Savaglio06chem}.  }
\end{minipage}
\label{fig-Metalicity}
\end{figure}
GRBs, however, are likely to be unique as beacons for probing the high
redshift Universe. They are detectable with current gamma-ray, X-ray and
infra-red detectors out to distances corresponding to the earliest stars 
formation epoch \cite{Lamb+00hiz,Ciardi+00hizgrb}, and they may provide
possible redshift signatures \cite{Meszaros+03gauge,Gou+04hiz} extending
into the $10\siml z\siml 20$ range. Their strong, featureless power law continuum 
spectrum shining through the intergalactic medium and intervening young galaxies
or protogalaxies provide a sensitive probe of the ionization state,
velocity distribution and chemical composition at those redshifts 
\cite{Loeb+01reioniz,Savaglio10grbchem,Hartmann10grbchem}. 
An example is shown in Fig. 6, 
indicating the change of
the metalicity (given by the oxygen to hydrogen abundance ratio) as a function of
redshift. These abundances are determined from spectral absorption lines in
the continuum radiation of GRBs and quasars. The GRB lines come mainly from the
host galaxy gas in the star forming region where the explosion occurred, while
the quasar lines arise in random intervening galaxies along the line of sight.
One sees that the GRB provide information out to higher redshifts than quasars,
and indicate systematically higher metalicities. This is expected, since the GRBs
are sampling gas in the star forming regions, which have been enriched by
nucleosynthesis and supernova explosions.

\begin{figure}[htb]
\begin{minipage}{0.6\textwidth}
\includegraphics[width=1.0\textwidth,height=3.5in,angle=0.0]{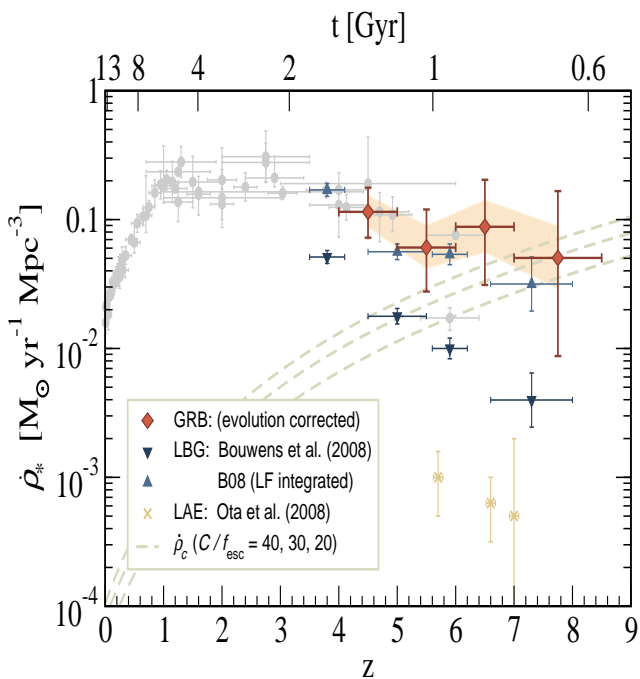}
\end{minipage}
\hfill
\hspace{2mm}
\begin{minipage}[t]{0.36\textwidth}
\vspace*{-1.8in}
\caption{Cosmic star formation history \cite{Kistler+09sfrgrb}. Shown are
the data compiled in \cite{Hopkins+06sfr} (light circles) and from
Ly-$\alpha$ emitters (LAE) \cite{Ota+08reioniz}. Also recent LBG data for two 
UV LF integrations: down to $0.2 L_\ast$ (down triangles, \cite{Bouwens+08hiz}),
and complete up to $z=3$ (up triangles).  Swift GRB-inferred rates are diamonds, 
the shaded band shows the range resulting from varying evolutionary parameters.  
Also shown is the critical $\rho_\ast$ from \cite{Madau+99reioniz} for $C/f_{esc} 
= 40, 30, 20$ (dashed lines, top to bottom), see also \cite{Robertson+12sfr}.
}
\end{minipage}
\label{fig-SFR}
\end{figure}
GRBs also provide an excellent tool for investigating the cosmic star formation
rate (SFR) of the high redshift universe, and thereby also the rate of large scale
structure (LSS) formation, out to (so far) redshifts in the 8-10 range. This is
exemplified in Fig. 7, 
which shows the star formation rate determined
through various techniques. The long GRBs are the endpoints of the lives of massive 
stars, and their rate is therefore approximately proportional to the star formation
rate in general. However, at high redshifts the rates are very uncertain, and may
be subject to various observational biases. There may also be evolutionary biases,
such as a dependence of the long GRB formation on the metalicity of the host galaxy,
which needs to be taken into account, e.g. \cite{Savaglio10grbchem,Hartmann10grbchem}.

The most distant GRBs may also provide the only possible probes of the era of 
the first generation of (Population III) stars formed in the Universe 
\cite{Komissarov+10pop3,Meszaros+10pop3,Toma+11pop3,Campisi+11pop3grb}. These
relics of the infant universe could be the most sensitive probes of the redshift
for the start of large scale structure formation, with significant implications for
the properties of the dark matter.

\section{The future}
\label{sec:future}

Both \swift and \fermi are likely to be functional and return GRB data for many years 
to come.  They have orbital lifetimes extending beyond 2025 and no critical expendables.  
The {\it SVOM} mission \cite{Paul+11svom} is an approved Chinese-French mission to 
observe GRBs and their afterglow.  It has a wide-field instrument to image the bursts 
and one to study the spectrum.  The spacecraft rapidly slews like \swift to point X-ray 
and optical telescopes for afterglow observation.  There are other concepts in 
consideration, such as {\it Lobster} \cite{Gehrels+12-lobster}, which performs the 
wide-field observations in the X-ray band suitable for high-redshift GRBs.
In addition, combined with such electromagnetic detection programs, increasingly
sensitive multi-messenger detection attempts will continue to be pursued using 
high energy neutrinos (\S \ref{sec:nu}) and gravitational waves (\S \ref{sec:gw}). 
These expanded observational efforts will require more detailed theoretical 
interpretation and models, extending well beyond what has been achieved so far.
Based on past experience, GRBs may be counted on to provide further exciting
surprises during the next decade.

\bigskip
\noindent
{\bf Acknowledgment:} We are grateful to NASA for support, and to our colleagues
for many useful discussions.

%


\begin{thebibliography}{100}
\providecommand*{\bibinfo}[2]{#2}
\providecommand*{\eprint}[1]{#1}
\providecommand*{\url}[1]{#1}
\def\jcap{Jour. Cosmology and Astro-Particle Phys.\,}
\def\mnras{M.N.R.A.S.\,}
\def\inprep{in~prep.\,}
\def\apj{Astrophys.J.\,}
\def\apjl{Astrophys.J.Lett.\,}
\def\apjs{Astrophys.J.Supp.\,}
\def\aj{Astron.J.\,}
\def\nat{Nature\,}
\def\na{New Ast.\,}
\def\nar{New~Ast.Rev.\,}
\def\nup{Nucl. Phys.\,}
\def\cmp{Comm. Math. Phys.\,}
\def\prl{Phys. Rev. Lett.\,}
\def\pl{Phys. Lett.\,}
\def\rmp{Rev. Mod. Phys.\,}
\def\ijmp{Int. Jour. Mod. Phys.\,}
\def\mpl{Mod. Phys. Lett.\,}
\def\pr{Phys. Rev.\,}
\def\prd{Phys.Rev.D\,}
\def\araa{Annu.Rev.Astron.Astrophys.\,}
\def\aap{Astron.Astrophys.\,}
\def\aaps{Astron.Astrophys.Supp.\,}
\def\aa{Astron.Astrophys.\,}
\def\pasj{Pub.Astr.Soc.Japan\,}
\def\physrep{Phys.Rep.\,}

\vspace{-8pt}\bibitem{Gehrels+04swift}
\bibinfo{author}{N.~{Gehrels}}, \bibinfo{author}{G.~{Chincarini}},
  \bibinfo{author}{P.~{Giommi}}, \bibinfo{author}{K.~O. {Mason}},
  \bibinfo{author}{J.~A. {Nousek}}, \bibinfo{author}{A.~A. {Wells}},
  \bibinfo{author}{N.~E. {White}}, \bibinfo{author}{S.~D. {Barthelmy}},
  \bibinfo{author}{D.~N. {Burrows}}, \bibinfo{author}{L.~R. {Cominsky}},
  \emph{et~al.}, \bibinfo{journal}{\apj} \bibinfo{volume}{\textbf{611}},
  \bibinfo{pages}{1005} (\bibinfo{date}{Aug. 2004}),
  \eprint{arXiv:astro-ph/0405233}.
\vspace{-8pt}\bibitem{Michelson+10fermi}
\bibinfo{author}{P.~F. {Michelson}}, \bibinfo{author}{W.~B. {Atwood}}, and
  \bibinfo{author}{S.~{Ritz}}, \bibinfo{journal}{Reports on Progress in
  Physics} \bibinfo{volume}{\textbf{73}}(7), \bibinfo{pages}{074901}
  (\bibinfo{date}{Jul. 2010}), \eprint{1011.0213}.
\vspace{-8pt}\bibitem{Meegan+09GBM}
\bibinfo{author}{C.~{Meegan}}, \bibinfo{author}{G.~{Lichti}},
  \bibinfo{author}{P.~N. {Bhat}}, \bibinfo{author}{E.~{Bissaldi}},
  \bibinfo{author}{M.~S. {Briggs}}, \bibinfo{author}{V.~{Connaughton}},
  \bibinfo{author}{R.~{Diehl}}, \bibinfo{author}{G.~{Fishman}},
  \bibinfo{author}{J.~{Greiner}}, \bibinfo{author}{A.~S. {Hoover}},
  \emph{et~al.}, \bibinfo{journal}{\apj} \bibinfo{volume}{\textbf{702}},
  \bibinfo{pages}{791} (\bibinfo{date}{Sep. 2009}), \eprint{0908.0450}.
\vspace{-8pt}\bibitem{Paul+11svom}
\bibinfo{author}{J.~{Paul}}, \bibinfo{author}{J.~{Wei}},
  \bibinfo{author}{S.~{Basa}}, and \bibinfo{author}{S.-N. {Zhang}},
  \bibinfo{journal}{Comptes Rendus Physique} \bibinfo{volume}{\textbf{12}},
  \bibinfo{pages}{298} (\bibinfo{date}{Apr. 2011}), \eprint{1104.0606}.
\vspace{-8pt}\bibitem{Woosley+06araa}
\bibinfo{author}{S.~E. {Woosley}} and \bibinfo{author}{J.~S. {Bloom}},
  \bibinfo{journal}{\araa} \bibinfo{volume}{\textbf{44}}, \bibinfo{pages}{507}
  (\bibinfo{date}{Sep. 2006}), \eprint{arXiv:astro-ph/0609142}.
\vspace{-8pt}\bibitem{Kouveliotou+93}
\bibinfo{author}{C.~{Kouveliotou}}, \bibinfo{author}{C.~A. {Meegan}},
  \bibinfo{author}{G.~J. {Fishman}}, \bibinfo{author}{N.~P. {Bhat}},
  \bibinfo{author}{M.~S. {Briggs}}, \bibinfo{author}{T.~M. {Koshut}},
  \bibinfo{author}{W.~S. {Paciesas}}, and \bibinfo{author}{G.~N. {Pendleton}},
  \bibinfo{journal}{\apjl} \bibinfo{volume}{\textbf{413}},
  \bibinfo{pages}{L101} (\bibinfo{date}{Aug. 1993}).
\vspace{-8pt}\bibitem{Gehrels+09araa}
\bibinfo{author}{N.~{Gehrels}}, \bibinfo{author}{E.~{Ramirez-Ruiz}}, and
  \bibinfo{author}{D.~B. {Fox}}, \bibinfo{journal}{\araa}
  \bibinfo{volume}{\textbf{47}}, \bibinfo{pages}{567} (\bibinfo{date}{Sep.
  2009}), \eprint{0909.1531}.
\vspace{-8pt}\bibitem{Vedrenne+09book}
\bibinfo{author}{G.~{Vedrenne}} and \bibinfo{author}{J.-L. {Atteia}},
  \bibinfo{title}{\emph{{Gamma-Ray Bursts}}} (\bibinfo{year}{2009}).
\vspace{-8pt}\bibitem{Band+93}
\bibinfo{author}{D.~{Band}}, \bibinfo{author}{J.~{Matteson}},
  \bibinfo{author}{L.~{Ford}}, \bibinfo{author}{B.~{Schaefer}},
  \bibinfo{author}{D.~{Palmer}}, \bibinfo{author}{B.~{Teegarden}},
  \bibinfo{author}{T.~{Cline}}, \bibinfo{author}{M.~{Briggs}},
  \bibinfo{author}{W.~{Paciesas}}, \bibinfo{author}{G.~{Pendleton}},
  \emph{et~al.}, \bibinfo{journal}{\apj} \bibinfo{volume}{\textbf{413}},
  \bibinfo{pages}{281} (\bibinfo{date}{Aug. 1993}).
\vspace{-8pt}\bibitem{Fishman+95cgro}
\bibinfo{author}{G.~J. {Fishman}} and \bibinfo{author}{C.~A. {Meegan}},
  \bibinfo{journal}{\araa} \bibinfo{volume}{\textbf{33}}, \bibinfo{pages}{415}
  (\bibinfo{date}{1995}).
\vspace{-8pt}\bibitem{Baring+97}
\bibinfo{author}{M.~G. {Baring}} and \bibinfo{author}{A.~K. {Harding}},
  \bibinfo{journal}{\apj} \bibinfo{volume}{\textbf{491}}, \bibinfo{pages}{663}
  (\bibinfo{date}{Dec. 1997}), \eprint{arXiv:astro-ph/9711217}.
\vspace{-8pt}\bibitem{Meszaros06}
\bibinfo{author}{P.~{M\'esz\'aros}}, \bibinfo{journal}{Reports of Progress in
  Physics} \bibinfo{volume}{\textbf{69}}, \bibinfo{pages}{2259}
  (\bibinfo{date}{2006}), \eprint{arXiv:astro-ph/0605208}.
\vspace{-8pt}\bibitem{Drenkhahn02}
\bibinfo{author}{G.~{Drenkhahn}}, \bibinfo{journal}{\aap}
  \bibinfo{volume}{\textbf{387}}, \bibinfo{pages}{714} (\bibinfo{date}{May
  2002}), \eprint{arXiv:astro-ph/0112509}.
\vspace{-8pt}\bibitem{Metzger+11grbmag}
\bibinfo{author}{B.~D. {Metzger}}, \bibinfo{author}{D.~{Giannios}},
  \bibinfo{author}{T.~A. {Thompson}}, \bibinfo{author}{N.~{Bucciantini}}, and
  \bibinfo{author}{E.~{Quataert}}, \bibinfo{journal}{\mnras}
  \bibinfo{volume}{\textbf{413}}, \bibinfo{pages}{2031} (\bibinfo{date}{May
  2011}), \eprint{1012.0001}.
\vspace{-8pt}\bibitem{Meszaros+11gevmag}
\bibinfo{author}{P.~{M{\'e}sz{\'a}ros}} and \bibinfo{author}{M.~J. {Rees}},
  \bibinfo{journal}{\apjl} \bibinfo{volume}{\textbf{733}},
  \bibinfo{pages}{L40+} (\bibinfo{date}{Jun. 2011}), \eprint{1104.5025}.
\vspace{-8pt}\bibitem{Komissarov+09maggrb}
\bibinfo{author}{S.~S. {Komissarov}}, \bibinfo{author}{N.~{Vlahakis}},
  \bibinfo{author}{A.~{K{\"o}nigl}}, and \bibinfo{author}{M.~V. {Barkov}},
  \bibinfo{journal}{\mnras} \bibinfo{volume}{\textbf{394}},
  \bibinfo{pages}{1182} (\bibinfo{date}{Apr. 2009}), \eprint{0811.1467}.
\vspace{-8pt}\bibitem{Narayan+10numjet}
\bibinfo{author}{R.~{Narayan}}, \bibinfo{author}{A.~{Tchekhovskoy}}, and
  \bibinfo{author}{J.~{McKinney}}, in \bibinfo{editors}{{L.~Maraschi,
  G.~Ghisellini, R.~Della Ceca, \& F.~Tavecchio}}, ed., \emph{Accretion and
  Ejection in AGN: a Global View} (\bibinfo{date}{Oct. 2010}),
  \bibinfo{volume}{vol. 427 of \emph{Astronomical Society of the Pacific
  Conference Series}}, \bibinfo{pages}{pp. 127--+}, \eprint{1001.1355}.
\vspace{-8pt}\bibitem{Frail+02collim}
\bibinfo{author}{D.~A. {Frail}}, \bibinfo{author}{S.~R. {Kulkarni}},
  \bibinfo{author}{R.~{Sari}}, \bibinfo{author}{S.~G. {Djorgovski}},
  \bibinfo{author}{J.~S. {Bloom}}, \bibinfo{author}{T.~J. {Galama}},
  \bibinfo{author}{D.~E. {Reichart}}, \bibinfo{author}{E.~{Berger}},
  \bibinfo{author}{F.~A. {Harrison}}, \bibinfo{author}{P.~A. {Price}},
  \emph{et~al.}, \bibinfo{journal}{\apjl} \bibinfo{volume}{\textbf{562}},
  \bibinfo{pages}{L55} (\bibinfo{date}{Nov. 2001}),
  \eprint{arXiv:astro-ph/0102282}.
\vspace{-8pt}\bibitem{Fong+12-111020sgrb}
\bibinfo{author}{W.-f. {Fong}}, \bibinfo{author}{E.~{Berger}},
  \bibinfo{author}{R.~{Margutti}}, \bibinfo{author}{B.~A. {Zauderer}},
  \bibinfo{author}{E.~{Troja}}, \bibinfo{author}{I.~{Czekala}},
  \bibinfo{author}{R.~{Chornock}}, \bibinfo{author}{N.~{Gehrels}},
  \bibinfo{author}{T.~{Sakamoto}}, \bibinfo{author}{D.~B. {Fox}},
  \emph{et~al.}, \bibinfo{journal}{ArXiv e-prints}  (\bibinfo{date}{Apr.
  2012}), \eprint{1204.5475}.
\vspace{-8pt}\bibitem{Paczynski86}
\bibinfo{author}{B.~{Pacz\'ynski}}, \bibinfo{journal}{\apjl}
  \bibinfo{volume}{\textbf{308}}, \bibinfo{pages}{L43} (\bibinfo{date}{Sep.
  1986}).
\vspace{-8pt}\bibitem{Shemi+90}
\bibinfo{author}{A.~{Shemi}} and \bibinfo{author}{T.~{Piran}},
  \bibinfo{journal}{\apjl} \bibinfo{volume}{\textbf{365}}, \bibinfo{pages}{L55}
  (\bibinfo{date}{Dec. 1990}).
\vspace{-8pt}\bibitem{Rees+92fbal}
\bibinfo{author}{M.~J. {Rees}} and \bibinfo{author}{P.~{M\'esz\'aros}},
  \bibinfo{journal}{\mnras} \bibinfo{volume}{\textbf{258}},
  \bibinfo{pages}{41P} (\bibinfo{date}{Sep. 1992}).
\vspace{-8pt}\bibitem{Meszaros+93impact}
\bibinfo{author}{P.~{M\'esz\'aros}} and \bibinfo{author}{M.~J. {Rees}},
  \bibinfo{journal}{\apj} \bibinfo{volume}{\textbf{405}}, \bibinfo{pages}{278}
  (\bibinfo{date}{Mar. 1993}).
\vspace{-8pt}\bibitem{Rees+94is}
\bibinfo{author}{M.~J. {Rees}} and \bibinfo{author}{P.~{M\'esz\'aros}},
  \bibinfo{journal}{\apjl} \bibinfo{volume}{\textbf{430}}, \bibinfo{pages}{L93}
  (\bibinfo{date}{Aug. 1994}), \eprint{arXiv:astro-ph/9404038}.
\vspace{-8pt}\bibitem{Sari+98spectra}
\bibinfo{author}{R.~{Sari}}, \bibinfo{author}{T.~{Piran}}, and
  \bibinfo{author}{R.~{Narayan}}, \bibinfo{journal}{\apjl}
  \bibinfo{volume}{\textbf{497}}, \bibinfo{pages}{L17+} (\bibinfo{date}{Apr.
  1998}), \eprint{arXiv:astro-ph/9712005}.
\vspace{-8pt}\bibitem{Waxman97grbfbal}
\bibinfo{author}{E.~{Waxman}}, \bibinfo{journal}{\apjl}
  \bibinfo{volume}{\textbf{485}}, \bibinfo{pages}{L5} (\bibinfo{date}{Aug.
  1997}), \eprint{arXiv:astro-ph/9704116}.
\vspace{-8pt}\bibitem{Meszaros+93multi}
\bibinfo{author}{P.~{Meszaros}} and \bibinfo{author}{M.~J. {Rees}},
  \bibinfo{journal}{\apjl} \bibinfo{volume}{\textbf{418}},
  \bibinfo{pages}{L59+} (\bibinfo{date}{Dec. 1993}),
  \eprint{arXiv:astro-ph/9309011}.
\vspace{-8pt}\bibitem{Meszaros+97ag}
\bibinfo{author}{P.~{M\'esz\'aros}} and \bibinfo{author}{M.~J. {Rees}},
  \bibinfo{journal}{\apj} \bibinfo{volume}{\textbf{476}}, \bibinfo{pages}{232}
  (\bibinfo{date}{Feb. 1997}), \eprint{arXiv:astro-ph/9606043}.
\vspace{-8pt}\bibitem{Costa+97}
\bibinfo{author}{E.~{Costa}}, \bibinfo{author}{F.~{Frontera}},
  \bibinfo{author}{J.~{Heise}}, \bibinfo{author}{M.~{Feroci}},
  \bibinfo{author}{J.~{in't Zand}}, \bibinfo{author}{F.~{Fiore}},
  \bibinfo{author}{M.~N. {Cinti}}, \bibinfo{author}{D.~{Dal Fiume}},
  \bibinfo{author}{L.~{Nicastro}}, \bibinfo{author}{M.~{Orlandini}},
  \emph{et~al.}, \bibinfo{journal}{\nat} \bibinfo{volume}{\textbf{387}},
  \bibinfo{pages}{783} (\bibinfo{date}{Jun. 1997}),
  \eprint{arXiv:astro-ph/9706065}.
\vspace{-8pt}\bibitem{Vanparadijs+97}
\bibinfo{author}{J.~{van Paradijs}}, \bibinfo{author}{P.~J. {Groot}},
  \bibinfo{author}{T.~{Galama}}, \bibinfo{author}{C.~{Kouveliotou}},
  \bibinfo{author}{R.~G. {Strom}}, \bibinfo{author}{J.~{Telting}},
  \bibinfo{author}{R.~G.~M. {Rutten}}, \bibinfo{author}{G.~J. {Fishman}},
  \bibinfo{author}{C.~A. {Meegan}}, \bibinfo{author}{M.~{Pettini}},
  \emph{et~al.}, \bibinfo{journal}{\nat} \bibinfo{volume}{\textbf{386}},
  \bibinfo{pages}{686} (\bibinfo{date}{Apr. 1997}).
\vspace{-8pt}\bibitem{Metzger+97}
\bibinfo{author}{M.~R. {Metzger}}, \bibinfo{author}{S.~G. {Djorgovski}},
  \bibinfo{author}{S.~R. {Kulkarni}}, \bibinfo{author}{C.~C. {Steidel}},
  \bibinfo{author}{K.~L. {Adelberger}}, \bibinfo{author}{D.~A. {Frail}},
  \bibinfo{author}{E.~{Costa}}, and \bibinfo{author}{F.~{Frontera}},
  \bibinfo{journal}{\nat} \bibinfo{volume}{\textbf{387}}, \bibinfo{pages}{878}
  (\bibinfo{date}{Jun. 1997}).
\vspace{-8pt}\bibitem{Frail+97}
\bibinfo{author}{D.~A. {Frail}}, \bibinfo{author}{S.~R. {Kulkarni}},
  \bibinfo{author}{L.~{Nicastro}}, \bibinfo{author}{M.~{Feroci}}, and
  \bibinfo{author}{G.~B. {Taylor}}, \bibinfo{journal}{\nat}
  \bibinfo{volume}{\textbf{389}}, \bibinfo{pages}{261} (\bibinfo{date}{Sep.
  1997}).
\vspace{-8pt}\bibitem{Akerlof+99}
\bibinfo{author}{C.~{Akerlof}}, \bibinfo{author}{R.~{Balsano}},
  \bibinfo{author}{S.~{Barthelmy}}, \bibinfo{author}{J.~{Bloch}},
  \bibinfo{author}{P.~{Butterworth}}, \bibinfo{author}{D.~{Casperson}},
  \bibinfo{author}{T.~{Cline}}, \bibinfo{author}{S.~{Fletcher}},
  \bibinfo{author}{F.~{Frontera}}, \bibinfo{author}{G.~{Gisler}},
  \emph{et~al.}, \bibinfo{journal}{\nat} \bibinfo{volume}{\textbf{398}},
  \bibinfo{pages}{400} (\bibinfo{date}{Apr. 1999}),
  \eprint{arXiv:astro-ph/9903271}.
\vspace{-8pt}\bibitem{Meszaros+94gev}
\bibinfo{author}{P.~{M\'esz\'aros}} and \bibinfo{author}{M.~J. {Rees}},
  \bibinfo{journal}{\mnras} \bibinfo{volume}{\textbf{269}},
  \bibinfo{pages}{L41+} (\bibinfo{date}{Jul. 1994}),
  \eprint{arXiv:astro-ph/9404056}.
\vspace{-8pt}\bibitem{Hurley+94-GRB940217}
\bibinfo{author}{K.~{Hurley}}, \bibinfo{author}{B.~L. {Dingus}},
  \bibinfo{author}{R.~{Mukherjee}}, \bibinfo{author}{P.~{Sreekumar}},
  \bibinfo{author}{C.~{Kouveliotou}}, \bibinfo{author}{C.~{Meegan}},
  \bibinfo{author}{G.~J. {Fishman}}, \bibinfo{author}{D.~{Band}},
  \bibinfo{author}{L.~{Ford}}, \bibinfo{author}{D.~{Bertsch}}, \emph{et~al.},
  \bibinfo{journal}{\nat} \bibinfo{volume}{\textbf{372}}, \bibinfo{pages}{652}
  (\bibinfo{date}{Dec. 1994}).
\vspace{-8pt}\bibitem{Abdo+09-080916}
\bibinfo{author}{A.~{Abdo}} and \bibinfo{author}{{the Fermi Collaboration}},
  \bibinfo{journal}{Science} \bibinfo{volume}{\textbf{323}},
  \bibinfo{pages}{1688} (\bibinfo{date}{Feb. 2009}).
\vspace{-8pt}\bibitem{Spada+00is}
\bibinfo{author}{M.~{Spada}}, \bibinfo{author}{A.~{Panaitescu}}, and
  \bibinfo{author}{P.~{M{\'e}sz{\'a}ros}}, \bibinfo{journal}{\apj}
  \bibinfo{volume}{\textbf{537}}, \bibinfo{pages}{824} (\bibinfo{date}{Jul.
  2000}), \eprint{arXiv:astro-ph/9908097}.
\vspace{-8pt}\bibitem{Beloborodov00is}
\bibinfo{author}{A.~M. {Beloborodov}}, \bibinfo{journal}{\apjl}
  \bibinfo{volume}{\textbf{539}}, \bibinfo{pages}{L25} (\bibinfo{date}{Aug.
  2000}), \eprint{arXiv:astro-ph/0004360}.
\vspace{-8pt}\bibitem{Kobayashi+01is}
\bibinfo{author}{S.~{Kobayashi}} and \bibinfo{author}{R.~{Sari}},
  \bibinfo{journal}{\apj} \bibinfo{volume}{\textbf{551}}, \bibinfo{pages}{934}
  (\bibinfo{date}{Apr. 2001}), \eprint{arXiv:astro-ph/0101006}.
\vspace{-8pt}\bibitem{Papathanassiou+96is}
\bibinfo{author}{H.~{Papathanassiou}} and \bibinfo{author}{P.~{Meszaros}},
  \bibinfo{journal}{\apjl} \bibinfo{volume}{\textbf{471}},
  \bibinfo{pages}{L91+} (\bibinfo{date}{Nov. 1996}),
  \eprint{arXiv:astro-ph/9609039}.
\vspace{-8pt}\bibitem{Pilla+98is}
\bibinfo{author}{R.~P. {Pilla}} and \bibinfo{author}{A.~{Loeb}},
  \bibinfo{journal}{\apjl} \bibinfo{volume}{\textbf{494}},
  \bibinfo{pages}{L167+} (\bibinfo{date}{Feb. 1998}),
  \eprint{arXiv:astro-ph/9710219}.
\vspace{-8pt}\bibitem{Kumar99}
\bibinfo{author}{P.~{Kumar}}, \bibinfo{journal}{\apjl}
  \bibinfo{volume}{\textbf{523}}, \bibinfo{pages}{L113} (\bibinfo{date}{Oct.
  1999}), \eprint{arXiv:astro-ph/9907096}.
\vspace{-8pt}\bibitem{Guetta+01is}
\bibinfo{author}{D.~{Guetta}}, \bibinfo{author}{M.~{Spada}}, and
  \bibinfo{author}{E.~{Waxman}}, \bibinfo{journal}{\apj}
  \bibinfo{volume}{\textbf{557}}, \bibinfo{pages}{399} (\bibinfo{date}{Aug.
  2001}), \eprint{arXiv:astro-ph/0011170}.
\vspace{-8pt}\bibitem{Panaitescu+00ssc}
\bibinfo{author}{A.~{Panaitescu}} and \bibinfo{author}{P.~{M{\'e}sz{\'a}ros}},
  \bibinfo{journal}{\apjl} \bibinfo{volume}{\textbf{544}}, \bibinfo{pages}{L17}
  (\bibinfo{date}{Nov. 2000}), \eprint{arXiv:astro-ph/0009309}.
\vspace{-8pt}\bibitem{Ghisellini+00grb-emis}
\bibinfo{author}{G.~{Ghisellini}}, \bibinfo{author}{A.~{Celotti}}, and
  \bibinfo{author}{D.~{Lazzati}}, \bibinfo{journal}{\mnras}
  \bibinfo{volume}{\textbf{313}}, \bibinfo{pages}{L1} (\bibinfo{date}{Mar.
  2000}), \eprint{arXiv:astro-ph/9912461}.
\vspace{-8pt}\bibitem{Narayan+09turb}
\bibinfo{author}{R.~{Narayan}} and \bibinfo{author}{P.~{Kumar}},
  \bibinfo{journal}{\mnras} \bibinfo{volume}{\textbf{394}},
  \bibinfo{pages}{L117} (\bibinfo{date}{Mar. 2009}), \eprint{0812.0018}.
\vspace{-8pt}\bibitem{Kumar+09turb}
\bibinfo{author}{P.~{Kumar}} and \bibinfo{author}{R.~{Narayan}},
  \bibinfo{journal}{\mnras} \bibinfo{volume}{\textbf{395}},
  \bibinfo{pages}{472} (\bibinfo{date}{May 2009}), \eprint{0812.0021}.
\vspace{-8pt}\bibitem{Lazar+09turb}
\bibinfo{author}{A.~{Lazar}}, \bibinfo{author}{E.~{Nakar}}, and
  \bibinfo{author}{T.~{Piran}}, \bibinfo{journal}{\apjl}
  \bibinfo{volume}{\textbf{695}}, \bibinfo{pages}{L10} (\bibinfo{date}{Apr.
  2009}), \eprint{0901.1133}.
\vspace{-8pt}\bibitem{Zhang+09turb}
\bibinfo{author}{W.~{Zhang}} and \bibinfo{author}{A.~{MacFadyen}},
  \bibinfo{journal}{\apj} \bibinfo{volume}{\textbf{698}}, \bibinfo{pages}{1261}
  (\bibinfo{date}{Jun. 2009}), \eprint{0902.2396}.
\vspace{-8pt}\bibitem{Preece+00batse}
\bibinfo{author}{R.~D. {Preece}}, \bibinfo{author}{M.~S. {Briggs}},
  \bibinfo{author}{R.~S. {Mallozzi}}, \bibinfo{author}{G.~N. {Pendleton}},
  \bibinfo{author}{W.~S. {Paciesas}}, and \bibinfo{author}{D.~L. {Band}},
  \bibinfo{journal}{\apjs} \bibinfo{volume}{\textbf{126}}, \bibinfo{pages}{19}
  (\bibinfo{date}{Jan. 2000}), \eprint{arXiv:astro-ph/9908119}.
\vspace{-8pt}\bibitem{Granot+01sync}
\bibinfo{author}{J.~{Granot}}, \bibinfo{author}{T.~{Piran}}, and
  \bibinfo{author}{R.~{Sari}}, \bibinfo{journal}{\apjl}
  \bibinfo{volume}{\textbf{534}}, \bibinfo{pages}{L163} (\bibinfo{date}{May
  2000}), \eprint{arXiv:astro-ph/0001160}.
\vspace{-8pt}\bibitem{Medvedev00jitter}
\bibinfo{author}{M.~V. {Medvedev}}, \bibinfo{journal}{\apj}
  \bibinfo{volume}{\textbf{540}}, \bibinfo{pages}{704} (\bibinfo{date}{Sep.
  2000}), \eprint{arXiv:astro-ph/0001314}.
\vspace{-8pt}\bibitem{Medvedev06jitter}
\bibinfo{author}{M.~V. {Medvedev}}, \bibinfo{journal}{\apj}
  \bibinfo{volume}{\textbf{637}}, \bibinfo{pages}{869} (\bibinfo{date}{Feb.
  2006}), \eprint{arXiv:astro-ph/0510472}.
\vspace{-8pt}\bibitem{Lloyd+02timeres}
\bibinfo{author}{N.~M. {Lloyd-Ronning}} and \bibinfo{author}{V.~{Petrosian}},
  \bibinfo{journal}{\apj} \bibinfo{volume}{\textbf{565}}, \bibinfo{pages}{182}
  (\bibinfo{date}{Jan. 2002}), \eprint{arXiv:astro-ph/0109340}.
\vspace{-8pt}\bibitem{Meszaros+00phot}
\bibinfo{author}{P.~{M{\'e}sz{\'a}ros}} and \bibinfo{author}{M.~J. {Rees}},
  \bibinfo{journal}{\apj} \bibinfo{volume}{\textbf{530}}, \bibinfo{pages}{292}
  (\bibinfo{date}{Feb. 2000}), \eprint{arXiv:astro-ph/9908126}.
\vspace{-8pt}\bibitem{Meszaros+02phot}
\bibinfo{author}{P.~{M{\'e}sz{\'a}ros}}, \bibinfo{author}{E.~{Ramirez-Ruiz}},
  \bibinfo{author}{M.~J. {Rees}}, and \bibinfo{author}{B.~{Zhang}},
  \bibinfo{journal}{\apj} \bibinfo{volume}{\textbf{578}}, \bibinfo{pages}{812}
  (\bibinfo{date}{Oct. 2002}), \eprint{arXiv:astro-ph/0205144}.
\vspace{-8pt}\bibitem{Ryde05}
\bibinfo{author}{F.~{Ryde}}, \bibinfo{journal}{\apjl}
  \bibinfo{volume}{\textbf{625}}, \bibinfo{pages}{L95} (\bibinfo{date}{Jun.
  2005}), \eprint{arXiv:astro-ph/0504450}.
\vspace{-8pt}\bibitem{Peer+04waxb}
\bibinfo{author}{A.~{Pe'er}} and \bibinfo{author}{E.~{Waxman}},
  \bibinfo{journal}{\apj} \bibinfo{volume}{\textbf{613}}, \bibinfo{pages}{448}
  (\bibinfo{date}{Sep. 2004}), \eprint{arXiv:astro-ph/0311252}.
\vspace{-8pt}\bibitem{Madau+00pair}
\bibinfo{author}{P.~{Madau}} and \bibinfo{author}{C.~{Thompson}},
  \bibinfo{journal}{\apj} \bibinfo{volume}{\textbf{534}}, \bibinfo{pages}{239}
  (\bibinfo{date}{May 2000}), \eprint{arXiv:astro-ph/9909060}.
\vspace{-8pt}\bibitem{Thompson+00pair}
\bibinfo{author}{C.~{Thompson}} and \bibinfo{author}{P.~{Madau}},
  \bibinfo{journal}{\apj} \bibinfo{volume}{\textbf{538}}, \bibinfo{pages}{105}
  (\bibinfo{date}{Jul. 2000}), \eprint{arXiv:astro-ph/9909111}.
\vspace{-8pt}\bibitem{Meszaros+01pair}
\bibinfo{author}{P.~{M{\'e}sz{\'a}ros}}, \bibinfo{author}{E.~{Ramirez-Ruiz}},
  and \bibinfo{author}{M.~J. {Rees}}, \bibinfo{journal}{Astrophys.J.}
  \bibinfo{volume}{\textbf{554}}, \bibinfo{pages}{660} (\bibinfo{date}{Jun.
  2001}), \eprint{arXiv:astro-ph/0011284}.
\vspace{-8pt}\bibitem{Beloborodov02pair}
\bibinfo{author}{A.~M. {Beloborodov}}, \bibinfo{journal}{\apj}
  \bibinfo{volume}{\textbf{565}}, \bibinfo{pages}{808} (\bibinfo{date}{Feb.
  2002}), \eprint{arXiv:astro-ph/0103321}.
\vspace{-8pt}\bibitem{Thompson06pair}
\bibinfo{author}{C.~{Thompson}}, \bibinfo{journal}{\apj}
  \bibinfo{volume}{\textbf{651}}, \bibinfo{pages}{333} (\bibinfo{date}{Nov.
  2006}), \eprint{arXiv:astro-ph/0507387}.
\vspace{-8pt}\bibitem{Eichler+00thermal}
\bibinfo{author}{D.~{Eichler}} and \bibinfo{author}{A.~{Levinson}},
  \bibinfo{journal}{\apj} \bibinfo{volume}{\textbf{529}}, \bibinfo{pages}{146}
  (\bibinfo{date}{Jan. 2000}), \eprint{arXiv:astro-ph/9903103}.
\vspace{-8pt}\bibitem{Thompson94}
\bibinfo{author}{C.~{Thompson}}, \bibinfo{journal}{\mnras}
  \bibinfo{volume}{\textbf{270}}, \bibinfo{pages}{480} (\bibinfo{date}{Oct.
  1994}).
\vspace{-8pt}\bibitem{Rees+05phot}
\bibinfo{author}{M.~J. {Rees}} and \bibinfo{author}{P.~{M{\'e}sz{\'a}ros}},
  \bibinfo{journal}{Astrophys.J.} \bibinfo{volume}{\textbf{628}},
  \bibinfo{pages}{847} (\bibinfo{date}{Aug. 2005}),
  \eprint{arXiv:astro-ph/0412702}.
\vspace{-8pt}\bibitem{Peer+05peak}
\bibinfo{author}{A.~{Pe'er}}, \bibinfo{author}{P.~{M{\'e}sz{\'a}ros}}, and
  \bibinfo{author}{M.~J. {Rees}}, \bibinfo{journal}{\apj}
  \bibinfo{volume}{\textbf{635}}, \bibinfo{pages}{476} (\bibinfo{date}{Dec.
  2005}), \eprint{arXiv:astro-ph/0504346}.
\vspace{-8pt}\bibitem{Peer+06phot}
\bibinfo{author}{A.~{Pe'er}}, \bibinfo{author}{P.~{M{\'e}sz{\'a}ros}}, and
  \bibinfo{author}{M.~J. {Rees}}, \bibinfo{journal}{\apj}
  \bibinfo{volume}{\textbf{642}}, \bibinfo{pages}{995} (\bibinfo{date}{May
  2006}), \eprint{arXiv:astro-ph/0510114}.
\vspace{-8pt}\bibitem{Amati+02episo}
\bibinfo{author}{L.~{Amati}}, \bibinfo{author}{F.~{Frontera}},
  \bibinfo{author}{M.~{Tavani}}, \bibinfo{author}{J.~J.~M. {in't Zand}},
  \bibinfo{author}{A.~{Antonelli}}, \bibinfo{author}{E.~{Costa}},
  \bibinfo{author}{M.~{Feroci}}, \bibinfo{author}{C.~{Guidorzi}},
  \bibinfo{author}{J.~{Heise}}, \bibinfo{author}{N.~{Masetti}}, \emph{et~al.},
  \bibinfo{journal}{\aap} \bibinfo{volume}{\textbf{390}}, \bibinfo{pages}{81}
  (\bibinfo{date}{Jul. 2002}), \eprint{arXiv:astro-ph/0205230}.
\vspace{-8pt}\bibitem{Ghirlanda+04corr}
\bibinfo{author}{G.~{Ghirlanda}}, \bibinfo{author}{G.~{Ghisellini}}, and
  \bibinfo{author}{D.~{Lazzati}}, \bibinfo{journal}{\apj}
  \bibinfo{volume}{\textbf{616}}, \bibinfo{pages}{331} (\bibinfo{date}{Nov.
  2004}), \eprint{arXiv:astro-ph/0405602}.
\vspace{-8pt}\bibitem{Thompson+07phot}
\bibinfo{author}{C.~{Thompson}}, \bibinfo{author}{P.~{M{\'e}sz{\'a}ros}}, and
  \bibinfo{author}{M.~J. {Rees}}, \bibinfo{journal}{\apj}
  \bibinfo{volume}{\textbf{666}}, \bibinfo{pages}{1012} (\bibinfo{date}{Sep.
  2007}), \eprint{arXiv:astro-ph/0608282}.
\vspace{-8pt}\bibitem{Usov94}
\bibinfo{author}{V.~V. {Usov}}, \bibinfo{journal}{\mnras}
  \bibinfo{volume}{\textbf{267}}, \bibinfo{pages}{1035} (\bibinfo{date}{Apr.
  1994}), \eprint{arXiv:astro-ph/9312024}.
\vspace{-8pt}\bibitem{Drenkhahn+02}
\bibinfo{author}{G.~{Drenkhahn}} and \bibinfo{author}{H.~C. {Spruit}},
  \bibinfo{journal}{\aap} \bibinfo{volume}{\textbf{391}}, \bibinfo{pages}{1141}
  (\bibinfo{date}{Sep. 2002}), \eprint{arXiv:astro-ph/0202387}.
\vspace{-8pt}\bibitem{Lyutikov+03grbmag}
\bibinfo{author}{M.~{Lyutikov}} and \bibinfo{author}{R.~{Blandford}},
  \bibinfo{journal}{ArXiv Astrophysics e-prints}  (\bibinfo{date}{Dec. 2003}),
  \eprint{arXiv:astro-ph/0312347}.
\vspace{-8pt}\bibitem{Meszaros+94ext}
\bibinfo{author}{P.~{Meszaros}}, \bibinfo{author}{M.~J. {Rees}}, and
  \bibinfo{author}{H.~{Papathanassiou}}, \bibinfo{journal}{\apj}
  \bibinfo{volume}{\textbf{432}}, \bibinfo{pages}{181} (\bibinfo{date}{Sep.
  1994}), \eprint{arXiv:astro-ph/9311071}.
\vspace{-8pt}\bibitem{Zhang+05magrev}
\bibinfo{author}{B.~{Zhang}} and \bibinfo{author}{S.~{Kobayashi}},
  \bibinfo{journal}{\apj} \bibinfo{volume}{\textbf{628}}, \bibinfo{pages}{315}
  (\bibinfo{date}{Jul. 2005}), \eprint{arXiv:astro-ph/0404140}.
\vspace{-8pt}\bibitem{Giannios+08revshock}
\bibinfo{author}{D.~{Giannios}}, \bibinfo{author}{P.~{Mimica}}, and
  \bibinfo{author}{M.~A. {Aloy}}, \bibinfo{journal}{\aap}
  \bibinfo{volume}{\textbf{478}}, \bibinfo{pages}{747} (\bibinfo{date}{Feb.
  2008}), \eprint{0711.1980}.
\vspace{-8pt}\bibitem{Narayan+11magshock}
\bibinfo{author}{R.~{Narayan}}, \bibinfo{author}{P.~{Kumar}}, and
  \bibinfo{author}{A.~{Tchekhovskoy}}, \bibinfo{journal}{\mnras}
  \bibinfo{volume}{\textbf{416}}, \bibinfo{pages}{2193} (\bibinfo{date}{Sep.
  2011}), \eprint{1105.0003}.
\vspace{-8pt}\bibitem{Fan+04magis}
\bibinfo{author}{Y.~Z. {Fan}}, \bibinfo{author}{D.~M. {Wei}}, and
  \bibinfo{author}{B.~{Zhang}}, \bibinfo{journal}{\mnras}
  \bibinfo{volume}{\textbf{354}}, \bibinfo{pages}{1031} (\bibinfo{date}{Nov.
  2004}), \eprint{arXiv:astro-ph/0407581}.
\vspace{-8pt}\bibitem{Zhang+11icmart}
\bibinfo{author}{B.~{Zhang}} and \bibinfo{author}{H.~{Yan}},
  \bibinfo{journal}{\apj} \bibinfo{volume}{\textbf{726}}, \bibinfo{pages}{90}
  (\bibinfo{date}{Jan. 2011}), \eprint{1011.1197}.
\vspace{-8pt}\bibitem{Blandford+77znajek}
\bibinfo{author}{R.~D. {Blandford}} and \bibinfo{author}{R.~L. {Znajek}},
  \bibinfo{journal}{\mnras} \bibinfo{volume}{\textbf{179}},
  \bibinfo{pages}{433} (\bibinfo{date}{May 1977}).
\vspace{-8pt}\bibitem{Meszaros+97poynting}
\bibinfo{author}{P.~{M\'esz\'aros}} and \bibinfo{author}{M.~J. {Rees}},
  \bibinfo{journal}{\apjl} \bibinfo{volume}{\textbf{482}},
  \bibinfo{pages}{L29+} (\bibinfo{date}{Jun. 1997}),
  \eprint{arXiv:astro-ph/9609065}.
\vspace{-8pt}\bibitem{McKinney+12magphot}
\bibinfo{author}{J.~C. {McKinney}} and \bibinfo{author}{D.~A. {Uzdensky}},
  \bibinfo{journal}{\mnras} \bibinfo{volume}{\textbf{419}},
  \bibinfo{pages}{573} (\bibinfo{date}{Jan. 2012}), \eprint{1011.1904}.
\vspace{-8pt}\bibitem{Giannios+07photspec}
\bibinfo{author}{D.~{Giannios}} and \bibinfo{author}{H.~C. {Spruit}},
  \bibinfo{journal}{\aap} \bibinfo{volume}{\textbf{469}}, \bibinfo{pages}{1}
  (\bibinfo{date}{Jul. 2007}), \eprint{arXiv:astro-ph/0611385}.
\vspace{-8pt}\bibitem{Abdo+09-090902}
\bibinfo{author}{A.~A. {Abdo}} and \bibinfo{author}{the {Fermi}~collaboration},
  \bibinfo{journal}{\apjl} \bibinfo{volume}{\textbf{706}},
  \bibinfo{pages}{L138} (\bibinfo{date}{Nov. 2009}), \eprint{0909.2470}.
\vspace{-8pt}\bibitem{Ackermann+11-090926}
\bibinfo{author}{M.~{Ackermann}} and \bibinfo{author}{the
  {Fermi}~collaboration}, \bibinfo{journal}{\apj}
  \bibinfo{volume}{\textbf{729}}, \bibinfo{pages}{114} (\bibinfo{date}{Mar.
  2011}), \eprint{1101.2082}.
\vspace{-8pt}\bibitem{Ackermann+10-090510}
\bibinfo{author}{M.~{Ackermann}} and \bibinfo{author}{the Fermi~collab.},
  \bibinfo{journal}{\apj} \bibinfo{volume}{\textbf{716}}, \bibinfo{pages}{1178}
  (\bibinfo{date}{Jun. 2010}).
\vspace{-8pt}\bibitem{Ghisellini+10grbrad}
\bibinfo{author}{G.~{Ghisellini}}, \bibinfo{author}{G.~{Ghirlanda}},
  \bibinfo{author}{L.~{Nava}}, and \bibinfo{author}{A.~{Celotti}},
  \bibinfo{journal}{\mnras} \bibinfo{volume}{\textbf{403}},
  \bibinfo{pages}{926} (\bibinfo{date}{Apr. 2010}), \eprint{0910.2459}.
\vspace{-8pt}\bibitem{Kumar+09gevfs}
\bibinfo{author}{P.~{Kumar}} and \bibinfo{author}{R.~{Barniol Duran}},
  \bibinfo{journal}{\mnras} \bibinfo{volume}{\textbf{400}},
  \bibinfo{pages}{L75} (\bibinfo{date}{Nov. 2009}), \eprint{0905.2417}.
\vspace{-8pt}\bibitem{Corsi+10-090510}
\bibinfo{author}{A.~{Corsi}}, \bibinfo{author}{D.~{Guetta}}, and
  \bibinfo{author}{L.~{Piro}}, \bibinfo{journal}{\apj}
  \bibinfo{volume}{\textbf{720}}, \bibinfo{pages}{1008} (\bibinfo{date}{Sep.
  2010}), \eprint{0911.4453}.
\vspace{-8pt}\bibitem{DePasquale+09-090510ag}
\bibinfo{author}{M.~{De Pasquale}}, \bibinfo{author}{P.~{Schady}},
  \bibinfo{author}{N.~P.~M. {Kuin}}, \bibinfo{author}{M.~J. {Page}},
  \bibinfo{author}{P.~A. {Curran}}, \bibinfo{author}{S.~{Zane}},
  \bibinfo{author}{S.~R. {Oates}}, \bibinfo{author}{S.~T. {Holland}},
  \bibinfo{author}{A.~A. {Breeveld}}, \bibinfo{author}{E.~A. {Hoversten}},
  \emph{et~al.}, \bibinfo{journal}{\apjl} \bibinfo{volume}{\textbf{709}},
  \bibinfo{pages}{L146} (\bibinfo{date}{Feb. 2010}), \eprint{0910.1629}.
\vspace{-8pt}\bibitem{He+11-090510}
\bibinfo{author}{H.-N. {He}}, \bibinfo{author}{X.-F. {Wu}},
  \bibinfo{author}{K.~{Toma}}, \bibinfo{author}{X.-Y. {Wang}}, and
  \bibinfo{author}{P.~{M{\'e}sz{\'a}ros}}, \bibinfo{journal}{\apj}
  \bibinfo{volume}{\textbf{733}}, \bibinfo{pages}{22}, \bibinfo{eid}{22}
  (\bibinfo{date}{May 2011}), \eprint{1009.1432}.
\vspace{-8pt}\bibitem{Toma+11phot}
\bibinfo{author}{K.~{Toma}}, \bibinfo{author}{X.-F. {Wu}}, and
  \bibinfo{author}{P.~{M{\'e}sz{\'a}ros}}, \bibinfo{journal}{\mnras}
  \bibinfo{volume}{\textbf{415}}, \bibinfo{pages}{1663} (\bibinfo{date}{Aug.
  2011}), \eprint{1002.2634}.
\vspace{-8pt}\bibitem{Veres+12mag}
\bibinfo{author}{P.~{Veres}} and \bibinfo{author}{P.~{M{\'e}sz{\'a}ros}},
  \bibinfo{journal}{\apj} \bibinfo{volume}{\textbf{755}}, \bibinfo{pages}{12},
  \bibinfo{eid}{12} (\bibinfo{date}{Aug. 2012}), \eprint{1202.2821}.
\vspace{-8pt}\bibitem{Razzaque10-090510}
\bibinfo{author}{S.~{Razzaque}}, \bibinfo{journal}{\apjl}
  \bibinfo{volume}{\textbf{724}}, \bibinfo{pages}{L109} (\bibinfo{date}{Nov.
  2010}), \eprint{1004.3330}.
\vspace{-8pt}\bibitem{Asano+09-090510}
\bibinfo{author}{K.~{Asano}}, \bibinfo{author}{S.~{Guiriec}}, and
  \bibinfo{author}{P.~{M{\'e}sz{\'a}ros}}, \bibinfo{journal}{\apjl}
  \bibinfo{volume}{\textbf{705}}, \bibinfo{pages}{L191} (\bibinfo{date}{Nov.
  2009}), \eprint{0909.0306}.
\vspace{-8pt}\bibitem{Murase+12reac}
\bibinfo{author}{K.~{Murase}}, \bibinfo{author}{K.~{Asano}},
  \bibinfo{author}{T.~{Terasawa}}, and \bibinfo{author}{P.~{M{\'e}sz{\'a}ros}},
  \bibinfo{journal}{\apj} \bibinfo{volume}{\textbf{746}}, \bibinfo{pages}{164},
  \bibinfo{eid}{164} (\bibinfo{date}{Feb. 2012}), \eprint{1107.5575}.
\vspace{-8pt}\bibitem{Asano+12grbhad}
\bibinfo{author}{K.~{Asano}} and \bibinfo{author}{P.~{M{\'e}sz{\'a}ros}},
  \bibinfo{journal}{ArXiv e-prints}  (\bibinfo{date}{Jun. 2012}),
  \eprint{1206.0347}.
\vspace{-8pt}\bibitem{Omodei+11fermigrb}
\bibinfo{author}{N.~{Omodei}} and \bibinfo{author}{{the Fermi LAT
  collaboration}}, \bibinfo{journal}{Talk at Fermi Meeting, Stanford U.}
  (\bibinfo{date}{Feb. 2011}).
\vspace{-8pt}\bibitem{Bouvier+11-CTA}
\bibinfo{author}{A.~{Bouvier}}, \bibinfo{author}{R.~{Gilmore}},
  \bibinfo{author}{V.~{Connaughton}}, \bibinfo{author}{N.~{Otte}},
  \bibinfo{author}{J.~R. {Primack}}, and \bibinfo{author}{D.~A. {Williams}},
  \bibinfo{journal}{ArXiv e-prints}  (\bibinfo{date}{Sep. 2011}),
  \eprint{1109.5680}.
\vspace{-8pt}\bibitem{Funk+12-CTA}
\bibinfo{author}{S.~{Funk}} and \bibinfo{author}{J.~{Hinton}},
  \bibinfo{journal}{ArXiv e-prints}  (\bibinfo{date}{May 2012}),
  \eprint{1205.0832}.
\vspace{-8pt}\bibitem{Guetta+11-latgbm}
\bibinfo{author}{D.~{Guetta}}, \bibinfo{author}{E.~{Pian}}, and
  \bibinfo{author}{E.~{Waxman}}, \bibinfo{journal}{\aap}
  \bibinfo{volume}{\textbf{525}}, \bibinfo{pages}{A53+} (\bibinfo{date}{Jan.
  2011}), \eprint{1003.0566}.
\vspace{-8pt}\bibitem{Beniamini+11-latgbm}
\bibinfo{author}{P.~{Beniamini}}, \bibinfo{author}{D.~{Guetta}},
  \bibinfo{author}{E.~{Nakar}}, and \bibinfo{author}{T.~{Piran}},
  \bibinfo{journal}{\mnras} \bibinfo{volume}{\textbf{416}},
  \bibinfo{pages}{3089} (\bibinfo{date}{Oct. 2011}), \eprint{1103.0745}.
\vspace{-8pt}\bibitem{Razzaque+04gev}
\bibinfo{author}{S.~{Razzaque}}, \bibinfo{author}{P.~{M{\'e}sz{\'a}ros}}, and
  \bibinfo{author}{B.~{Zhang}}, \bibinfo{journal}{\apj}
  \bibinfo{volume}{\textbf{613}}, \bibinfo{pages}{1072} (\bibinfo{date}{Oct.
  2004}), \eprint{arXiv:astro-ph/0404076}.
\vspace{-8pt}\bibitem{Beloborodov10pn}
\bibinfo{author}{A.~M. {Beloborodov}}, \bibinfo{journal}{\mnras}
  \bibinfo{volume}{\textbf{407}}, \bibinfo{pages}{1033} (\bibinfo{date}{Sep.
  2010}), \eprint{0907.0732}.
\vspace{-8pt}\bibitem{Wang+10kn}
\bibinfo{author}{X.-Y. {Wang}}, \bibinfo{author}{H.-N. {He}},
  \bibinfo{author}{Z.~{Li}}, \bibinfo{author}{X.-F. {Wu}}, and
  \bibinfo{author}{Z.-G. {Dai}}, \bibinfo{journal}{\apj}
  \bibinfo{volume}{\textbf{712}}, \bibinfo{pages}{1232} (\bibinfo{date}{Apr.
  2010}), \eprint{0911.4189}.
\vspace{-8pt}\bibitem{Zhang+11-latgrb}
\bibinfo{author}{B.-B. {Zhang}}, \bibinfo{author}{B.~{Zhang}},
  \bibinfo{author}{E.-W. {Liang}}, \bibinfo{author}{Y.-Z. {Fan}},
  \bibinfo{author}{X.-F. {Wu}}, \bibinfo{author}{A.~{Pe'er}},
  \bibinfo{author}{A.~{Maxham}}, \bibinfo{author}{H.~{Gao}}, and
  \bibinfo{author}{Y.-M. {Dong}}, \bibinfo{journal}{\apj}
  \bibinfo{volume}{\textbf{730}}, \bibinfo{pages}{141} (\bibinfo{date}{Apr.
  2011}), \eprint{1009.3338}.
\vspace{-8pt}\bibitem{Coppi+97ebl}
\bibinfo{author}{P.~S. {Coppi}} and \bibinfo{author}{F.~A. {Aharonian}},
  \bibinfo{journal}{\apjl} \bibinfo{volume}{\textbf{487}}, \bibinfo{pages}{L9+}
  (\bibinfo{date}{Sep. 1997}), \eprint{arXiv:astro-ph/9610176}.
\vspace{-8pt}\bibitem{Finke+10ebl}
\bibinfo{author}{J.~D. {Finke}}, \bibinfo{author}{S.~{Razzaque}}, and
  \bibinfo{author}{C.~D. {Dermer}}, \bibinfo{journal}{\apj}
  \bibinfo{volume}{\textbf{712}}, \bibinfo{pages}{238} (\bibinfo{date}{Mar.
  2010}), \eprint{0905.1115}.
\vspace{-8pt}\bibitem{Primack+11ebl}
\bibinfo{author}{J.~R. {Primack}}, \bibinfo{author}{A.~{Dom{\'{\i}}nguez}},
  \bibinfo{author}{R.~C. {Gilmore}}, and \bibinfo{author}{R.~S. {Somerville}},
  in \bibinfo{editors}{{F.~A.~Aharonian, W.~Hofmann, \& F.~M.~Rieger}}, ed.,
  \emph{American Institute of Physics Conference Series} (\bibinfo{date}{Sep.
  2011}), \bibinfo{volume}{vol. 1381 of \emph{American Institute of Physics
  Conference Series}}, \bibinfo{pages}{pp. 72--83}, \eprint{1107.2566}.
\vspace{-8pt}\bibitem{Gilmore+10gamgam}
\bibinfo{author}{R.~{Gilmore}} and \bibinfo{author}{E.~{Ramirez-Ruiz}},
  \bibinfo{journal}{\apj} \bibinfo{volume}{\textbf{721}}, \bibinfo{pages}{709}
  (\bibinfo{date}{Sep. 2010}), \eprint{1006.3897}.
\vspace{-8pt}\bibitem{Wang+06gevflare}
\bibinfo{author}{X.-Y. {Wang}}, \bibinfo{author}{Z.~{Li}}, and
  \bibinfo{author}{P.~{M{\'e}sz{\'a}ros}}, \bibinfo{journal}{\apjl}
  \bibinfo{volume}{\textbf{641}}, \bibinfo{pages}{L89} (\bibinfo{date}{Apr.
  2006}), \eprint{arXiv:astro-ph/0601229}.
\vspace{-8pt}\bibitem{Asano+09grb}
\bibinfo{author}{K.~{Asano}}, \bibinfo{author}{S.~{Inoue}}, and
  \bibinfo{author}{P.~{M{\'e}sz{\'a}ros}}, \bibinfo{journal}{\apj}
  \bibinfo{volume}{\textbf{699}}, \bibinfo{pages}{953} (\bibinfo{date}{Jul.
  2009}), \eprint{0807.0951}.
\vspace{-8pt}\bibitem{Asano+11grbtemp}
\bibinfo{author}{K.~{Asano}} and \bibinfo{author}{P.~{M{\'e}sz{\'a}ros}},
  \bibinfo{journal}{\apj} \bibinfo{volume}{\textbf{739}}, \bibinfo{pages}{103},
  \bibinfo{eid}{103} (\bibinfo{date}{Oct. 2011}), \eprint{1107.4825}.
\vspace{-8pt}\bibitem{Asano+10optex}
\bibinfo{author}{K.~{Asano}}, \bibinfo{author}{S.~{Inoue}}, and
  \bibinfo{author}{P.~{M{\'e}sz{\'a}ros}}, \bibinfo{journal}{\apjl}
  \bibinfo{volume}{\textbf{725}}, \bibinfo{pages}{L121} (\bibinfo{date}{Dec.
  2010}), \eprint{1009.5178}.
\vspace{-8pt}\bibitem{Racusin+08nakedeye}
\bibinfo{author}{J.~L. {Racusin}}, \bibinfo{author}{S.~V. {Karpov}},
  \bibinfo{author}{M.~{Sokolowski}}, \bibinfo{author}{J.~{Granot}},
  \bibinfo{author}{X.~F. {Wu}}, \bibinfo{author}{V.~{Pal'Shin}},
  \bibinfo{author}{S.~{Covino}}, \bibinfo{author}{A.~J. {van der Horst}},
  \bibinfo{author}{S.~R. {Oates}}, \bibinfo{author}{P.~{Schady}},
  \emph{et~al.}, \bibinfo{journal}{\nat} \bibinfo{volume}{\textbf{455}},
  \bibinfo{pages}{183} (\bibinfo{date}{Sep. 2008}), \eprint{0805.1557}.
\vspace{-8pt}\bibitem{Bahcall+00pn}
\bibinfo{author}{J.~N. {Bahcall}} and \bibinfo{author}{P.~{M{\'e}sz{\'a}ros}},
  \bibinfo{journal}{Physical Review Letters} \bibinfo{volume}{\textbf{85}},
  \bibinfo{pages}{1362} (\bibinfo{date}{Aug. 2000}),
  \eprint{arXiv:hep-ph/0004019}.
\vspace{-8pt}\bibitem{Meszaros+00gevnu}
\bibinfo{author}{P.~{M{\'e}sz{\'a}ros}} and \bibinfo{author}{M.~J. {Rees}},
  \bibinfo{journal}{\apjl} \bibinfo{volume}{\textbf{541}}, \bibinfo{pages}{L5}
  (\bibinfo{date}{Sep. 2000}), \eprint{arXiv:astro-ph/0007102}.
\vspace{-8pt}\bibitem{Rees+05photdis}
\bibinfo{author}{M.~J. {Rees}} and \bibinfo{author}{P.~{M{\'e}sz{\'a}ros}},
  \bibinfo{journal}{\apj} \bibinfo{volume}{\textbf{628}}, \bibinfo{pages}{847}
  (\bibinfo{date}{Aug. 2005}), \eprint{arXiv:astro-ph/0412702}.
\vspace{-8pt}\bibitem{Ryde+11phot}
\bibinfo{author}{F.~{Ryde}}, \bibinfo{author}{A.~{Pe'Er}},
  \bibinfo{author}{T.~{Nymark}}, \bibinfo{author}{M.~{Axelsson}},
  \bibinfo{author}{E.~{Moretti}}, \bibinfo{author}{C.~{Lundman}},
  \bibinfo{author}{M.~{Battelino}}, \bibinfo{author}{E.~{Bissaldi}},
  \bibinfo{author}{J.~{Chiang}}, \bibinfo{author}{M.~S. {Jackson}},
  \emph{et~al.}, \bibinfo{journal}{\mnras} \bibinfo{volume}{\textbf{415}},
  \bibinfo{pages}{3693} (\bibinfo{date}{Aug. 2011}), \eprint{1103.0708}.
\vspace{-8pt}\bibitem{Peer11fermirev}
\bibinfo{author}{A.~{Pe'er}}, \bibinfo{journal}{ArXiv e-prints}
  (\bibinfo{date}{Nov. 2011}), \eprint{1111.3378}.
\vspace{-8pt}\bibitem{Vurm+11phot}
\bibinfo{author}{I.~{Vurm}}, \bibinfo{author}{A.~M. {Beloborodov}}, and
  \bibinfo{author}{J.~{Poutanen}}, \bibinfo{journal}{\apj}
  \bibinfo{volume}{\textbf{738}}, \bibinfo{pages}{77} (\bibinfo{date}{Sep.
  2011}), \eprint{1104.0394}.
\vspace{-8pt}\bibitem{Meszaros+11col}
\bibinfo{author}{P.~{M{\'e}sz{\'a}ros}} and \bibinfo{author}{M.~J. {Rees}},
  \bibinfo{journal}{\apjl} \bibinfo{volume}{\textbf{733}},
  \bibinfo{pages}{L40+} (\bibinfo{date}{May 2011}), \eprint{1104.5025}.
\vspace{-8pt}\bibitem{Bosnjak+12delay}
\bibinfo{author}{{\v Z}.~{Bo{\v s}njak}} and \bibinfo{author}{P.~{Kumar}},
  \bibinfo{journal}{\mnras} \bibinfo{volume}{\textbf{421}},
  \bibinfo{pages}{L39} (\bibinfo{date}{Mar. 2012}), \eprint{1108.0929}.
\vspace{-8pt}\bibitem{Murase+11reac}
\bibinfo{author}{K.~{Murase}}, \bibinfo{author}{K.~{Asano}},
  \bibinfo{author}{T.~{Terasawa}}, and \bibinfo{author}{P.~{Meszaros}},
  \bibinfo{journal}{ArXiv e-prints}  (\bibinfo{date}{Jul. 2011}),
  \eprint{1107.5575}.
\vspace{-8pt}\bibitem{Centrella11gwrev}
\bibinfo{author}{J.~{Centrella}}, in \bibinfo{editors}{{F.~A.~Aharonian,
  W.~Hofmann, \& F.~M.~Rieger}}, ed., \emph{American Institute of Physics
  Conference Series} (\bibinfo{date}{Sep. 2011}), \bibinfo{volume}{vol. 1381 of
  \emph{American Institute of Physics Conference Series}}, \bibinfo{pages}{pp.
  98--116}, \eprint{1109.3492}.
\vspace{-8pt}\bibitem{Leonor+09-ligogrb}
\bibinfo{author}{I.~{Leonor}}, \bibinfo{author}{P.~J. {Sutton}},
  \bibinfo{author}{R.~{Frey}}, \bibinfo{author}{G.~{Jones}},
  \bibinfo{author}{S.~{M{\'a}rka}}, and \bibinfo{author}{Z.~{M{\'a}rka}},
  \bibinfo{journal}{Classical and Quantum Gravity}
  \bibinfo{volume}{\textbf{26}}(20), \bibinfo{pages}{204017}
  (\bibinfo{date}{Oct. 2009}).
\vspace{-8pt}\bibitem{Fryer+02gwcol}
\bibinfo{author}{C.~L. {Fryer}}, \bibinfo{author}{D.~E. {Holz}}, and
  \bibinfo{author}{S.~A. {Hughes}}, \bibinfo{journal}{\apj}
  \bibinfo{volume}{\textbf{565}}, \bibinfo{pages}{430} (\bibinfo{date}{Jan.
  2002}), \eprint{arXiv:astro-ph/0106113}.
\vspace{-8pt}\bibitem{Kobayashi+03gwgrb}
\bibinfo{author}{S.~{Kobayashi}} and \bibinfo{author}{P.~{M{\'e}sz{\'a}ros}},
  \bibinfo{journal}{\apj} \bibinfo{volume}{\textbf{589}}, \bibinfo{pages}{861}
  (\bibinfo{date}{Jun. 2003}), \eprint{arXiv:astro-ph/0210211}.
\vspace{-8pt}\bibitem{Corsi+09mag}
\bibinfo{author}{A.~{Corsi}} and \bibinfo{author}{P.~{M{\'e}sz{\'a}ros}},
  \bibinfo{journal}{\apj} \bibinfo{volume}{\textbf{702}}, \bibinfo{pages}{1171}
  (\bibinfo{date}{Sep. 2009}), \eprint{0907.2290}.
\vspace{-8pt}\bibitem{Ott+11gwcoll}
\bibinfo{author}{C.~D. {Ott}}, \bibinfo{author}{C.~{Reisswig}},
  \bibinfo{author}{E.~{Schnetter}}, \bibinfo{author}{E.~{O'Connor}},
  \bibinfo{author}{U.~{Sperhake}}, \bibinfo{author}{F.~{L{\"o}ffler}},
  \bibinfo{author}{P.~{Diener}}, \bibinfo{author}{E.~{Abdikamalov}},
  \bibinfo{author}{I.~{Hawke}}, and \bibinfo{author}{A.~{Burrows}},
  \bibinfo{journal}{Physical Review Letters}
  \bibinfo{volume}{\textbf{106}}(16), \bibinfo{pages}{161103},
  \bibinfo{eid}{161103} (\bibinfo{date}{Apr. 2011}), \eprint{1012.1853}.
\vspace{-8pt}\bibitem{Kiuchi+11gwcoll}
\bibinfo{author}{K.~{Kiuchi}}, \bibinfo{author}{M.~{Shibata}},
  \bibinfo{author}{P.~J. {Montero}}, and \bibinfo{author}{J.~A. {Font}},
  \bibinfo{journal}{Physical Review Letters}
  \bibinfo{volume}{\textbf{106}}(25), \bibinfo{pages}{251102},
  \bibinfo{eid}{251102} (\bibinfo{date}{Jun. 2011}), \eprint{1105.5035}.
\vspace{-8pt}\bibitem{Waxman11-crnu}
\bibinfo{author}{E.~{Waxman}}, \bibinfo{journal}{ArXiv e-prints}
  (\bibinfo{date}{Jan. 2011}), \eprint{1101.1155}.
\vspace{-8pt}\bibitem{Waxman+97grbnu}
\bibinfo{author}{E.~Waxman} and \bibinfo{author}{J.~Bahcall},
  \bibinfo{journal}{{\prl}} \bibinfo{volume}{\textbf{78}},
  \bibinfo{pages}{2292} (\bibinfo{date}{1997}).
\vspace{-8pt}\bibitem{Murase+06grbnu}
\bibinfo{author}{K.~{Murase}} and \bibinfo{author}{S.~{Nagataki}},
  \bibinfo{journal}{\prd} \bibinfo{volume}{\textbf{73}}(6),
  \bibinfo{pages}{063002} (\bibinfo{date}{Mar. 2006}),
  \eprint{arXiv:astro-ph/0512275}.
\vspace{-8pt}\bibitem{Waxman+00nuag}
\bibinfo{author}{E.~{Waxman}} and \bibinfo{author}{J.~N. {Bahcall}},
  \bibinfo{journal}{\apj} \bibinfo{volume}{\textbf{541}}, \bibinfo{pages}{707}
  (\bibinfo{date}{Oct. 2000}), \eprint{arXiv:hep-ph/9909286}.
\vspace{-8pt}\bibitem{Meszaros+01choked}
\bibinfo{author}{P.~{M{\'e}sz{\'a}ros}} and \bibinfo{author}{E.~{Waxman}},
  \bibinfo{journal}{Physical Review Letters} \bibinfo{volume}{\textbf{87}}(17),
  \bibinfo{pages}{171102} (\bibinfo{date}{Oct. 2001}),
  \eprint{arXiv:astro-ph/0103275}.
\vspace{-8pt}\bibitem{Murase+08photonu}
\bibinfo{author}{K.~{Murase}}, \bibinfo{journal}{\prd}
  \bibinfo{volume}{\textbf{78}}(10), \bibinfo{pages}{101302},
  \bibinfo{eid}{101302} (\bibinfo{date}{Nov. 2008}), \eprint{0807.0919}.
\vspace{-8pt}\bibitem{Wang+09photonu}
\bibinfo{author}{X.-Y. {Wang}} and \bibinfo{author}{Z.-G. {Dai}},
  \bibinfo{journal}{\apjl} \bibinfo{volume}{\textbf{691}}, \bibinfo{pages}{L67}
  (\bibinfo{date}{Feb. 2009}), \eprint{0807.0290}.
\vspace{-8pt}\bibitem{Gao+11pop3nu}
\bibinfo{author}{S.~{Gao}}, \bibinfo{author}{K.~{Toma}}, and
  \bibinfo{author}{P.~{M{\'e}sz{\'a}ros}}, \bibinfo{journal}{\prd}
  \bibinfo{volume}{\textbf{83}}(10), \bibinfo{pages}{103004}
  (\bibinfo{date}{May 2011}), \eprint{1103.5477}.
\vspace{-8pt}\bibitem{Ahlers+11-grbprob}
\bibinfo{author}{M.~{Ahlers}}, \bibinfo{author}{M.~C. {Gonzalez-Garcia}}, and
  \bibinfo{author}{F.~{Halzen}}, \bibinfo{journal}{Astroparticle Physics}
  \bibinfo{volume}{\textbf{35}}, \bibinfo{pages}{87} (\bibinfo{date}{Sep.
  2011}), \eprint{1103.3421}.
\vspace{-8pt}\bibitem{Abbasi+11-ic40nugrb}
\bibinfo{author}{R.~{Abbasi}}, \bibinfo{author}{Y.~{Abdou}},
  \bibinfo{author}{T.~{Abu-Zayyad}}, \bibinfo{author}{J.~{Adams}},
  \bibinfo{author}{J.~A. {Aguilar}}, \bibinfo{author}{M.~{Ahlers}},
  \bibinfo{author}{K.~{Andeen}}, \bibinfo{author}{J.~{Auffenberg}},
  \bibinfo{author}{X.~{Bai}}, \bibinfo{author}{M.~{Baker}}, \emph{et~al.},
  \bibinfo{journal}{Physical Review Letters}
  \bibinfo{volume}{\textbf{106}}(14), \bibinfo{pages}{141101}
  (\bibinfo{date}{Apr. 2011}), \eprint{1101.1448}.
\vspace{-8pt}\bibitem{Abbasi+12grbnu-nat}
\bibinfo{author}{R.~{Abbasi}}, \bibinfo{author}{Y.~{Abdou}},
  \bibinfo{author}{T.~{Abu-Zayyad}}, \bibinfo{author}{M.~{Ackermann}},
  \bibinfo{author}{J.~{Adams}}, \bibinfo{author}{J.~A. {Aguilar}},
  \bibinfo{author}{M.~{Ahlers}}, \bibinfo{author}{D.~{Altmann}},
  \bibinfo{author}{K.~{Andeen}}, \bibinfo{author}{J.~{Auffenberg}},
  \emph{et~al.}, \bibinfo{journal}{\nat} \bibinfo{volume}{\textbf{484}},
  \bibinfo{pages}{351} (\bibinfo{date}{Apr. 2012}), \eprint{1204.4219}.
\vspace{-8pt}\bibitem{Hummer+11nu-ic3}
\bibinfo{author}{S.~{H{\"u}mmer}}, \bibinfo{author}{P.~{Baerwald}}, and
  \bibinfo{author}{W.~{Winter}}, \bibinfo{journal}{Physical Review Letters}
  \bibinfo{volume}{\textbf{108}}(23), \bibinfo{pages}{231101},
  \bibinfo{eid}{231101} (\bibinfo{date}{Jun. 2012}), \eprint{1112.1076}.
\vspace{-8pt}\bibitem{Li11-nu-ic3}
\bibinfo{author}{Z.~{Li}}, \bibinfo{journal}{\prd}
  \bibinfo{volume}{\textbf{85}}(2), \bibinfo{pages}{027301},
  \bibinfo{eid}{027301} (\bibinfo{date}{Jan. 2012}), \eprint{1112.2240}.
\vspace{-8pt}\bibitem{He+12grbnu}
\bibinfo{author}{H.-N. {He}}, \bibinfo{author}{R.-Y. {Liu}},
  \bibinfo{author}{X.-Y. {Wang}}, \bibinfo{author}{S.~{Nagataki}},
  \bibinfo{author}{K.~{Murase}}, and \bibinfo{author}{Z.-G. {Dai}},
  \bibinfo{journal}{\apj} \bibinfo{volume}{\textbf{752}}, \bibinfo{pages}{29},
  \bibinfo{eid}{29} (\bibinfo{date}{Jun. 2012}), \eprint{1204.0857}.
\vspace{-8pt}\bibitem{Eichler+89ns}
\bibinfo{author}{D.~{Eichler}}, \bibinfo{author}{M.~{Livio}},
  \bibinfo{author}{T.~{Piran}}, and \bibinfo{author}{D.~N. {Schramm}},
  \bibinfo{journal}{\nat} \bibinfo{volume}{\textbf{340}}, \bibinfo{pages}{126}
  (\bibinfo{date}{Jul. 1989}).
\vspace{-8pt}\bibitem{Narayan+92merg}
\bibinfo{author}{R.~{Narayan}}, \bibinfo{author}{B.~{Paczynski}}, and
  \bibinfo{author}{T.~{Piran}}, \bibinfo{journal}{\apjl}
  \bibinfo{volume}{\textbf{395}}, \bibinfo{pages}{L83} (\bibinfo{date}{Aug.
  1992}), \eprint{arXiv:astro-ph/9204001}.
\vspace{-8pt}\bibitem{Meszaros+92tidal}
\bibinfo{author}{P.~{Meszaros}} and \bibinfo{author}{M.~J. {Rees}},
  \bibinfo{journal}{\apj} \bibinfo{volume}{\textbf{397}}, \bibinfo{pages}{570}
  (\bibinfo{date}{Oct. 1992}).
\vspace{-8pt}\bibitem{Woosley93col}
\bibinfo{author}{S.~E. {Woosley}}, \bibinfo{journal}{\apj}
  \bibinfo{volume}{\textbf{405}}, \bibinfo{pages}{273} (\bibinfo{date}{Mar.
  1993}).
\vspace{-8pt}\bibitem{Paczynski98}
\bibinfo{author}{B.~{Pacz\'ynski}}, \bibinfo{journal}{\apjl}
  \bibinfo{volume}{\textbf{494}}, \bibinfo{pages}{L45+} (\bibinfo{date}{Feb.
  1998}), \eprint{arXiv:astro-ph/9710086}.
\vspace{-8pt}\bibitem{Macfadyen+99col}
\bibinfo{author}{A.~I. {MacFadyen}} and \bibinfo{author}{S.~E. {Woosley}},
  \bibinfo{journal}{\apj} \bibinfo{volume}{\textbf{524}}, \bibinfo{pages}{262}
  (\bibinfo{date}{Oct. 1999}), \eprint{arXiv:astro-ph/9810274}.
\vspace{-8pt}\bibitem{Wheeler+00grbmag}
\bibinfo{author}{J.~C. {Wheeler}}, \bibinfo{author}{I.~{Yi}},
  \bibinfo{author}{P.~{H{\"o}flich}}, and \bibinfo{author}{L.~{Wang}},
  \bibinfo{journal}{\apj} \bibinfo{volume}{\textbf{537}}, \bibinfo{pages}{810}
  (\bibinfo{date}{Jul. 2000}), \eprint{arXiv:astro-ph/9909293}.
\vspace{-8pt}\bibitem{Ruffert+99merg}
\bibinfo{author}{M.~{Ruffert}} and \bibinfo{author}{H.-T. {Janka}},
  \bibinfo{journal}{\aap} \bibinfo{volume}{\textbf{344}}, \bibinfo{pages}{573}
  (\bibinfo{date}{Apr. 1999}), \eprint{arXiv:astro-ph/9809280}.
\vspace{-8pt}\bibitem{Rosswog05nsbh}
\bibinfo{author}{S.~{Rosswog}}, \bibinfo{journal}{\apj}
  \bibinfo{volume}{\textbf{634}}, \bibinfo{pages}{1202} (\bibinfo{date}{Dec.
  2005}), \eprint{arXiv:astro-ph/0508138}.
\vspace{-8pt}\bibitem{Rezzolla+11sgrb-mag}
\bibinfo{author}{L.~{Rezzolla}}, \bibinfo{author}{B.~{Giacomazzo}},
  \bibinfo{author}{L.~{Baiotti}}, \bibinfo{author}{J.~{Granot}},
  \bibinfo{author}{C.~{Kouveliotou}}, and \bibinfo{author}{M.~A. {Aloy}},
  \bibinfo{journal}{\apjl} \bibinfo{volume}{\textbf{732}}, \bibinfo{pages}{L6},
  \bibinfo{eid}{L6} (\bibinfo{date}{May 2011}), \eprint{1101.4298}.
\vspace{-8pt}\bibitem{Macfadyen+01sncol}
\bibinfo{author}{A.~I. {MacFadyen}}, \bibinfo{author}{S.~E. {Woosley}}, and
  \bibinfo{author}{A.~{Heger}}, \bibinfo{journal}{\apj}
  \bibinfo{volume}{\textbf{550}}, \bibinfo{pages}{410} (\bibinfo{date}{Mar.
  2001}), \eprint{arXiv:astro-ph/9910034}.
\vspace{-8pt}\bibitem{Zhang+04jetnum}
\bibinfo{author}{W.~{Zhang}}, \bibinfo{author}{S.~E. {Woosley}}, and
  \bibinfo{author}{A.~{Heger}}, \bibinfo{journal}{\apj}
  \bibinfo{volume}{\textbf{608}}, \bibinfo{pages}{365} (\bibinfo{date}{Jun.
  2004}), \eprint{arXiv:astro-ph/0308389}.
\vspace{-8pt}\bibitem{Paczynski98grbsfr}
\bibinfo{author}{B.~{Pacz\'ynski}}, \bibinfo{journal}{\apjl}
  \bibinfo{volume}{\textbf{494}}, \bibinfo{pages}{L45+} (\bibinfo{date}{Feb.
  1998}), \eprint{arXiv:astro-ph/9710086}.
\vspace{-8pt}\bibitem{Fryer+99}
\bibinfo{author}{C.~L. {Fryer}}, \bibinfo{author}{S.~E. {Woosley}}, and
  \bibinfo{author}{D.~H. {Hartmann}}, \bibinfo{journal}{\apj}
  \bibinfo{volume}{\textbf{526}}, \bibinfo{pages}{152} (\bibinfo{date}{Nov.
  1999}), \eprint{arXiv:astro-ph/9904122}.
\vspace{-8pt}\bibitem{Fryer06col}
\bibinfo{author}{C.~L. {Fryer}}, \bibinfo{journal}{New Astronomy Review}
  \bibinfo{volume}{\textbf{50}}, \bibinfo{pages}{492} (\bibinfo{date}{Oct.
  2006}).
\vspace{-8pt}\bibitem{Woosley11prog}
\bibinfo{author}{S.~E. {Woosley}}, \bibinfo{journal}{ArXiv e-prints}
  (\bibinfo{date}{May 2011}), \eprint{1105.4193}.
\vspace{-8pt}\bibitem{Stanek+06}
\bibinfo{author}{K.~Z. {Stanek}}, \bibinfo{author}{O.~Y. {Gnedin}},
  \bibinfo{author}{J.~F. {Beacom}}, \bibinfo{author}{A.~P. {Gould}},
  \bibinfo{author}{J.~A. {Johnson}}, \bibinfo{author}{J.~A. {Kollmeier}},
  \bibinfo{author}{M.~{Modjaz}}, \bibinfo{author}{M.~H. {Pinsonneault}},
  \bibinfo{author}{R.~{Pogge}}, and \bibinfo{author}{D.~H. {Weinberg}},
  \bibinfo{journal}{Acta Astronomica} \bibinfo{volume}{\textbf{56}},
  \bibinfo{pages}{333} (\bibinfo{date}{Dec. 2006}),
  \eprint{arXiv:astro-ph/0604113}.
\vspace{-8pt}\bibitem{Galama+98-980425}
\bibinfo{author}{T.~J. {Galama}}, \bibinfo{author}{P.~M. {Vreeswijk}},
  \bibinfo{author}{J.~{van Paradijs}}, \bibinfo{author}{C.~{Kouveliotou}},
  \bibinfo{author}{T.~{Augusteijn}}, \bibinfo{author}{H.~{B{\"o}hnhardt}},
  \bibinfo{author}{J.~P. {Brewer}}, \bibinfo{author}{V.~{Doublier}},
  \bibinfo{author}{J.-F. {Gonzalez}}, \bibinfo{author}{B.~{Leibundgut}},
  \emph{et~al.}, \bibinfo{journal}{\nat} \bibinfo{volume}{\textbf{395}},
  \bibinfo{pages}{670} (\bibinfo{date}{Oct. 1998}),
  \eprint{arXiv:astro-ph/9806175}.
\vspace{-8pt}\bibitem{Hjorth+03-030329sn}
\bibinfo{author}{J.~{Hjorth}}, \bibinfo{author}{J.~{Sollerman}},
  \bibinfo{author}{P.~{M{\o}ller}}, \bibinfo{author}{J.~P.~U. {Fynbo}},
  \bibinfo{author}{S.~E. {Woosley}}, \bibinfo{author}{C.~{Kouveliotou}},
  \bibinfo{author}{N.~R. {Tanvir}}, \bibinfo{author}{J.~{Greiner}},
  \bibinfo{author}{M.~I. {Andersen}}, \bibinfo{author}{A.~J. {Castro-Tirado}},
  \emph{et~al.}, \bibinfo{journal}{\nat} \bibinfo{volume}{\textbf{423}},
  \bibinfo{pages}{847} (\bibinfo{date}{Jun. 2003}),
  \eprint{arXiv:astro-ph/0306347}.
\vspace{-8pt}\bibitem{Dellavalle11grbsn}
\bibinfo{author}{M.~{Della Valle}}, \bibinfo{journal}{International Journal of
  Modern Physics D} \bibinfo{volume}{\textbf{20}}, \bibinfo{pages}{1745}
  (\bibinfo{date}{2011}).
\vspace{-8pt}\bibitem{Hjorth+11grbsn}
\bibinfo{author}{J.~{Hjorth}} and \bibinfo{author}{J.~S. {Bloom}},
  \bibinfo{journal}{ArXiv e-prints}  (\bibinfo{date}{Apr. 2011}),
  \eprint{1104.2274}.
\vspace{-8pt}\bibitem{Soderberg+07sne}
\bibinfo{author}{A.~M. {Soderberg}}, in \bibinfo{editors}{S.~{Immler},
  K.~{Weiler}, and R.~{McCray}}, eds., \emph{Supernova 1987A: 20 Years After:
  Supernovae and Gamma-Ray Bursters} (\bibinfo{date}{Oct. 2007}),
  \bibinfo{volume}{vol. 937 of \emph{American Institute of Physics Conference
  Series}}, \bibinfo{pages}{pp. 492--499}.
\vspace{-8pt}\bibitem{Soderberg+10hn}
\bibinfo{author}{A.~M. {Soderberg}}, \bibinfo{author}{S.~{Chakraborti}},
  \bibinfo{author}{G.~{Pignata}}, \bibinfo{author}{R.~A. {Chevalier}},
  \bibinfo{author}{P.~{Chandra}}, \bibinfo{author}{A.~{Ray}},
  \bibinfo{author}{M.~H. {Wieringa}}, \bibinfo{author}{A.~{Copete}},
  \bibinfo{author}{V.~{Chaplin}}, \bibinfo{author}{V.~{Connaughton}},
  \emph{et~al.}, \bibinfo{journal}{\nat} \bibinfo{volume}{\textbf{463}},
  \bibinfo{pages}{513} (\bibinfo{date}{Jan. 2010}), \eprint{0908.2817}.
\vspace{-8pt}\bibitem{Paczynski98grbhn}
\bibinfo{author}{B.~{Paczy{\'n}ski}}, in \bibinfo{editors}{C.~A. {Meegan},
  R.~D. {Preece}, and T.~M. {Koshut}}, eds., \emph{Gamma-Ray Bursts, 4th
  Hunstville Symposium} (\bibinfo{date}{May 1998}), \bibinfo{volume}{vol. 428
  of \emph{American Institute of Physics Conference Series}},
  \bibinfo{pages}{pp. 783--787}, \eprint{arXiv:astro-ph/9706232}.
\vspace{-8pt}\bibitem{Waxman+99-1998bw}
\bibinfo{author}{E.~{Waxman}} and \bibinfo{author}{A.~{Loeb}},
  \bibinfo{journal}{\apj} \bibinfo{volume}{\textbf{515}}, \bibinfo{pages}{721}
  (\bibinfo{date}{Apr. 1999}), \eprint{arXiv:astro-ph/9808135}.
\vspace{-8pt}\bibitem{Nomoto+10hn}
\bibinfo{author}{K.~{Nomoto}}, \bibinfo{author}{T.~{Moriya}},
  \bibinfo{author}{N.~{Tominaga}}, and \bibinfo{author}{T.~{Suzuki}}, in
  \bibinfo{editors}{N.~{Kawai} and S.~{Nagataki}}, eds., \emph{American
  Institute of Physics Conference Series} (\bibinfo{date}{Oct. 2010}),
  \bibinfo{volume}{vol. 1279 of \emph{American Institute of Physics Conference
  Series}}, \bibinfo{pages}{pp. 60--68}.
\vspace{-8pt}\bibitem{Thielemann+11sn}
\bibinfo{author}{F.-K. {Thielemann}}, \bibinfo{author}{R.~{Hirschi}},
  \bibinfo{author}{M.~{Liebend{\"o}rfer}}, and \bibinfo{author}{R.~{Diehl}}, in
  \bibinfo{editors}{R.~{Diehl}, D.~H. {Hartmann}, and N.~{Prantzos}}, eds.,
  \emph{Lecture Notes in Physics, Berlin Springer Verlag}
  (\bibinfo{date}{2011}), \bibinfo{volume}{vol. 812 of \emph{Lecture Notes in
  Physics, Berlin Springer Verlag}}, \bibinfo{pages}{pp. 153--232},
  \eprint{1008.2144}.
\vspace{-8pt}\bibitem{Lazzati+12hn}
\bibinfo{author}{D.~{Lazzati}}, \bibinfo{author}{B.~J. {Morsony}},
  \bibinfo{author}{C.~H. {Blackwell}}, and \bibinfo{author}{M.~C. {Begelman}},
  \bibinfo{journal}{\apj} \bibinfo{volume}{\textbf{750}}, \bibinfo{pages}{68},
  \bibinfo{eid}{68} (\bibinfo{date}{May 2012}), \eprint{1111.0970}.
\vspace{-8pt}\bibitem{Gehrels+05sgrb}
\bibinfo{author}{N.~{Gehrels}}, \bibinfo{author}{C.~L. {Sarazin}},
  \bibinfo{author}{P.~T. {O'Brien}}, \bibinfo{author}{B.~{Zhang}},
  \bibinfo{author}{L.~{Barbier}}, \bibinfo{author}{S.~D. {Barthelmy}},
  \bibinfo{author}{A.~{Blustin}}, \bibinfo{author}{D.~N. {Burrows}},
  \bibinfo{author}{J.~{Cannizzo}}, \bibinfo{author}{J.~R. {Cummings}},
  \emph{et~al.}, \bibinfo{journal}{\nat} \bibinfo{volume}{\textbf{437}},
  \bibinfo{pages}{851} (\bibinfo{date}{Oct. 2005}),
  \eprint{arXiv:astro-ph/0505630}.
\vspace{-8pt}\bibitem{Berger+05-050724}
\bibinfo{author}{E.~{Berger}}, \bibinfo{author}{P.~A. {Price}},
  \bibinfo{author}{S.~B. {Cenko}}, \bibinfo{author}{A.~{Gal-Yam}},
  \bibinfo{author}{A.~M. {Soderberg}}, \bibinfo{author}{M.~{Kasliwal}},
  \bibinfo{author}{D.~C. {Leonard}}, \bibinfo{author}{P.~B. {Cameron}},
  \bibinfo{author}{D.~A. {Frail}}, \bibinfo{author}{S.~R. {Kulkarni}},
  \emph{et~al.}, \bibinfo{journal}{\nat} \bibinfo{volume}{\textbf{438}},
  \bibinfo{pages}{988} (\bibinfo{date}{Dec. 2005}),
  \eprint{arXiv:astro-ph/0508115}.
\vspace{-8pt}\bibitem{Berger+06sgrb}
\bibinfo{author}{E.~{Berger}}, \bibinfo{author}{B.~E. {Penprase}},
  \bibinfo{author}{S.~B. {Cenko}}, \bibinfo{author}{S.~R. {Kulkarni}},
  \bibinfo{author}{D.~B. {Fox}}, \bibinfo{author}{C.~C. {Steidel}}, and
  \bibinfo{author}{N.~A. {Reddy}}, \bibinfo{journal}{\apj}
  \bibinfo{volume}{\textbf{642}}, \bibinfo{pages}{979} (\bibinfo{date}{May
  2006}), \eprint{arXiv:astro-ph/0511498}.
\vspace{-8pt}\bibitem{Nakar07sgrb}
\bibinfo{author}{E.~{Nakar}}, \bibinfo{journal}{\physrep}
  \bibinfo{volume}{\textbf{442}}, \bibinfo{pages}{166} (\bibinfo{date}{Apr.
  2007}), \eprint{arXiv:astro-ph/0701748}.
\vspace{-8pt}\bibitem{Berger+07sgrb}
\bibinfo{author}{E.~{Berger}}, \bibinfo{author}{D.~B. {Fox}},
  \bibinfo{author}{P.~A. {Price}}, \bibinfo{author}{E.~{Nakar}},
  \bibinfo{author}{A.~{Gal-Yam}}, \bibinfo{author}{D.~E. {Holz}},
  \bibinfo{author}{B.~P. {Schmidt}}, \bibinfo{author}{A.~{Cucchiara}},
  \bibinfo{author}{S.~B. {Cenko}}, \bibinfo{author}{S.~R. {Kulkarni}},
  \emph{et~al.}, \bibinfo{journal}{\apj} \bibinfo{volume}{\textbf{664}},
  \bibinfo{pages}{1000} (\bibinfo{date}{Aug. 2007}),
  \eprint{arXiv:astro-ph/0611128}.
\vspace{-8pt}\bibitem{Berger11sgrb}
\bibinfo{author}{E.~{Berger}}, \bibinfo{journal}{\nar}
  \bibinfo{volume}{\textbf{55}}, \bibinfo{pages}{1} (\bibinfo{date}{Jan.
  2011}), \eprint{1005.1068}.
\vspace{-8pt}\bibitem{Coward+12dnsmerg}
\bibinfo{author}{D.~{Coward}}, \bibinfo{author}{E.~{Howell}},
  \bibinfo{author}{T.~{Piran}}, \bibinfo{author}{G.~{Stratta}},
  \bibinfo{author}{M.~{Branchesi}}, \bibinfo{author}{O.~{Bromberg}},
  \bibinfo{author}{B.~{Gendre}}, \bibinfo{author}{R.~{Burman}}, and
  \bibinfo{author}{D.~{Guetta}}, \bibinfo{journal}{ArXiv e-prints}
  (\bibinfo{date}{Feb. 2012}), \eprint{1202.2179}.
\vspace{-8pt}\bibitem{Kiel+10dns-col-popsynt}
\bibinfo{author}{P.~D. {Kiel}}, \bibinfo{author}{J.~R. {Hurley}}, and
  \bibinfo{author}{M.~{Bailes}}, \bibinfo{journal}{\mnras}
  \bibinfo{volume}{\textbf{406}}, \bibinfo{pages}{656} (\bibinfo{date}{Jul.
  2010}), \eprint{1004.0131}.
\vspace{-8pt}\bibitem{Shibata+11sgrb-mag}
\bibinfo{author}{M.~{Shibata}}, \bibinfo{author}{Y.~{Suwa}},
  \bibinfo{author}{K.~{Kiuchi}}, and \bibinfo{author}{K.~{Ioka}},
  \bibinfo{journal}{\apjl} \bibinfo{volume}{\textbf{734}},
  \bibinfo{pages}{L36}, \bibinfo{eid}{L36} (\bibinfo{date}{Jun. 2011}),
  \eprint{1105.3302}.
\vspace{-8pt}\bibitem{Greiner+09-z6.7}
\bibinfo{author}{J.~{Greiner}}, \bibinfo{author}{T.~{Kr{\"u}hler}},
  \bibinfo{author}{J.~P.~U. {Fynbo}}, \bibinfo{author}{A.~{Rossi}},
  \bibinfo{author}{R.~{Schwarz}}, \bibinfo{author}{S.~{Klose}},
  \bibinfo{author}{S.~{Savaglio}}, \bibinfo{author}{N.~R. {Tanvir}},
  \bibinfo{author}{S.~{McBreen}}, \bibinfo{author}{T.~{Totani}}, \emph{et~al.},
  \bibinfo{journal}{\apj} \bibinfo{volume}{\textbf{693}}, \bibinfo{pages}{1610}
  (\bibinfo{date}{Mar. 2009}), \eprint{0810.2314}.
\vspace{-8pt}\bibitem{Tanvir+09-z8.2}
\bibinfo{author}{N.~R. {Tanvir}}, \bibinfo{author}{D.~B. {Fox}},
  \bibinfo{author}{A.~J. {Levan}}, \bibinfo{author}{E.~{Berger}},
  \bibinfo{author}{K.~{Wiersema}}, \bibinfo{author}{J.~P.~U. {Fynbo}},
  \bibinfo{author}{A.~{Cucchiara}}, \bibinfo{author}{T.~{Kr{\"u}hler}},
  \bibinfo{author}{N.~{Gehrels}}, \bibinfo{author}{J.~S. {Bloom}},
  \emph{et~al.}, \bibinfo{journal}{\nat} \bibinfo{volume}{\textbf{461}},
  \bibinfo{pages}{1254} (\bibinfo{date}{Oct. 2009}), \eprint{0906.1577}.
\vspace{-8pt}\bibitem{Salvaterra+09-z8.2}
\bibinfo{author}{R.~{Salvaterra}}, \bibinfo{author}{M.~{Della Valle}},
  \bibinfo{author}{S.~{Campana}}, \bibinfo{author}{G.~{Chincarini}},
  \bibinfo{author}{S.~{Covino}}, \bibinfo{author}{P.~{D'Avanzo}},
  \bibinfo{author}{A.~{Fern{\'a}ndez-Soto}}, \bibinfo{author}{C.~{Guidorzi}},
  \bibinfo{author}{F.~{Mannucci}}, \bibinfo{author}{R.~{Margutti}},
  \emph{et~al.}, \bibinfo{journal}{\nat} \bibinfo{volume}{\textbf{461}},
  \bibinfo{pages}{1258} (\bibinfo{date}{Oct. 2009}), \eprint{0906.1578}.
\vspace{-8pt}\bibitem{Cucchiara+11-z9}
\bibinfo{author}{A.~{Cucchiara}}, \bibinfo{author}{A.~J. {Levan}},
  \bibinfo{author}{D.~B. {Fox}}, \bibinfo{author}{N.~R. {Tanvir}},
  \bibinfo{author}{T.~N. {Ukwatta}}, \bibinfo{author}{E.~{Berger}},
  \bibinfo{author}{T.~{Kr{\"u}hler}}, \bibinfo{author}{A.~{K{\"u}pc{\"u}
  Yolda{\c s}}}, \bibinfo{author}{X.~F. {Wu}}, \bibinfo{author}{K.~{Toma}},
  \emph{et~al.}, \bibinfo{journal}{\apj} \bibinfo{volume}{\textbf{736}},
  \bibinfo{pages}{7}, \bibinfo{eid}{7} (\bibinfo{date}{Jul. 2011}),
  \eprint{1105.4915}.
\vspace{-8pt}\bibitem{Bromm+06hizgrb}
\bibinfo{author}{V.~{Bromm}} and \bibinfo{author}{A.~{Loeb}},
  \bibinfo{journal}{\apj} \bibinfo{volume}{\textbf{642}}, \bibinfo{pages}{382}
  (\bibinfo{date}{May 2006}), \eprint{arXiv:astro-ph/0509303}.
\vspace{-8pt}\bibitem{Komissarov+10pop3}
\bibinfo{author}{S.~S. {Komissarov}} and \bibinfo{author}{M.~V. {Barkov}},
  \bibinfo{journal}{\mnras} \bibinfo{volume}{\textbf{402}},
  \bibinfo{pages}{L25} (\bibinfo{date}{Feb. 2010}), \eprint{0909.0857}.
\vspace{-8pt}\bibitem{Stacy+09pop3}
\bibinfo{author}{A.~{Stacy}}, \bibinfo{author}{T.~H. {Greif}}, and
  \bibinfo{author}{V.~{Bromm}}, in \bibinfo{editors}{{S.~Jogee, I.~Marinova,
  L.~Hao, \& G.~A.~Blanc}}, ed., \emph{Galaxy Evolution: Emerging Insights and
  Future Challenges} (\bibinfo{date}{Dec. 2009}), \bibinfo{volume}{vol. 419 of
  \emph{Astronomical Society of the Pacific Conference Series}},
  \bibinfo{pages}{p. 339}.
\vspace{-8pt}\bibitem{Norman10pop3}
\bibinfo{author}{M.~L. {Norman}}, in \bibinfo{editors}{{D.~J.~Whalen, V.~Bromm,
  \& N.~Yoshida}}, ed., \emph{American Institute of Physics Conference Series}
  (\bibinfo{date}{Nov. 2010}), \bibinfo{volume}{vol. 1294 of \emph{American
  Institute of Physics Conference Series}}, \bibinfo{pages}{pp. 17--27},
  \eprint{1011.4624}.
\vspace{-8pt}\bibitem{Meszaros+10pop3}
\bibinfo{author}{P.~{M{\'e}sz{\'a}ros}} and \bibinfo{author}{M.~J. {Rees}},
  \bibinfo{journal}{\apj} \bibinfo{volume}{\textbf{715}}, \bibinfo{pages}{967}
  (\bibinfo{date}{Jun. 2010}), \eprint{1004.2056}.
\vspace{-8pt}\bibitem{Toma+11pop3}
\bibinfo{author}{K.~{Toma}}, \bibinfo{author}{T.~{Sakamoto}}, and
  \bibinfo{author}{P.~{M{\'e}sz{\'a}ros}}, \bibinfo{journal}{\apj}
  \bibinfo{volume}{\textbf{731}}, \bibinfo{pages}{127} (\bibinfo{date}{Apr.
  2011}), \eprint{1008.1269}.
\vspace{-8pt}\bibitem{Amati+06episo}
\bibinfo{author}{L.~{Amati}}, \bibinfo{journal}{\mnras}
  \bibinfo{volume}{\textbf{372}}, \bibinfo{pages}{233} (\bibinfo{date}{Oct.
  2006}), \eprint{arXiv:astro-ph/0601553}.
\vspace{-8pt}\bibitem{Yonetoku+04-EpkLum}
\bibinfo{author}{D.~{Yonetoku}}, \bibinfo{author}{T.~{Murakami}},
  \bibinfo{author}{T.~{Nakamura}}, \bibinfo{author}{R.~{Yamazaki}},
  \bibinfo{author}{A.~K. {Inoue}}, and \bibinfo{author}{K.~{Ioka}},
  \bibinfo{journal}{\apj} \bibinfo{volume}{\textbf{609}}, \bibinfo{pages}{935}
  (\bibinfo{date}{Jul. 2004}), \eprint{arXiv:astro-ph/0309217}.
\vspace{-8pt}\bibitem{Ghirlanda+04-EpkEjet}
\bibinfo{author}{G.~{Ghirlanda}}, \bibinfo{author}{G.~{Ghisellini}}, and
  \bibinfo{author}{D.~{Lazzati}}, \bibinfo{journal}{\apj}
  \bibinfo{volume}{\textbf{616}}, \bibinfo{pages}{331} (\bibinfo{date}{Nov.
  2004}), \eprint{arXiv:astro-ph/0405602}.
\vspace{-8pt}\bibitem{Ghirlanda07grbcosm}
\bibinfo{author}{G.~{Ghirlanda}}, \bibinfo{journal}{Royal Society of London
  Philosophical Transactions Series A} \bibinfo{volume}{\textbf{365}},
  \bibinfo{pages}{1385} (\bibinfo{date}{May 2007}),
  \eprint{arXiv:astro-ph/0702212}.
\vspace{-8pt}\bibitem{Dai+04cosm}
\bibinfo{author}{Z.~G. {Dai}}, \bibinfo{author}{E.~W. {Liang}}, and
  \bibinfo{author}{D.~{Xu}}, \bibinfo{journal}{\apjl}
  \bibinfo{volume}{\textbf{612}}, \bibinfo{pages}{L101} (\bibinfo{date}{Sep.
  2004}), \eprint{arXiv:astro-ph/0407497}.
\vspace{-8pt}\bibitem{Liang+06-EisoEpktbrk}
\bibinfo{author}{E.~{Liang}} and \bibinfo{author}{B.~{Zhang}},
  \bibinfo{journal}{\mnras} \bibinfo{volume}{\textbf{369}},
  \bibinfo{pages}{L37} (\bibinfo{date}{Jun. 2006}),
  \eprint{arXiv:astro-ph/0512177}.
\vspace{-8pt}\bibitem{Cardone+10grbcosm}
\bibinfo{author}{V.~F. {Cardone}}, \bibinfo{author}{M.~G. {Dainotti}},
  \bibinfo{author}{S.~{Capozziello}}, and \bibinfo{author}{R.~{Willingale}},
  \bibinfo{journal}{\mnras} \bibinfo{volume}{\textbf{408}},
  \bibinfo{pages}{1181} (\bibinfo{date}{Oct. 2010}), \eprint{1005.0122}.
\vspace{-8pt}\bibitem{Zhang+02break}
\bibinfo{author}{B.~{Zhang}} and \bibinfo{author}{P.~{M{\'e}sz{\'a}ros}},
  \bibinfo{journal}{\apj} \bibinfo{volume}{\textbf{581}}, \bibinfo{pages}{1236}
  (\bibinfo{date}{Dec. 2002}), \eprint{arXiv:astro-ph/0206158}.
\vspace{-8pt}\bibitem{Ghirlanda+12grbcorr}
\bibinfo{author}{G.~{Ghirlanda}}, \bibinfo{author}{L.~{Nava}},
  \bibinfo{author}{G.~{Ghisellini}}, \bibinfo{author}{A.~{Celotti}},
  \bibinfo{author}{D.~{Burlon}}, \bibinfo{author}{S.~{Covino}}, and
  \bibinfo{author}{A.~{Melandri}}, \bibinfo{journal}{\mnras}
  \bibinfo{volume}{\textbf{420}}, \bibinfo{pages}{483} (\bibinfo{date}{Feb.
  2012}), \eprint{1107.4096}.
\vspace{-8pt}\bibitem{Nakar+05corr}
\bibinfo{author}{E.~{Nakar}} and \bibinfo{author}{T.~{Piran}},
  \bibinfo{journal}{\mnras} \bibinfo{volume}{\textbf{360}},
  \bibinfo{pages}{L73} (\bibinfo{date}{Jun. 2005}),
  \eprint{arXiv:astro-ph/0412232}.
\vspace{-8pt}\bibitem{Ghirlanda+12grbcosm}
\bibinfo{author}{G.~{Ghirlanda}}, \bibinfo{author}{G.~{Ghisellini}},
  \bibinfo{author}{L.~{Nava}}, \bibinfo{author}{R.~{Salvaterra}},
  \bibinfo{author}{G.~{Tagliaferri}}, \bibinfo{author}{S.~{Campana}},
  \bibinfo{author}{S.~{Covino}}, \bibinfo{author}{P.~{D'Avanzo}},
  \bibinfo{author}{D.~{Fugazza}}, \bibinfo{author}{A.~{Melandri}},
  \emph{et~al.}, \bibinfo{journal}{\mnras} \bibinfo{pages}{p. 2753}
  (\bibinfo{date}{Mar. 2012}), \eprint{1203.0003}.
\vspace{-8pt}\bibitem{Graziani11grbcosm}
\bibinfo{author}{C.~{Graziani}}, \bibinfo{journal}{\na}
  \bibinfo{volume}{\textbf{16}}, \bibinfo{pages}{57} (\bibinfo{date}{Feb.
  2011}), \eprint{1002.3434}.
\vspace{-8pt}\bibitem{Liang+10grbcosm}
\bibinfo{author}{N.~{Liang}}, \bibinfo{author}{P.~{Wu}}, and
  \bibinfo{author}{S.~N. {Zhang}}, \bibinfo{journal}{\prd}
  \bibinfo{volume}{\textbf{81}}(8), \bibinfo{pages}{083518},
  \bibinfo{eid}{083518} (\bibinfo{date}{Apr. 2010}), \eprint{0911.5644}.
\vspace{-8pt}\bibitem{Demianski+11grbcosm}
\bibinfo{author}{M.~{Demianski}} and \bibinfo{author}{E.~{Piedipalumbo}},
  \bibinfo{journal}{\mnras} \bibinfo{volume}{\textbf{415}},
  \bibinfo{pages}{3580} (\bibinfo{date}{Aug. 2011}), \eprint{1104.5614}.
\vspace{-8pt}\bibitem{Savaglio06chem}
\bibinfo{author}{S.~{Savaglio}}, \bibinfo{journal}{New Journal of Physics}
  \bibinfo{volume}{\textbf{8}}, \bibinfo{pages}{195} (\bibinfo{date}{Sep.
  2006}), \eprint{arXiv:astro-ph/0609489}.
\vspace{-8pt}\bibitem{Lamb+00hiz}
\bibinfo{author}{D.~Q. {Lamb}} and \bibinfo{author}{D.~E. {Reichart}},
  \bibinfo{journal}{\apj} \bibinfo{volume}{\textbf{536}}, \bibinfo{pages}{1}
  (\bibinfo{date}{Jun. 2000}), \eprint{arXiv:astro-ph/0002035}.
\vspace{-8pt}\bibitem{Ciardi+00hizgrb}
\bibinfo{author}{B.~{Ciardi}} and \bibinfo{author}{A.~{Loeb}},
  \bibinfo{journal}{\apj} \bibinfo{volume}{\textbf{540}}, \bibinfo{pages}{687}
  (\bibinfo{date}{Sep. 2000}), \eprint{arXiv:astro-ph/0002412}.
\vspace{-8pt}\bibitem{Meszaros+03gauge}
\bibinfo{author}{P.~{M{\'e}sz{\'a}ros}} and \bibinfo{author}{M.~J. {Rees}},
  \bibinfo{journal}{\apjl} \bibinfo{volume}{\textbf{591}}, \bibinfo{pages}{L91}
  (\bibinfo{date}{Jul. 2003}), \eprint{arXiv:astro-ph/0305115}.
\vspace{-8pt}\bibitem{Gou+04hiz}
\bibinfo{author}{L.~J. {Gou}}, \bibinfo{author}{P.~{M{\'e}sz{\'a}ros}},
  \bibinfo{author}{T.~{Abel}}, and \bibinfo{author}{B.~{Zhang}},
  \bibinfo{journal}{\apj} \bibinfo{volume}{\textbf{604}}, \bibinfo{pages}{508}
  (\bibinfo{date}{Apr. 2004}), \eprint{arXiv:astro-ph/0307489}.
\vspace{-8pt}\bibitem{Loeb+01reioniz}
\bibinfo{author}{A.~{Loeb}} and \bibinfo{author}{R.~{Barkana}},
  \bibinfo{journal}{\araa} \bibinfo{volume}{\textbf{39}}, \bibinfo{pages}{19}
  (\bibinfo{date}{2001}), \eprint{arXiv:astro-ph/0010467}.
\vspace{-8pt}\bibitem{Savaglio10grbchem}
\bibinfo{author}{S.~{Savaglio}}, in \bibinfo{editors}{K.~{Cunha}, M.~{Spite},
  and B.~{Barbuy}}, eds., \emph{IAU Symposium} (\bibinfo{date}{Mar. 2010}),
  \bibinfo{volume}{vol. 265 of \emph{IAU Symposium}}, \bibinfo{pages}{pp.
  139--146}, \eprint{0911.2328}.
\vspace{-8pt}\bibitem{Hartmann10grbchem}
\bibinfo{author}{D.~{Hartmann}}, in \emph{AAS/High Energy Astrophysics Division
  \#11} (\bibinfo{date}{Feb. 2010}), \bibinfo{volume}{vol.~42 of \emph{Bulletin
  of the American Astronomical Society}}, \bibinfo{pages}{p. 677}.
\vspace{-8pt}\bibitem{Kistler+09sfrgrb}
\bibinfo{author}{M.~D. {Kistler}}, \bibinfo{author}{H.~{Y{\"u}ksel}},
  \bibinfo{author}{J.~F. {Beacom}}, \bibinfo{author}{A.~M. {Hopkins}}, and
  \bibinfo{author}{J.~S.~B. {Wyithe}}, \bibinfo{journal}{\apjl}
  \bibinfo{volume}{\textbf{705}}, \bibinfo{pages}{L104} (\bibinfo{date}{Nov.
  2009}), \eprint{0906.0590}.
\vspace{-8pt}\bibitem{Hopkins+06sfr}
\bibinfo{author}{A.~M. {Hopkins}} and \bibinfo{author}{J.~F. {Beacom}},
  \bibinfo{journal}{\apj} \bibinfo{volume}{\textbf{651}}, \bibinfo{pages}{142}
  (\bibinfo{date}{Nov. 2006}), \eprint{arXiv:astro-ph/0601463}.
\vspace{-8pt}\bibitem{Ota+08reioniz}
\bibinfo{author}{K.~{Ota}}, \bibinfo{author}{M.~{Iye}},
  \bibinfo{author}{N.~{Kashikawa}}, \bibinfo{author}{K.~{Shimasaku}},
  \bibinfo{author}{M.~{Kobayashi}}, \bibinfo{author}{T.~{Totani}},
  \bibinfo{author}{M.~{Nagashima}}, \bibinfo{author}{T.~{Morokuma}},
  \bibinfo{author}{H.~{Furusawa}}, \bibinfo{author}{T.~{Hattori}},
  \emph{et~al.}, \bibinfo{journal}{\apj} \bibinfo{volume}{\textbf{677}},
  \bibinfo{pages}{12} (\bibinfo{date}{Apr. 2008}), \eprint{0707.1561}.
\vspace{-8pt}\bibitem{Bouwens+08hiz}
\bibinfo{author}{R.~J. {Bouwens}}, \bibinfo{author}{G.~D. {Illingworth}},
  \bibinfo{author}{M.~{Franx}}, and \bibinfo{author}{H.~{Ford}},
  \bibinfo{journal}{\apj} \bibinfo{volume}{\textbf{686}}, \bibinfo{pages}{230}
  (\bibinfo{date}{Oct. 2008}), \eprint{0803.0548}.
\vspace{-8pt}\bibitem{Madau+99reioniz}
\bibinfo{author}{P.~{Madau}}, \bibinfo{author}{F.~{Haardt}}, and
  \bibinfo{author}{M.~J. {Rees}}, \bibinfo{journal}{\apj}
  \bibinfo{volume}{\textbf{514}}, \bibinfo{pages}{648} (\bibinfo{date}{Apr.
  1999}), \eprint{arXiv:astro-ph/9809058}.
\vspace{-8pt}\bibitem{Robertson+12sfr}
\bibinfo{author}{B.~E. {Robertson}} and \bibinfo{author}{R.~S. {Ellis}},
  \bibinfo{journal}{\apj} \bibinfo{volume}{\textbf{744}}, \bibinfo{pages}{95},
  \bibinfo{eid}{95} (\bibinfo{date}{Jan. 2012}), \eprint{1109.0990}.
\vspace{-8pt}\bibitem{Campisi+11pop3grb}
\bibinfo{author}{M.~A. {Campisi}}, \bibinfo{author}{U.~{Maio}},
  \bibinfo{author}{R.~{Salvaterra}}, and \bibinfo{author}{B.~{Ciardi}},
  \bibinfo{journal}{\mnras} \bibinfo{volume}{\textbf{416}},
  \bibinfo{pages}{2760} (\bibinfo{date}{Oct. 2011}), \eprint{1106.1439}.
\vspace{-8pt}\bibitem{Gehrels+12-lobster}
\bibinfo{author}{N.~{Gehrels}}, \bibinfo{author}{S.~D. {Barthelmy}}, and
  \bibinfo{author}{J.~K. {Cannizzo}}, in \emph{IAU Symposium}
  (\bibinfo{date}{Apr. 2012}), \bibinfo{volume}{vol. 285 of \emph{IAU
  Symposium}}, \bibinfo{pages}{pp. 41--46}.

\end{thebibliography}

\end{document}